\documentclass[twocolumn]{aastex62} % use larger type; default would be 10pt

\usepackage{xspace}
\usepackage{graphicx}
\usepackage{natbib}
\usepackage{amsmath}
\usepackage{booktabs}
\usepackage{multirow}
\usepackage{afterpage}
\usepackage{rotating}

\newcommand{\hide}[1]{}

\newcommand{\ra}{\ensuremath{\alpha}\xspace}
\newcommand{\dec}{\ensuremath{\delta}\xspace}
\newcommand{\radec}{\ensuremath{(\ra,\dec [J2000])}\xspace}

\newcommand{\kms}{\ensuremath{\,{\rm km\,s^{-1}}}\xspace}

\newcommand{\microns}{\ensuremath{\,\mu{\rm m}}\xspace}

\newcommand{\pc}{\ensuremath{\,{\rm pc}}\xspace}
\newcommand{\kpc}{\ensuremath{\,{\rm kpc}}\xspace}
\newcommand{\K}{\ensuremath{\,{\rm K}}\xspace}

\newcommand{\ghz}{\ensuremath{\,{\rm GHz}}\xspace}

\newcommand{\degree}{\ensuremath{^\circ}\xspace}

\newcommand{\mjy}{\ensuremath{\,{\rm mJy}}\xspace}

\newcommand{\ev}{\ensuremath{\,{\rm eV}}\xspace}
\newcommand{\hi}{{\rm H\,{\footnotesize I}}\xspace}
\newcommand{\hii}{{\rm H\,{\footnotesize II}}\xspace}
\newcommand{\cii}{{\rm [C\,{\footnotesize II}]}\xspace}
\newcommand{\nii}{{\rm [N\,{\footnotesize II}]}\xspace}

\newcommand{\cor}{\ensuremath{^{\rm 13}{\rm CO}}\xspace}

\def\fsec{{}^{\prime\prime}\mskip-9mu.\,}  % arc seconds over a period 
\def\fd{{}^{\circ}\mskip-9mu.\,}          % degrees over a period     
\shorttitle{\cii\ 158\,\microns emission in S235}
\shortauthors{Anderson et al.}
\begin{document}

\title{The Origin of \cii\ 158\,\micron\ Emission toward the \hii\ Region Complex S235}

\author[0000-0001-8800-1793]{L.~D.~Anderson}
\affiliation{Department of Physics and Astronomy, West Virginia University, Morgantown WV 26506}
\affiliation{Adjunct Astronomer at the Green Bank Observatory, P.O. Box 2, Green Bank WV 24944}
\affiliation{Center for Gravitational Waves and Cosmology, West Virginia University, Chestnut Ridge Research Building, Morgantown, WV 26505}

\author[0000-0002-1397-546X]{Z.~Makai}
\affiliation{Department of Physics and Astronomy, West Virginia University, Morgantown WV 26506}

\author[0000-0001-8061-216X]{M.~Luisi}
\affiliation{Department of Physics and Astronomy, West Virginia University, Morgantown WV 26506}
\affiliation{Center for Gravitational Waves and Cosmology, West Virginia University, Chestnut Ridge Research Building, Morgantown, WV 26505}

\author[0000-0002-5306-4089]{M.~Andersen}
\affiliation{Gemini South, Casilla 603, La Serena, Chile 0000-0002-5306-4089}

\author{D.~Russeil}
\affiliation{Aix Marseille University, CNRS, CNES, LAM, Marseille, France 13388}

\author[0000-0002-9431-6297]{M.~R.~Samal}
\affiliation{Physical Research Laboratory, Navrangpura, Ahmedabad, Gujarat 380009, India}

\author[0000-0003-3485-6678]{N.~Schneider}
\affiliation{I. Physikalisches Institut der Universit{\"a}t zu K{\"o}ln, Z{\"u}lpicher Stra$\beta$e 77, 50937, K{\"o}ln, Germany}

\author[0000-0001-6172-3403]{P.~Tremblin}
\affiliation{CEA-Saclay, Gif-sur-Yvette, France 91191}

\author[0000-0001-9509-7316]{A.~Zavagno}
\affiliation{Aix Marseille University, CNRS, CNES, LAM, Marseille, France 13388}

%\author{P.~Hennebelle}
%\affiliation{CEA-Saclay, Gif-sur-Yvette, France 91191}

\author[0000-0003-4338-9055]{M.~S.~Kirsanova}
\affiliation{Institute of Astronomy of the Russian Academy of Sciences, Moscow, Russia 119017}
\affiliation{Ural Federal University, Astronomical Observatory, Lenin 51, Ekaterinburg, Russia, 620083}

\author[0000-0002-8351-3877]{V.~Ossenkopf-Okada}
\affiliation{I. Physikalisches Institut der Universit{\"a}t zu K{\"o}ln, Z{\"u}lpicher Stra$\beta$e 77, 50937, K{\"o}ln, Germany}

\author{A.~M.~Sobolev}
\affiliation{Ural Federal University, Astronomical Observatory, Lenin 51, Ekaterinburg, Russia, 620083}

\correspondingauthor{L.D.~Anderson}
\email{loren.anderson@mail.wvu.edu}

\begin{abstract}
Although the $^2{\rm P}_{3/2}-^2\!{\rm P}_{1/2}$ transition of \cii\ at $\lambda\simeq\,158\,$\micron\ is known to be an excellent tracer of active star formation, we still do not have a complete understanding of where within star formation regions the emission 
originates.  Here, we use \textit{SOFIA} upGREAT observations of \cii\ emission toward the \hii region complex Sh2-235 (S235) to better understand in detail the origin of \cii emission. We complement these data with a fully-sampled Green Bank Telescope radio recombination line map tracing the ionized hydrogen gas. About half of the total \cii\ emission associated with S235 is spatially coincident with ionized hydrogen gas, although spectroscopic analysis shows little evidence that this emission is coming from the ionized hydrogen volume. Velocity-integrated \cii intensity is strongly correlated with \textit{WISE} 12\,\microns intensity across the entire complex, indicating that both trace ultra-violet radiation fields.  The 22\,\microns and radio continuum intensities are only correlated with \cii\ intensity in the ionized hydrogen portion of the S235 region and the correlations between the \cii\ and molecular 
gas tracers are poor across the region.
We find similar results for emission averaged over a sample of external galaxies, although the strength of the correlations is weaker.   Therefore, although many tracers are correlated with the strength of \cii\ emission, only {\it WISE} 12\,\micron\ emission is correlated on small-scales of the individual \hii\ region S235 and also has a decent correlation at the scale of entire galaxies.  Future studies of a larger sample of Galactic \hii regions would help to determine whether these results are truly representative. 
%Using the observed intensities, we found good correlations between \cii\ and dust 
%components traced by the $12$\microns and $22$\microns emission, 
%EXPLAIN!
%find that the \cii\ intensity increases faster than dust intensities. 

% For next paper?
% With the kinematics of the observed emission lines, we can distinguish between the ionized and the different 
%molecular gas components along the line of sight toward S235. 
%In agreement with previous studies, we find the main velocity at $\sim -20\,$km s$^{-1}$ 
%but we also find a velocity component at $\sim -12\,$km s$^{-1}$ which probably traces the diffuse molecular gas behind the star formation complex.
\end{abstract}


%-------------------------------------------------------------------------------------------
\section{Introduction}
\label{sec:intro}

% About origin of [CII]
The \cii\ line is one of the most important transitions in the interstellar medium (ISM).  This line arises from the 
$\mathrm{^{2}P_{3/2}-^{2}P_{1/2}}$ transition of ionized carbon at $\lambda \sim 158$\microns ($1.9\,$THz), at an equivalent temperature of 
$\Delta E / k_{\rm B} \simeq 91.2\,$K. Between 0.1 and $1\%$ of the total FIR-luminosity of galaxies is 
provided by this emission line \citep[][]{crawford1985, malhotra97,boselli2002}, a result that also holds for Galactic star formation regions \citep{stacey1991, schneider98}. We observe \cii\ emission from diffuse clouds, the warm ionized medium (WIM), 
the surface of molecular clouds, dense photodissociation regions (PDRs), and cold \hi clouds \citep{pineda2013,pabst2017}. Since carbon has a lower 
ionization potential than hydrogen ($11.3\,$eV vs. $13.6\,$eV), ionized carbon exists in a variety of environments, and can trace the 
H$^{+}$/H/H$_{2}$ transition layer.

%\cii\ and SFR
\cii emission is a good tracer of galactic star-formation rates (SFRs) in galaxies \citep{delooze2011}.
%found a tight correlation between the 
%\cii\ luminosities of $24$ normal star-forming galaxies and the SFR, with a dispersion of $0.27$ dex around the mean. 
%They concluded that the 
%origin of this strong correlation is either ongoing or future star formation. 
They suggested, however, that \cii\ emission cannot be used reliably as an SFR indicator for low-metallicity 
dwarf galaxies, and the scatter of the \cii/SFR relationship increases as the galactic metallicity decreases.   
% Previous correlations between \cii\ and CO
Since CO and \cii emission are both correlated with the SFR, the connection between CO emission and \cii\ intensities has been widely studied. \citet{crawford1985} showed a strong linear relationship between the intensities of \cii\ and $^{12}$CO within gas-rich galaxies. 
%Using different PDR models, where the gas density and radiation field strength were not fixed ($\mathrm{10^{2}\, cm^{-3} \leq n \leq 10^{6}\, cm^{-3}}$ 
%and $\mathrm{10\, \leq G_{0} \leq 10^{6}}$), and observations of both Galactic and extragalactic sources, 
\citet{wolfire1989} found 
a tight linear correlation between the intensities of \cii and CO in observations of both Galactic and extragalactic sources, suggesting a common origin of these lines. 
%However, in \hii regions illuminated 
%by FUV radiation fields, the emitting region of these two lines is slightly different, i.e. they arise from distinct depths of PDRs.

% GOT C+
Despite the strong \cii/SFR relationship, there is still some doubt about where exactly the \cii\ emission originates.  A detailed study by \citet{pabst2017} shows strong correlation between \cii\ and {\it Spitzer} 8.0\,\micron\ emission from PDRs.  Some additional information comes from the 
Galactic Observations of Terahertz C$^+$ \citep[``GOT C$^+$''][]{langer2011} survey, a \textit{Herschel} \citep{pilbratt2010} open time key project\footnote{More 
information can be found on the GOT C$^+$ web site: \url{https://irsa.ipac.caltech.edu/data/Herschel/GOT_Cplus/overview.html}}.
Using GOT C$^+$ data, \citet{pineda2013} 
found that about half of \cii emission ($\sim$47\%) is produced in regions of dense PDRs, 28\% in dark H$_2$ gas, 21\% in cold atomic gas, and just 4\% in ionized hydrogen gas.
The fraction of \cii\ emission originating from the 
ionized phases of the ISM varies widely, from $\sim 5\%$ to $50\%$ depending on the electron density and ionizing radiation strength \citep{abel2006}. \cii\ 
emission also arises in regions of diffuse neutral gas \citep{madden1993}.
%\footnote{A significant amount of molecular hydrogen located at the outer parts of the molecular clouds that is associated 
%with atomic and ionized carbon (C$^{0}$ and C$^{+}$) rather than with CO is also called as ``dark-'' or ``hidden-H$_{2}$'' \citep[e.g.,][]{langer2010,wolfire2010}.}. 
%They found smaller fractions of the observed \cii\ emission come from cold atomic ($\sim 21\%$) and ionized gas ($\sim 4\%$).  
GOT C$^+$ sparsely sampled the Galactic plane along 454 sight lines, in a variety of Galactic environments, but the lack of spatial information toward star formation regions makes their results difficult to generalize.
%in this way their results represent only some biased volume average while some intensity-weighted average is needed to interpret intensities integrated over whole galaxies.

% About Hii regions
Since they make the ultraviolet (UV) photons that create C$^+$, the locations of massive stars should be strongly correlated with the locations of intense \cii emission. 
The UV radiation from massive stars frequently creates ionized hydrogen, or ``\hii,'' regions.  
%The ultraviolet (UV), visible and 
%infrared (IR) spectra of \hii regions are rich in emission lines, including collisionally excited lines of metal ions and recombination 
%lines of H and He.  
Dust within the regions absorbs and scatters high-energy photons. This leads to dust grain heating and subsequent emission of thermal photons in the mid- and far-infrared \citep[MIR and FIR; e.g.,][]{jones04,relano16}. 
%Due to their larger mass, large dust grains emit at longer IR wavelengths than small grains \citep[e.g.,][]{draine07}. \citet{draine11} estimates that in a typical spiral galaxy, $\sim 1/3$ of the energy radiated by stars is absorbed by dust grains and re-emitted in the IR. 
Therefore, this process is sometimes referred to as ``photon-destruction" \citep[e.g.,][]{swamy67}. The resulting lack of available H-ionizing photons has been shown to reduce the size of ``dusty" \hii\ regions \citep{sarazin77}. 
Within an \hii\ region, radiation pressure from the central source accelerates the dust grains outwards \citep{draine11b, akimkin15, akimkin17}.
%, creating a dust cavity around the central source. 
%Since large dust grains ($\sim$0.1 $\mu$m) are accelerated more efficiently than small grains ($\sim$0.01 $\mu$m), the grain-size distribution changes such that the larger grains responsible for FIR emission are preferentially found at larger distances from the ionizing source than small grains \citep{ishiki18}. 

Outside the ionized hydrogen zone of \hii regions is a PDR, which is the boundary between the \hii region and the interstellar medium.
%Photons with energies $>11.3\,\ev$ can ionize carbon, and such photons are abundant in PDRs.
%The warm surfaces of molecular clouds exposed to interstellar radiation, including those surrounding \hii regions, are called photo-dissociation regions (PDRs)\footnote{In the 
%literature, PDR is also referred as photon-dominated region.}. 
%PDRs are regions where the chemical and physical conditions are dominated by 
%the radiation field. 
%Far-ultraviolet (FUV) photons with energies less than the ionization potential of hydrogen (13.6\,\ev) play an important role in the 
%thermal and chemical balance of gas in PDRs \citep{hollenbach1999}. \ldacom{give the general reference for PDR here }
% Layered structure
PDRs have a layered structure because interstellar dust shields species from far-UV (FUV) photons, and hence chemical stratifications are produced by 
the progressively weaker FUV-field \citep{ossenkopf2007}. The ``ionization front'' is the boundary of the \hii region, interior to which nearly all gas is ionized.  Beyond the ionization front, hydrogen is predominantly neutral but carbon may be mostly ionized due to photons with energies between 13.6\,\ev and 11.3\,\ev.
At the dissociation front, H$_{2}$ becomes the dominant species. Further from the ionizing source, where the material is 
more opaque to FUV-photons and the temperature is decreasing, ionized carbon recombines to produce atomic carbon and $^{12}$CO, creating a 
transition layer of C$^{+}$/C/CO \citep[cf.][their Figure~3]{hollenbach1999}.
Although the connection between \cii and PDRs is well-established, the origin and distribution of \cii\ emission toward individual \hii regions in the Milky Way has 
received relatively little study.  The few studies have have been done suggest that most of the \cii emission toward \hii\ regions arises from dense PDRs. 
%On Galactic scales, 
%using data from the \emph{Herschel} [C II] %Galactic plane survey. 
In a study of the Orion~B molecular cloud, \citet{pabst2017} report that nearly all \cii emission ($\sim$95\%) originates from the irradiated molecular cloud, with only a small ($\sim$5\%) contribution from the adjacent \hii\ region \citep[see also][]{pabst19}. Unlike \citet{pineda2013}, they do not make a clear distinction between PDRs and dark molecular gas.
%; theyand note that the bulk of the [CII] %emission arises from PDR surfaces. 
Their result is in rough agreement with \citet{goicoechea2015a} who found that 85\% of the \cii emission in the Orion molecular cloud 1 (OMC1) is produced on  the surface of the molecular cloud.  A smaller amount 
($\sim 15\%$) of the \cii\ emission comes from a gas component not associated with CO. 
\citet{goicoechea2018} also found support for \cii\ emission arising from dense PDR gas in OMC1. 
\citet{simon2012} observed the \hii region complex S106 with the {\it Stratospheric Observatory for Infrared Astronomy} 
\citep[{\it SOFIA};][]{young2012} and found 
%that the strongest \cii\ emission originates from the edge of the two clusters of water masers located in 
%the S106 \hii region. 
%They interpreted this result as a probable signature of evaporation or ablation of dense clumps due to wind and/or radiation from 
%the ionizing star. 
%\citet{simon2012} found that 
that part of the \cii emission comes from the ionized hydrogen region since the locations of \cii emission are similar to that of the cm continuum.  A more recent study of S106 with {\it SOFIA}, however, argued that the \cii\ emission is actually from the PDRs \citep{schneider18}.
\citet{graf2012} investigated the \cii\ emission toward NGC2024 in the Orion B complex with \textit{SOFIA} observations.
%of [$^{12}$C{\footnotesize II}] and [$^{13}$C{\footnotesize II}]. 
They concluded that the observed ionized carbon comes from a highly clumpy interface between the molecular cloud and the \hii region and shows a good 
spatial correlation with the $8$\microns continuum. %\citet{goicoechea2015a} observed OMC1 with the HIFI instrument 
%\citep{degraauw2010} on \textit{Herschel}  \citep{pilbratt2010}. They showed that the major contributor ($\sim 85\%$) of the \cii\ luminosity is the far 
%ultra-violet illuminated cloud and dense PDRs at the interface between the \hii region surrounding the Trapezium cluster and OMC1. 

Here, we present \textit{SOFIA} observations of the \cii\ $158$\microns line toward the massive star-forming complex Sh2-235, with the goal of understanding 
the origin of \cii emission. S235 is a rich complex, with three separate \hii\ regions and prominent PDRs.  It therefore has the environments associated with strong \cii emission.  We can thus use the results from S235 to provide context to the results from external galaxies.  We deal mainly with velocity-integrated \cii emission; most of the detailed kinematics of the region will be discussed in a forthcoming paper.
%-------------------------------------------------------------------------------------------
\begin{figure*}[!h]
   \centering
   \includegraphics[width=\textwidth]{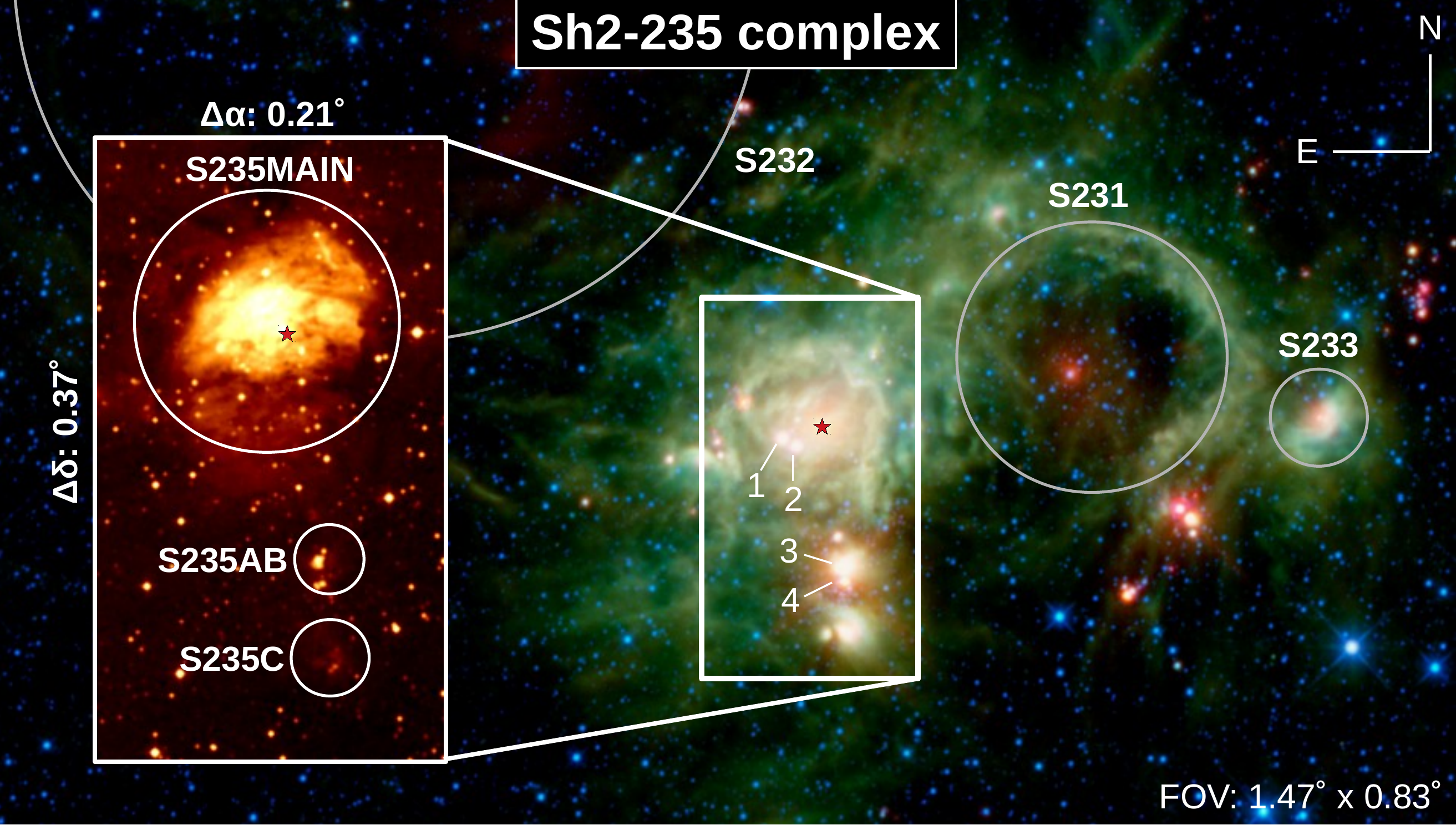}
   \caption{S235 star forming complex in a four-color \textit{WISE} image created by NASA/JPL-Caltech/{\it WISE} Team. Red, green, blue and cyan 
   correspond to infrared wavelengths of $22$\microns, $12$\microns, $4.6$\microns, and $3.4$\microns, respectively. 
   The total field of view (FOV) is $1\fd47 \times 0\fd83$, oriented in RA and Decl., and centered near $(\alpha, \delta ({\rm J2000}))~=$~(05:41:16,~+35:50:52).  All Sharpless \hii\ regions in the field are identified.  Numbers 1--4 indicate infrared IRS sources from \citet{evans1981} and the red star indicates the location of the ionizing source of S235.
   The inset shows DSS red data showing emission from ionized gas in the area we investigate 
   in this paper, which has a FOV of $0\fd21 \times 0\fd37$.  Circles denote the main S235 \hii region 
   (S235MAIN, upper circle) that contains the inner ionized hydrogen region and the surrounding PDRs, and the two smaller 
   \hii regions (S235AB and S235C, two lower circles).  S235AB contains the S235A \hii region and the S235B reflection nebula; these we treat as one star formation region. \label{fig:s235complex}}
\end{figure*}

\section{The Sh2-235 star formation complex}
\label{sec:source}

The Sh2-235 star forming complex \citep[hereafter ``S235;''][]{sharpless1959}
%(SIMBAD\footnote{\url{http://cdsweb.u-strasbg.fr/}} coordinates: %$\mathrm{\alpha_{2000}=5^{h}41^{m}6^{s}}$, 
%$\mathrm{\delta_{2000}=35^{\circ}50^{\}24^{\prime\prime}}$; $l=173\fd6198$, %$b=+2\fd8100$), 
is located toward the Galactic anti-center.  The distance to a water maser in the complex is
$1.56^{+0.09}_{-0.08}$\,\kpc \citep{burns15}.  This is roughly consistent with the recent GAIA DR2 parallax of the ionizing source of S235
BD+351201 \citep{brown18}, which corresponds to a distance of
$1.73 \pm 0.13$\,\kpc.  Here, we adopt a distance of 1.6\,\kpc for the region.
 Since its first appearance in scientific literature \citep{minkowski1946}, it has 
been extensively studied from the optical through radio regimes 
\citep[e.g.,][]{evans1981,nordh1984,allen2005,kirsanova2008,boley2009,camargo2011,kirsanova2014,bieging2016,dewangan2017}.

In this paper, we study three main regions of the S235 complex (Figure \ref{fig:s235complex}). The main S235 region \hii\ region (Sh2-235) which we call ``S235MAIN'' is ionized by an O9.5V star BD+351201 \citep{georgelin1973}. Active star formation is continuing in S235MAIN, as it hosts more than $200$ young stellar 
objects \citep[YSOs;][]{dewangan2011}.  S235MAIN hosts the IR sources IRS1 and IRS2 \citep{evans1981}, both of which are created by B-type stars.

There are two smaller \hii regions to the south of S235MAIN: S235A and S235C \citep{israel1978}.  
S235A is located $\sim 10^{\prime}$ ($\sim\!4\,\pc$) south of S235MAIN, and is also known as IRS3 \citep{evans1981} and  radio source G173.72+2.70 \citep{israel1978}. S235A has 
methanol and water masers \citep[see][and references therein]{chavarria2014}, and is known as an expanding \hii region ionized by stars of main sequence spectral types between B0 and O9.5 \citep[e.g.,][]{felli1997}.  
Near to S235A is the reflection nebula S235B (IRS4) caused by a B-type star \citep{boley2009}.  We refer to S235A and S235B combined as the star forming region ``S235AB'' since they are not separated at the angular resolution of our data.  
%\citet{evans1981} found that the molecular gas from S235AB and S235C is redshifted \ldacom{is this really redshifted?} relative 
%to that of S235MAIN, with velocity components of $-17\,\mathrm{km\,s^{-1}}$ and $-20\,\mathrm{km\,s^{-1}}$, respectively. 
S235C is located 
$\sim\!5\arcmin$ ($\sim\!2\,\pc$) south of S235A and is ionized by a B0.5 star \citep[see Table~1 in][]{bieging2016}. \citet{dewangan2017} showed that these smaller star formation regions are interacting with the surrounding molecular clouds, and that star formation may be triggered by the expansion of the \hii region \citep[e.g.,][]{kirsanova2008,camargo2011}. \citet{kirsanova2014} suggested that the star formation 
in S235AB is not related to the expansion of S235MAIN.

The S235 complex is an ideal target for studies of \cii emission.  It is nearby, bright, and has been the focus of many previous studies.  It also contains three separate \hii regions, two of which are compact.  The size of \hii regions depends on their age and the intensity of ionizing radiation. Therefore, we can examine differences in \cii\ emission for \hii\ regions of different ages and ionizing radiation fields.
%(S235A and S235C are thought to be younger and are ionized by less massive stars).
%-------------------------------------------------------------------------------------------
\section{Data}
\label{sec:obs}

% SOFIA
\subsection{{\it SOFIA} \cii\ and \nii\ data}
\label{subsec:cii_obs}

We observed \cii\ and \nii\ emission toward the S235 complex in {\it SOFIA} Cycles~4 and 5 in November 2016 and February 2017 using the \textit{SOFIA} upGREAT instrument \citep{risacher2016}.  upGREAT is an enhanced version of the German Receiver for Astronomy at Terahertz Frequencies \citep[GREAT;][]{heyminck2012}.  We used the upGREAT  LFA
channel (a 7 pixel array in 2016 and a 2$\times$7 pixel array in 2017)
to tune to \cii\ and the L1 (single pixel) channel to tune to the ${\rm ^3P_1 - ^3P_0}$ \nii\ 205\,\micron\ (1.46\,THz) line.  The total observing time for both cycles was 3.5\,hours.  We observed in total power on-the-fly (OTF) mapping mode and mapped a total area of 
%S235MAIN and the smaller regions, S235AB and 
%S235C were observed separately resulting two submaps. These submaps %were later combined into one coherent map with size of 
$0\fd21 \times 0\fd37$ 
($\alpha \times \delta$), centered at \radec = (5h41m02.5s, $35\degree51$m57s). 
We employed a fast mapping mode for S235MAIN, which resulted in an undersampled map for \nii, and a slow mapping mode for S235AB and S235C, which resulted in a fully-sampled \nii\ map.
The spatial resolution of the \cii\ data is $14.8\arcsec$, the velocity resolution is 0.385\,\kms, and the full velocity range is $-60$ to $20\,\kms$.  The spatial resolution of the \nii\ data is $20.2\arcsec$, the velocity resolution is 0.500\,\kms, and the full velocity range is $-60$ to $20\,\kms$.

The final data cubes provided by the SOFIA Science Center (in units of main beam temperature $T_{\rm MB}$) were processed using the Grenoble Image and Line Data Analysis Software (GILDAS)\footnote{\url{http://iram.fr/IRAMFR/GILDAS/} \citep{pety2005}}. The data 
were first scaled to $\mathrm{T_{A}^{\ast}}$, the antenna temperature corrected for atmospheric opacity, using a forward efficiency of 0.97. Antenna temperature values were converted to main beam temperatures using the main beam efficiency of $\eta_{\rm MB} = 0.69$. 
%The gridded \cii\ map was created using $1/\sigma^2_{rms}$ weighting, where $\sigma_{rms}$ is derived 
%from the line-free portions of the spectra. 
If the rms of an individual spectrum was higher than two times the radiometer noise, the spectrum was ignored. First order (if 
rms $<$ radiometer noise) or third order (if rms $<$ 2 $\times$ radiometer noise) baselines were removed from all spectra.

\begin{figure}
   \centering
   \includegraphics[width=0.45\textwidth]{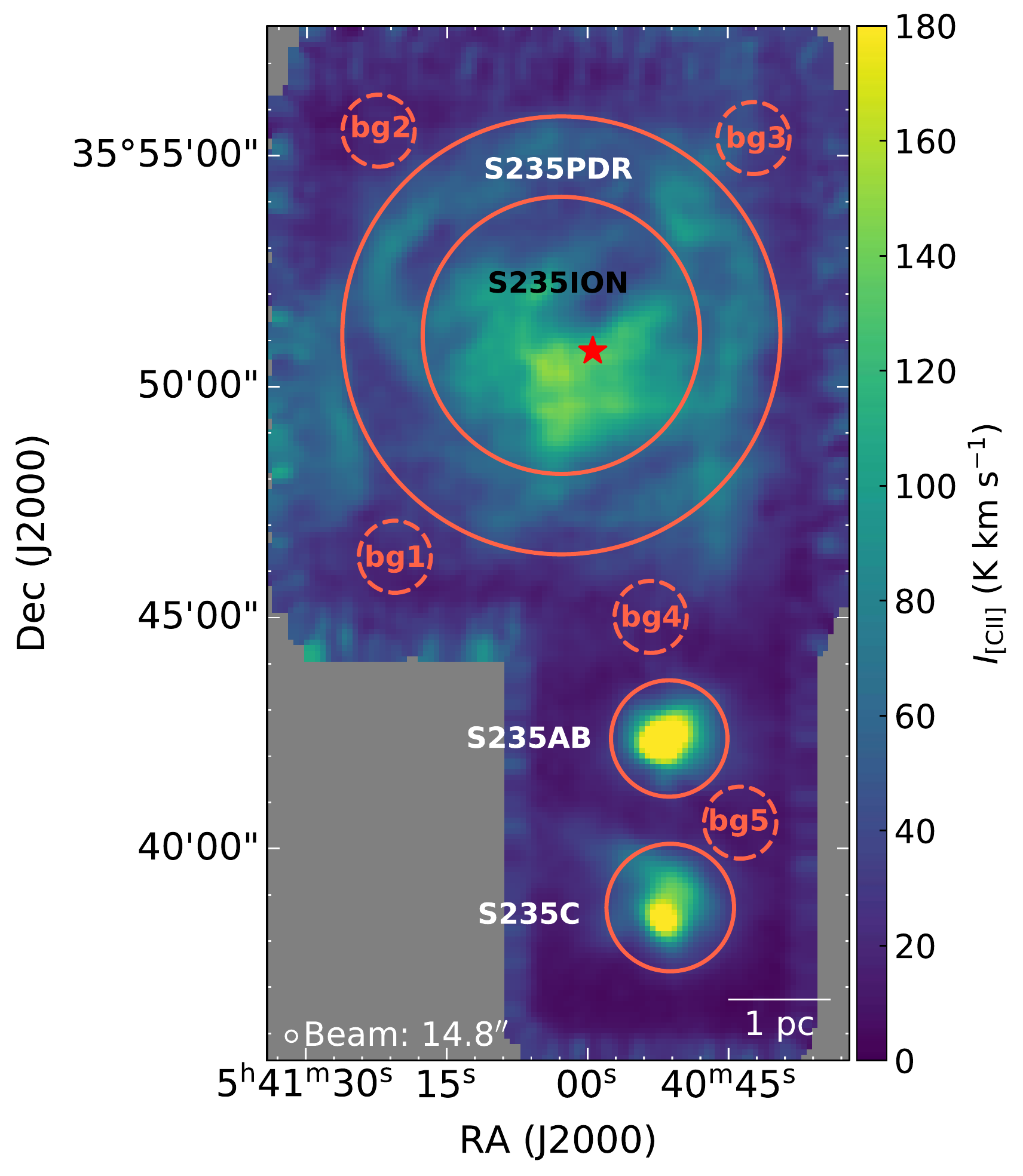}
   \caption{Integrated ($-29~{\rm to} -12$\,\kms) intensity map of {\it SOFIA} \cii\  $158$\microns data, smoothed by a $3\times3$ pixel Gaussian kernel (approximately the beam size).  The red circles enclose regions of interest, with dashed 
   circles denoting background regions. The filled red star marks the position of the ionizing source BD$+351201$. The smaller \hii regions show strong and compact 
   \cii\ emission.  \label{fig:cii_m0map}}
\end{figure}

Here, we frequently use the integrated \cii\ intensity, ``moment~0,'' from the velocity range $-29$ to $-12$\,\kms (Figure~\ref{fig:cii_m0map}).
All significant \cii\ emission associated with S235 is found within this velocity range (see Figure~\ref{fig:av_specs}). 
This figure shows strong \cii\ emission from the PDRs surrounding S235, but the most intense emission in the field is found toward S235AB and S235C. 
%Some \cii\ emission is coincident with the infrared sources of IRS1 and IRS2  \citep[e.g.,][]{evans1981}. 
%The observed \cii\ intensity
%in the PDR is inhomogeneously distributed. In the southern and southeastern portions of the PDR, there is significantly less ionized carbon emission than in the rest of the PDR area where \cii\ emission shows filamentary structures. 

\nii\ is only weakly detected when averaged over the entire ionized hydrogen region.  Because there is little spatial information on the distribution of \nii, we limit analyses using these data.

\subsubsection{Regions of interest}
\label{subsec:regions}

Using the \cii\ moment~0 map as a guide, we determine regions of interest in the S235 field. For S235MAIN, we define one region of interest spatially 
coincident with the ionized hydrogen gas (``S235ION''), and an annular region surrounding the ionized hydrogen gas that contains most of the plane-of-sky PDR emission (``S235PDR'').  We 
define regions of interest for the two smaller \hii regions located to the south of S235MAIN (``S235AB'' and ``S235C'').  We also sample diffuse emission in the 
field using five smaller background regions (``bg1--5''). We list the parameters of these regions of interest in  Table~\ref{tab:regions}.

\begin{deluxetable}{lccc}
\tablecaption{Parameters of the regions of interest \label{tab:regions}}
\tablehead{
\colhead{Region} & \colhead{$\alpha_{J2000}$} & \colhead{$\delta_{J2000}$} & \colhead{Radius}\\
\colhead{} & \colhead{h:m:s} & \colhead{d:m:s} & \colhead{$\arcsec$}}
\decimals
\startdata
S235PDR & 5:41:02.76 & 35:51:06.48 & $284.4$\tablenotemark{a} \\
S235ION & 5:41:02.76 & 35:51:06.48 & $187.2$ \\
S235AB  & 5:40:51.24 & 35:42:23.04 & $75.6$ \\
S235C   & 5:40:51.24 & 35:38:42.36 & $82.8$ \\
bg1     & 5:41:20.40 & 35:46:19.20 & $46.8$ \\
bg2     & 5:41:22.20 & 35:55:32.52 & $46.8$ \\
bg3     & 5:40:42.24 & 35:55:22.80 & $46.8$ \\
bg4     & 5:40:53.40 & 35:45:00.72 & $46.8$ \\
bg5     & 5:40:43.68 & 35:40:33.60 & $46.8$ \\
\enddata
\tablenotetext{a}{The S235PDR region is annular, with its inner radius equal to the radius of S235ION.  The value listed here represents the outer radius.}
\end{deluxetable}

\subsection{Green Bank Telescope Radio Recombination Line Maps \label{sec:rrl}}
To understand the distribution and velocity structure of the ionized gas, we create a fully-sampled map in 6\,\ghz\ radio recombination line (RRL) emission using the Green Bank Telescope (GBT).  The region has been observed previously in H-$\alpha$ by \citet{lafon83}, who found a strong velocity gradient in the ionized hydrogen gas; the velocity is $\sim -15\,\kms$ in the south-east of the region (near our ``bg1'' region) and $\sim -25\,\kms$ in the north-west (near our ``bg3'' region).  This strong gradient is unusual for an \hii\ region and may be due in part to absorption of H-$\alpha$ emission or to a bulk ``champagne flow'' of ionized gas.

We follow the same observing setup as in \citet{anderson18} and \citet{luisi18}, tuning to 64 different frequencies
at two polarizations within the 4$-$8 GHz receiver bandpass. Of
these 64 tunings, 22 are Hn$\alpha$ transitions from $n = 95$ to $n = 117$.  Carbon and helium RRLs fall within the same bandpass as that of hydrogen, shifted by $-149.56$\,\kms\ and $-122.15$\,\kms\ from hydrogen, respectively.  Carbon RRLs are thought to arise from PDRs surrounding \hii\ regions, whereas helium RRLs come from the ionized hydrogen region.
%We also tune to the \form\ $1_{10}-1_{11}$ transition at 4.829\,\ghz.  
%25 are Hn$\beta$ lines, 8 are Hn$\gamma$ lines, and 9 are
%molecular lines. We use only the Hn$\alpha$ transitions for
%further analysis here; 

There are 15 usable Hn$\alpha$ lines after removing those spoiled by radio frequency interference (RFI) or instrumental effects.  Over the frequencies of the usable Hn$\alpha$ lines, the beam size ranges from $98\arcsec$ to $183\arcsec$, with an average of $141\arcsec$.
We calibrate the intensity scale
of our spectra using noise diodes fired during data acquisition and assume a main beam efficiency of 0.94.
%By observing the primary flux calibrator 3C286
%we confirmed that this calibrates the data to within
%10%. In addition, we periodically observed the Galactic
%H II regions W3 and W43 as test sources to verify the
%RRL intensity calibration scale and found agreement at
%the 10% level with the results of Balser et al. (2011).
%The GBT C-band gain to convert from antenna temperature
%to flux density is $2\,\K\jy^{-1}$ at these frequencies \citep{ghigo01}.

The map is $0.8\degree \times 0.7\degree$, centered at \radec = (5h40m42s, $+35\degree$48m52s).  
%We perform our observations in ``On-the-Fly" (OTF) mode, slewing at 54\arcsec\,s$^{-1}$, and sampling every 0.38\,s, or 20\arcsec. Rows are spaced every 40\arcsec, which is the Nyquist rate at the highest frequency.  We observe a reference position $\sim 3$\degree off the map center every 16 rows, or $\sim 20$\,minutes, such that the time between any on-source scan and a reference scan is less than 10 minutes.  
We create four complete maps, two in RA and two in Dec., to mitigate in-scan artifacts, and average all together.

%The advantage of observing a large number of hydrogen-$\alpha$ transitions simultaneously is that 
We average all lines at a given position to make one sensitive spectrum \citep{balser06}, a technique that is well-understood \citep{anderson11,liu13,alves15,luisi18}. After removing transient RFI, we remove a polynomial baseline for each transition and shift the spectra so that they are aligned in velocity \citep{balser06}. We re-grid the 15 good Hn$\alpha$ lines to a velocity resolution of 0.5\kms and a spatial resolution of 1\arcmin. We then average the individual maps using a weighting factor of $t_{\rm intg} T^{-2}_{\rm sys}$ where $t_{\rm intg}$ is the integration time and $T_{\rm sys}$ is the system temperature.

\begin{figure}
    \centering
    \includegraphics[width=3.3in]{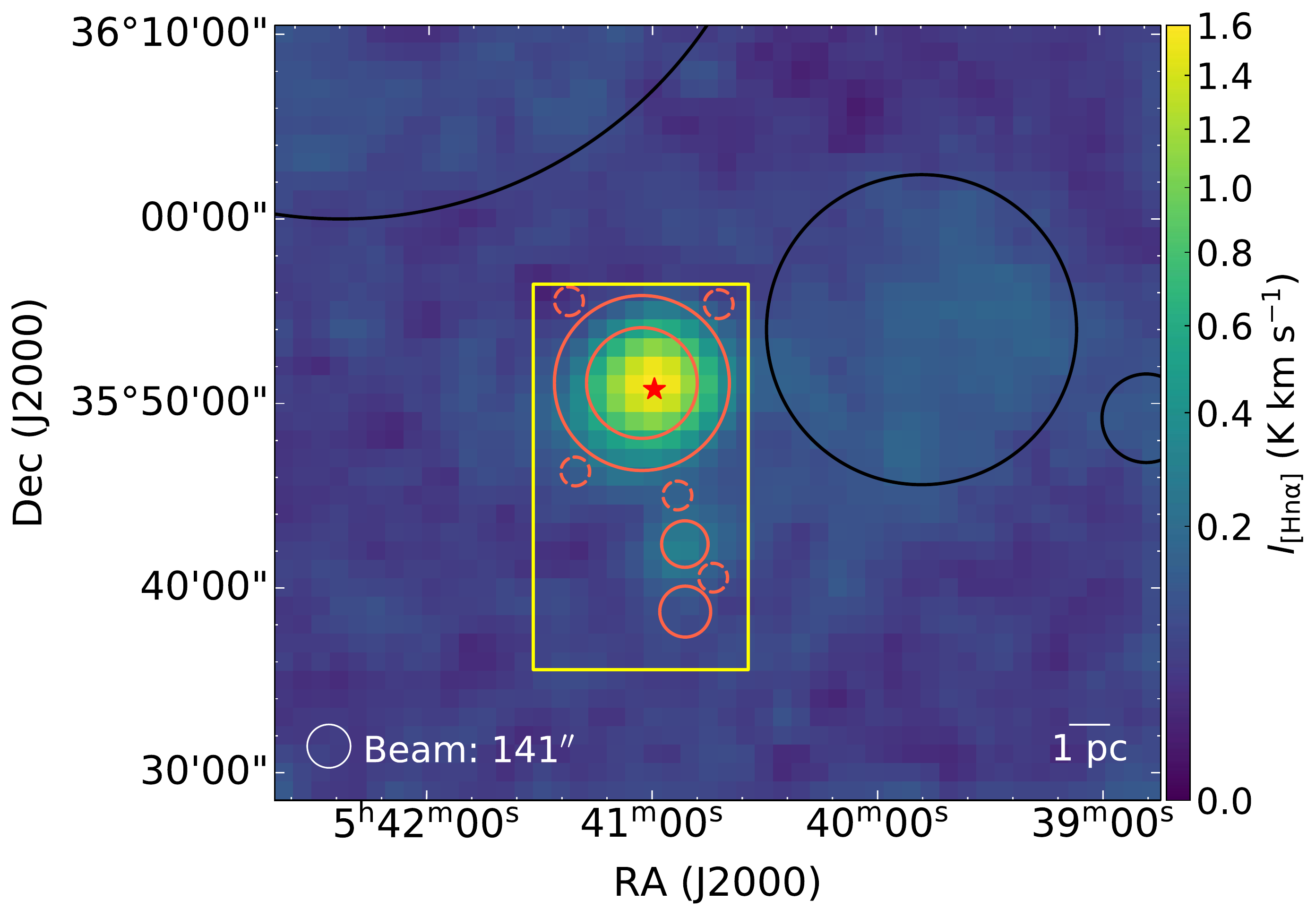}
    \caption{Integrated ($-29~{\rm to} -12$\,\kms) intensity map of GBT RRL data.  The yellow rectangle shows the extent of the {\it SOFIA} \cii\ data (Figure~\ref{fig:cii_m0map}).  As in Figure~\ref{fig:cii_m0map}, the red circles enclose regions of interest, with dashed 
   circles denoting background regions. Black circles enclose other \hii\ regions in the field (cf. Figure~\ref{fig:s235complex}).  The filled red star marks the position of the ionizing source BD$+351201$.  The strongest RRL emission comes from S235MAIN, although S235AB is also bright and there is detectable emission from the other regions in the field.}
    \label{fig:rrl}
\end{figure}

We show the integrated intensity moment zero hydrogen RRL map in Figure~\ref{fig:rrl}.  The emission is strongest for S235MAIN and peaked at the location of the ionizing source.  
%Although it appears there is emission from the location of the PDR, this is likely just emission from within the S235ION region of interest that is smoothed into the S235PDR region by the rather coarse angular resolution.  
S235AB also shows detected emission, but S235C does not.  The other \hii\ regions in the field have some detected RRL emission.  We show the peak line velocity derived from Gaussian fits in Figure~\ref{fig:av_specs}, although due to the relatively poor spatial resolution we do not show the fit for S235PDR.  For S235ION, the fit is to the average spectrum from our RRL map integrated over the entire region of interest.  For S235AB and S235C, we use the pointed RRL results from \citet{anderson15}, because the signal to noise in the RRL map is rather poor and the observational setup in \citet{anderson15} is the same as ours used here.

%We show the integrated intensity from two velocity ranges of \form\ in Figure~\ref{fig:h2co_m0map}.  These two velocity ranges together contain all significant \form\ emission (see Section~\ref{sec:velocity}).  The emission from \form\ and CO have a very good morphological correlation.

%\begin{figure}
%    \centering
%    \includegraphics[width=0.45\textwidth]{H2CO_m0_smap2_-23--12.eps}\\
%    \includegraphics[width=0.45\textwidth]{H2CO_m0_smap2_-12--4.eps}
%    \caption{Integrated intensity maps of \form\ emission over $-23~{\rm to} -12$\,\kms (top)  and $-12~{\rm to} -4$\,\kms (bottom).  As in Figure~\ref{fig:cii_m0map}, the red circles enclose regions of interest, with dashed circles denoting background regions. Black circles enclose other \hii\ regions in the field (cf. Figure~\ref{fig:s235complex}).  The filled red star marks the position of the ionizing source BD$+351201$.  White contours are of integrated $^{12}$CO 2$-$1 intensity (see Section~\ref{sec:co}), integrated over the same velocity ranges. There is excellent agreement between H$_2$CO and CO emission. \label{fig:h2co_m0map}}
%\end{figure}

% WISE 12 micron data
\subsection{Ancillary data}
We also use {\it WISE} MIR, 1.4\,\ghz\ NVSS radio continuum, and CO rotational transitions to investigate the S235 region (Figure~\ref{fig:int_maps}). For later comparisons, we regrid all ancillary data to $10\arcsec$ pixels, and do the same for the {\it SOFIA} \cii\ data.  The CO data in panels (d), (e), and (f) of this figure are of integrated intensity, integrated over $-29$ to $-12$\,\kms\ as we did for \cii\ (Figure~\ref{fig:cii_m0map}).

\subsubsection{{\it WISE} 12\microns and 22\microns data}
\label{subsec:dust_data}

\textit{WISE} \citep{wright2010} mapped the entire sky at four wavelengths: $3.4$\microns, $4.6$\microns, 
$12$\microns and $22$\microns. The angular resolutions are $6\fsec1$, $6\fsec4$, $6\fsec5$ and $12^{\prime\prime}$ with $5\sigma$ point-source
sensitivities of $0.08$ mJy, $0.11$ mJy, $1$ mJy and $6$ mJy at the native pixel scales, respectively. We use here the $12$\microns and $22$\microns bands. The $12$\microns band is sensitive to the emission from polycyclic aromatic hydrocarbon (PAH) features at 
$11.2$\microns, $12.7$\microns, and $16.4$\microns \citep[e.g.,][]{roser2015,tielens2008}.  Due to high expected optical depth in this wavelength range, the {\it WISE} 12\,\micron\ band is likely a {\it surface}, rather than volume tracer, in contrast to the some of the other ancillary data used here.
The W4 $22$\microns bandpass is sensitive to 
stochastically-heated very small grains (VSGs) within the \hii region plasma, and also to dust grains within the PDRs \citep[PAHs are prominent 
contributors of $24$\microns emission; see][]{robitaille2012}.

The {\it WISE} data have ``Digital Number'' (DN) units and we used the DN-to-mJy conversion factors of $2.9045 \times 10^{-3}$ and $5.2269 \times 10^{-2}$ for the $12$\microns 
and $22$\microns data, respectively\footnote{For more information, see \url{http://wise2.ipac.caltech.edu/docs/release/prelim/expsup/sec2_3f.html}}. We 
use the color-correction of \citet{wright2010}, assuming a spectral index of $\alpha=0$. This correction increases the 12 and 22\,\micron\ flux densities by 
$9.1\%$ and $1.0\%$, respectively.

The $12$\microns data show the same spatial distribution as \cii throughout the field, including for the smaller \hii regions to the south (see panel (a) in Figure~\ref{fig:int_maps}). %Within this ionized region, the major contributor to the emission ($\sim 52\%$) seems to be the ionizing star and the two small dense clumps \ldacom{is dense clumps accurate?}, 
%reported by \citet{kirsanova2008}. 
%This suggests that the PAHs are not destroyed yet by the young embedded infrared sources (IRS1 and IRS2). 
In the PDR of S235, 
the $12$\microns emission follows the filamentary distribution of the ionized carbon. 
%The average flux density is higher in S235ION than in S235PDR by $\sim 27\%$. 
%The S235C possesses $\sim 5\%$ higher total flux density than S235AB, while the mean flux density is $\sim 16\%$ higher in S235AB.
The \textit{WISE} $22$\microns emission (panel (b) in Figure~\ref{fig:int_maps}) is centrally concentrated for S235, with comparatively faint emission in the PDRs.  The 22\,\micron\ emission for S235AB and S235C is compact.
%The main contributors to the $22$\microns emission are PAHs and VSGs. %However, in our case the main source 
%of this dust emission is likely the PAHs that is not yet destroyed or blown away by the central star and the embedded IR sources. 
%In the surrounding S235PDR region, the 
%$22$\microns emission generally comes from a young star cluster, called ``S235East2'' in e.g. \citet{kirsanova2014}, reported first by \citet{kumar2006}. 
%The average 
%$22$\microns flux density is stronger in S235ION than in S235PDR by $\sim 25\%$. 
%The 22\,\microns flux density of S235AB is $\sim\!25\%$ higher than that of S235C.

\subsubsection{NVSS 1.4\,\ghz\ continuum data}
\label{subsec:rad_cont}

The ionized gas of \hii regions emits in radio continuum due to free-free (Bremsstrahlung) radiation. We obtained $1.4\,$GHz radio continuum data 
from the NRAO VLA Sky Survey\footnote{\url{https://www.cv.nrao.edu/nvss/}} \citep[NVSS;][]{condon1998}.
%The entire survey covers $82\%$ of the celestial 
%sphere (i.e., J2000 $\delta \geq -40^{\circ}$) and detected almost 2 %million of discrete sources.  
The NVSS FWHM synthesized beam size is $\sim45\arcsec$ 
 and its rms uncertainty is $\approx 0.45\,$mJy/beam. During the regridding process, we convert the NVSS data from their native units of Jy\,beam$^{-1}$ to mJy\,pixel$^{-1}$, the same as we use for our {\it WISE} data.
 
Similar to the RRL and 22\,\micron\ band data, the radio continuum emission (panel (c) in Figure~\ref{fig:int_maps}) is concentrated around the ionizing source.  That the \cii\ and radio continuum 
distributions differ shows that we observe two different ionized phases of the ISM along the line of sight.

 % Auxiliary data, molecular lines
\subsubsection{CO data\label{sec:co}}
\label{subsec:aux_data}

The ancillary molecular line data used here of $^{12}\mathrm{CO}$ $2-1$, $^{12}\mathrm{CO}$ $3-2$, and $^{13}\mathrm{CO}$ $2-1$
are from \citet{bieging2016}. The observations were made with the Heinrich Hertz Submillimeter Telescope\footnote{For more information, see: 
\url{http://aro.as.arizona.edu/smt_docs/smt_telescope_specs.htm}} in March to April, $2010$. 
%The receiver for the 
% $^{12}$CO$J=2-1$ observations was the dual-polarization ALMA Band 6 protoype sideband-separating mixer system \citep{ediss2004}. 
The two $J=2-1$ lines were observed simultaneously. The 
$\mathrm{^{12}CO}$ emission line 
%at $230.538\,\mathrm{GHz}$\footnote{The frequencies of the transitions from the Cologne Database for Molecular 
%Spectroscopy \citep[CDMS;][]{mueller2001}: \url{http://www.astro.uni-koeln.de/cdms}} 
was observed in the upper sideband, while the $\mathrm{^{13}CO}$ line 
at $220.399\,\mathrm{GHz}$ was observed in the lower sideband.  %The OTF-maps averaged the two orthogonal polarizations. 
The maps cover $\sim 1\fd2 \times 0\fd8$ ($\alpha, \delta$), have a velocity 
resolution of $0.3\,$km s$^{-1}$, and have a spatial resolution of $38\arcsec$.

The $^{12}\mathrm{CO}$ $3-2$ line was observed in April, 2014, also with the 
Heinrich Hertz Submillimeter Telescope, using the multi-pixel focal plane array of superconducting 
mixers \citep[``SuperCam'';][]{kloosterman2012}. The FWHM beam width is $24\arcsec$. 
%for the $J=3-2$ transition at $345.796\,\mathrm{GHz}$.
The total area covered 
is $\sim 0\fd3 \times 1\fd0$ ($\alpha, \delta$) with $0.23\,$\kms velocity resolution.

\citet{kirsanova2008} distinguished three main velocity components in the S235MAIN region using $^{13}$CO $1-0$ and CS $2-1$ observations: a ``red'' 
component ($-18 < \mathrm{V_{lsr}} < -15\,$km s$^{-1}$), a ``central'' component ($-21 < \mathrm{V_{lsr}} < -18\,$km s$^{-1}$) and a ``blue'' component 
($-25 < \mathrm{V_{lsr}} < -21\,$km s$^{-1}$). With additional observations of NH$_3$, \citet{kirsanova2014} confirmed the existence of these different velocity 
components.  Using $^{12}$CO $1-0$ data, 
\citet{dewangan2017} reported that the emission toward S235MAIN peaks in the velocity range $-22$ to $-20$, while the smaller \hii regions, S235A (``S235AB'') and S235C, 
are at $\sim -17\,$\kms.

The $^{12}$CO $2-1$ data have a broad velocity component (centered at $\sim -11.5\,$km s$^{-1}$) that is 
well-separated from the aforementioned velocity components, but is absent in the other molecular and \cii\ line data (see Figure~\ref{fig:av_specs}). 
This redshifted component is possibly related to more diffuse
gas not associated with S235; the component is also prominent in EBHIS \hi emission \citep{winkel2016}.
As this component is redshifted relative 
to the overall velocity of the star-forming region S235, we likely observe the diffuse molecular gas component foreground or background to the star formation 
complex. 
%Further studies of the \cii\ kinematics of S235 will be discussed in a forthcoming paper.

\begin{figure*}
   \centering
\includegraphics[height=3.7in]{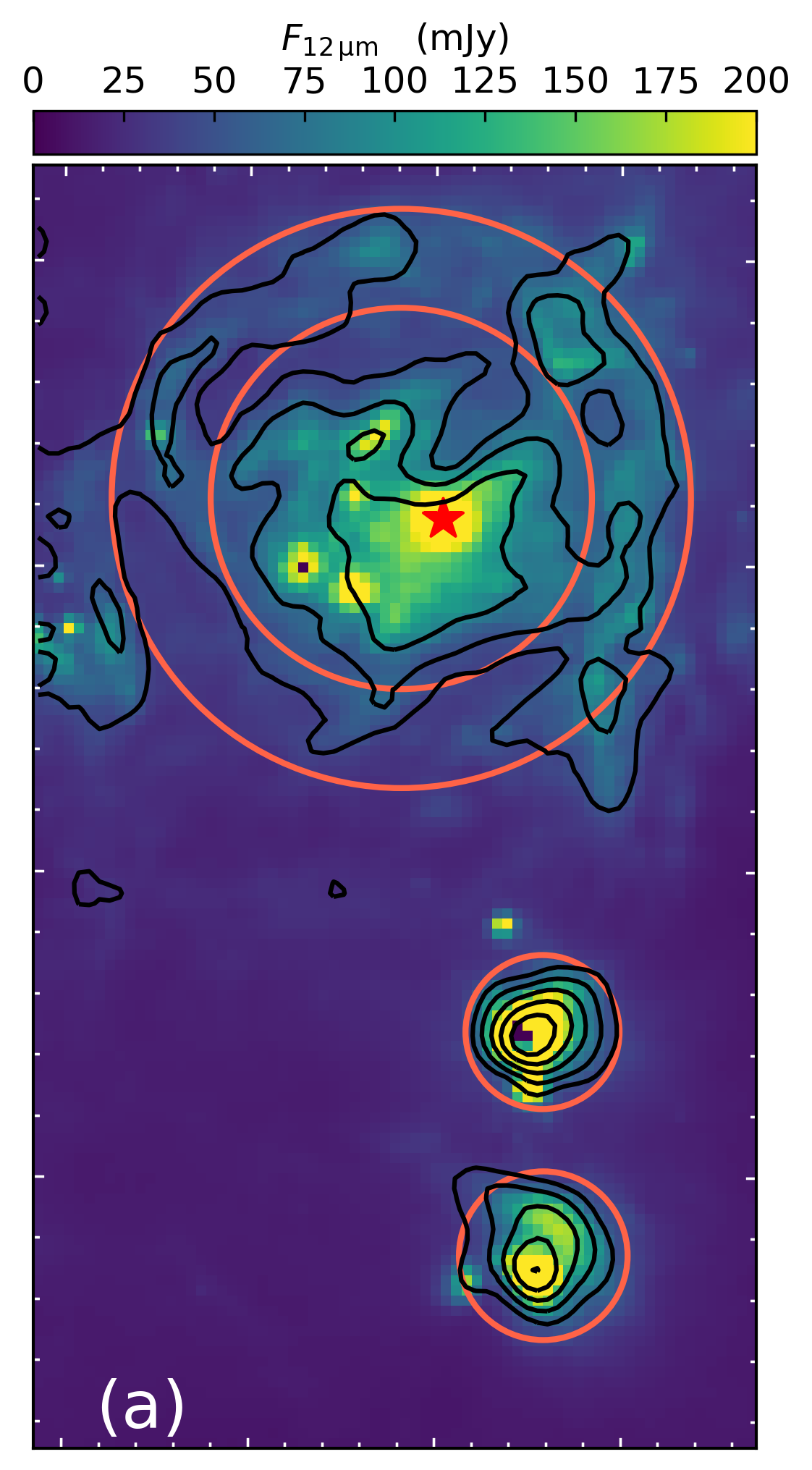}
   \includegraphics[height=3.7in]{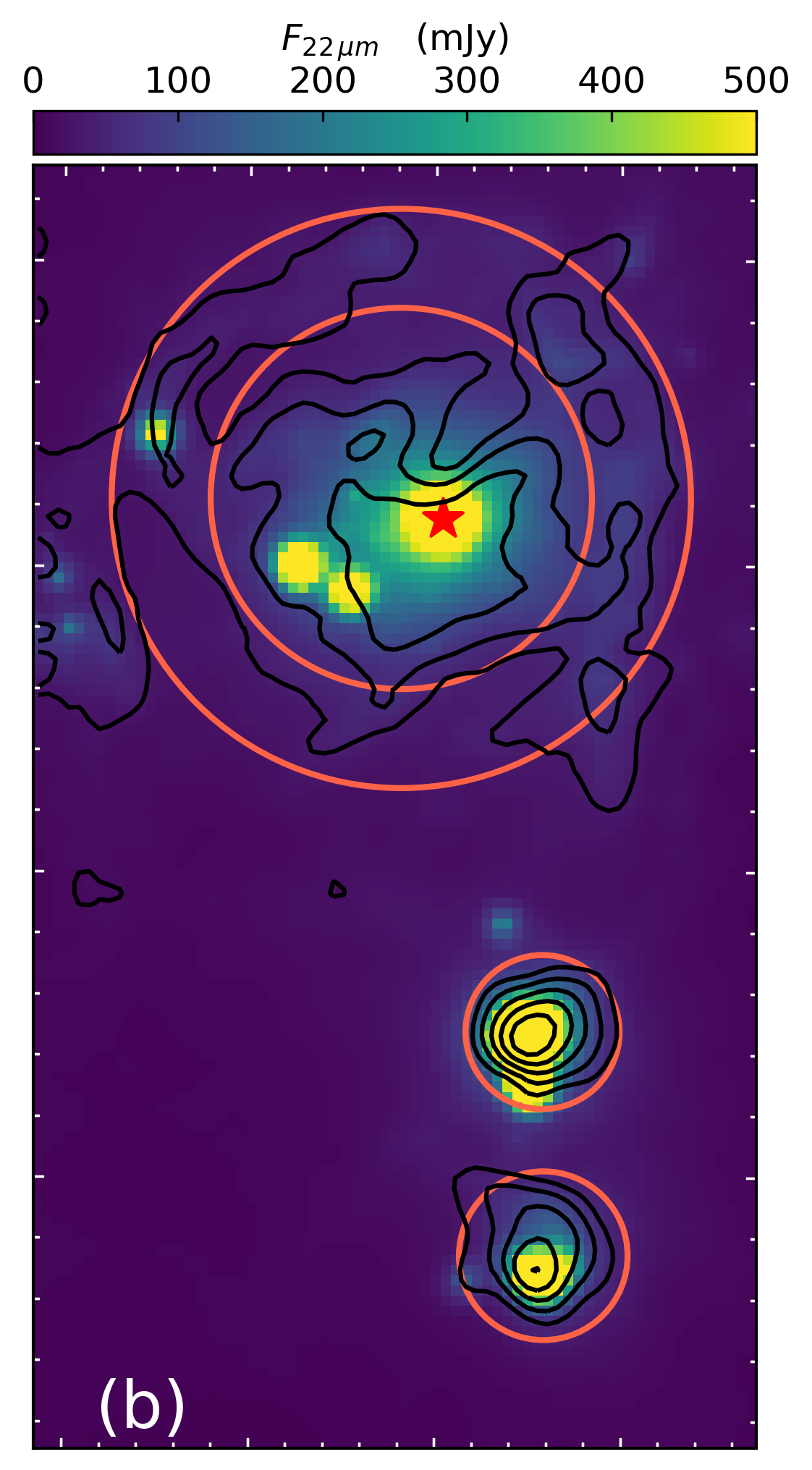}
   \includegraphics[height=3.7in]{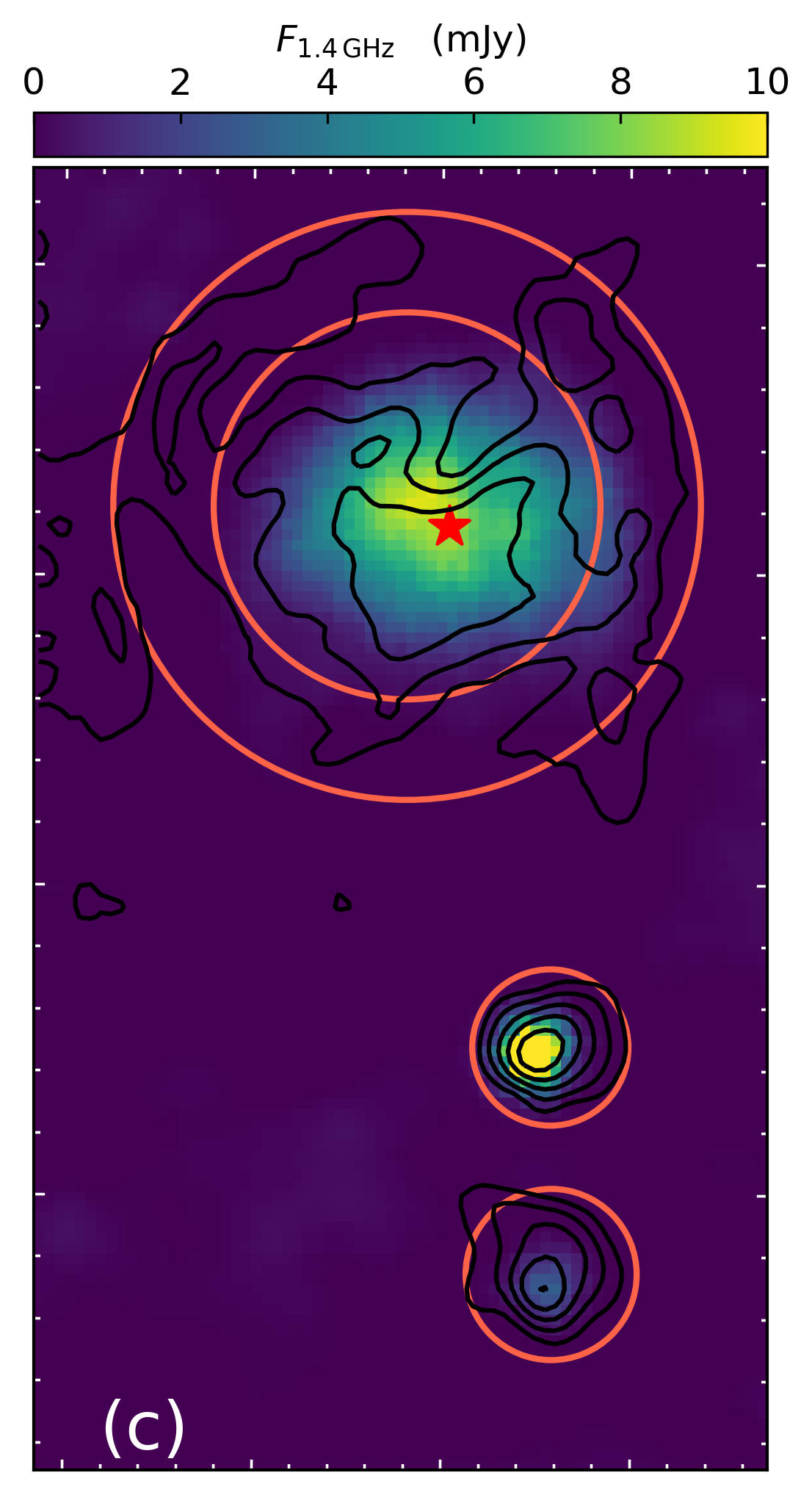}
    \vskip10pt
    \includegraphics[height=3.7in]{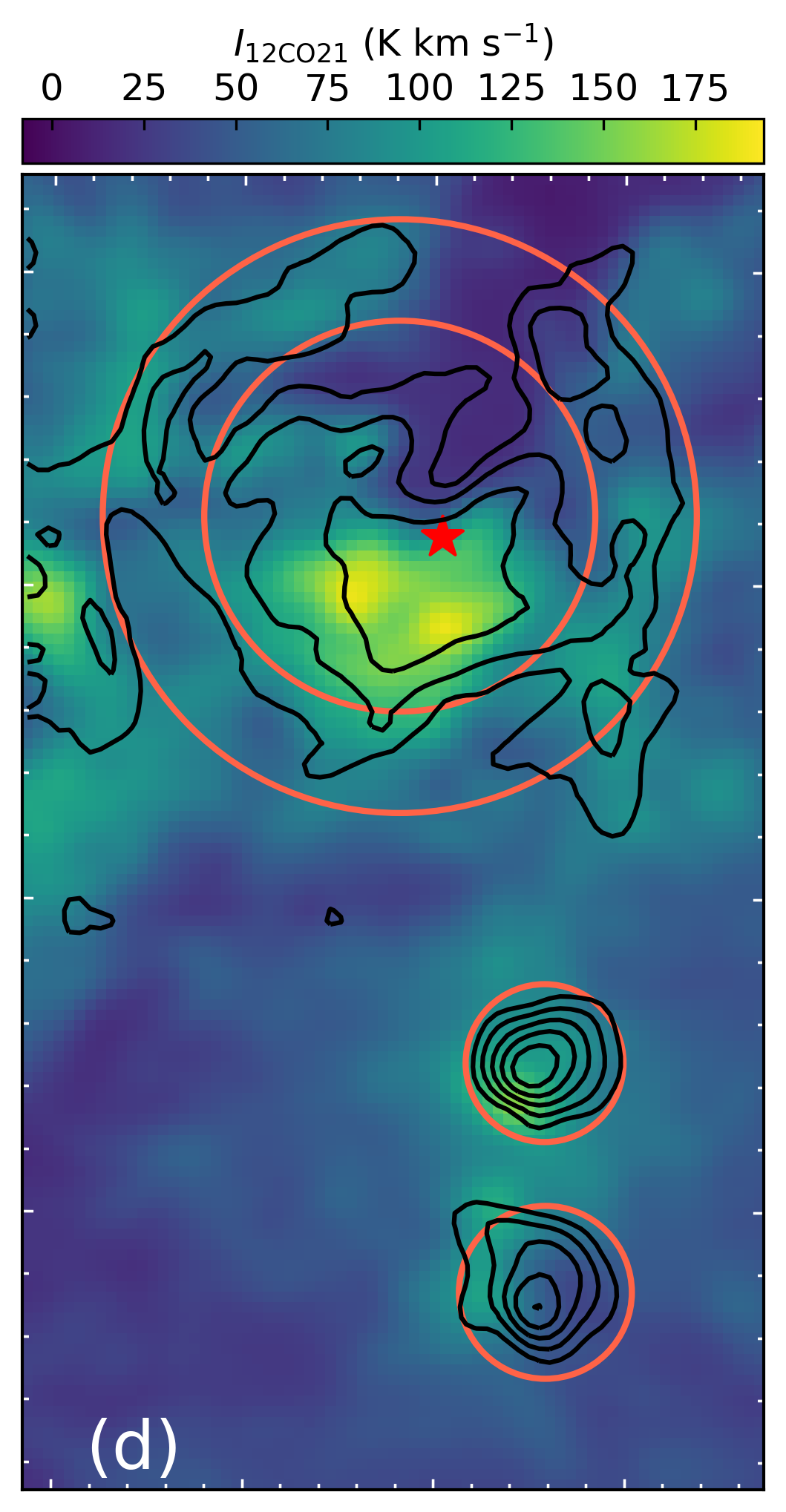}~~~
   \includegraphics[height=3.7in]{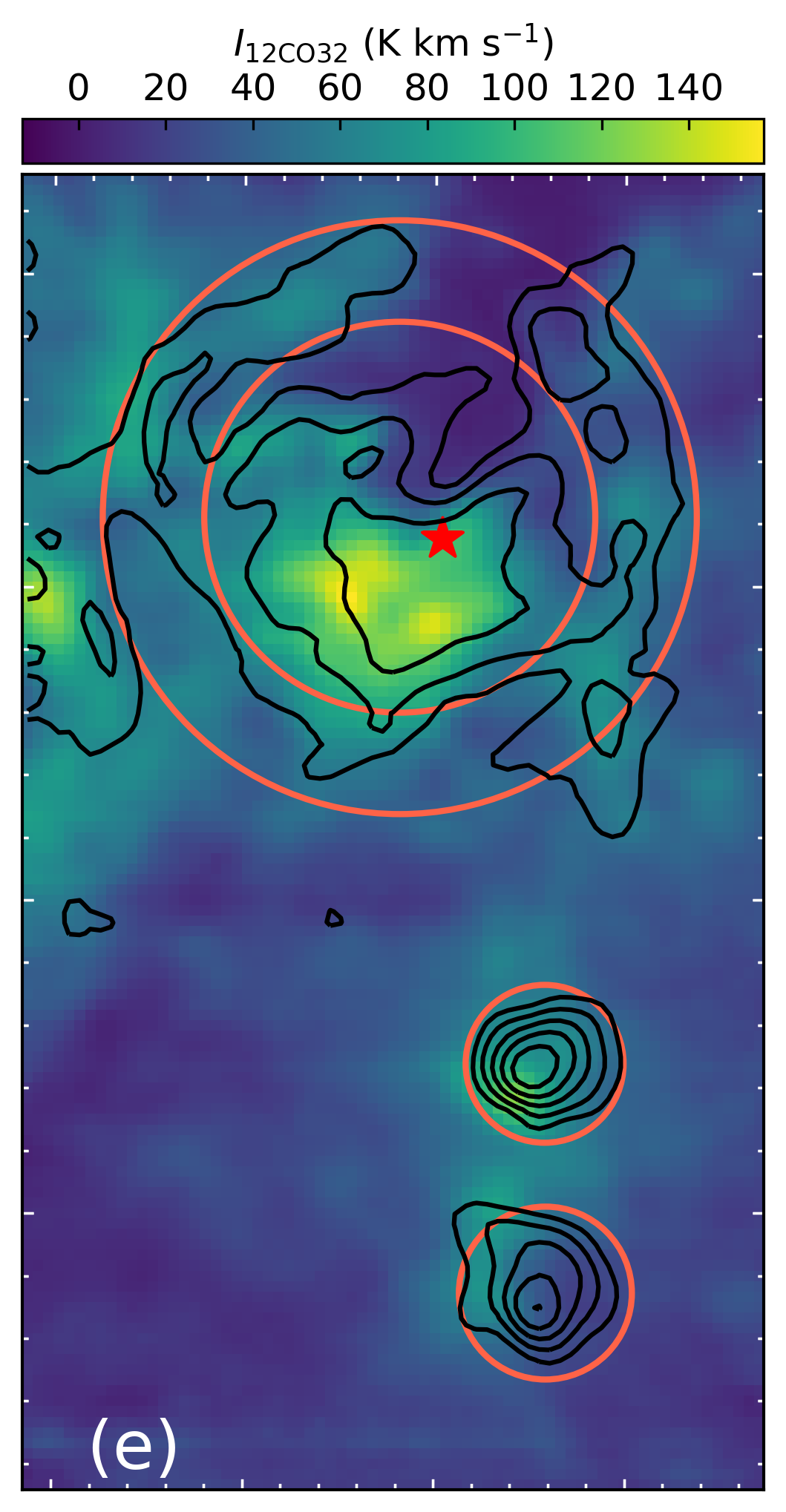}~~~
   \includegraphics[height=3.7in]{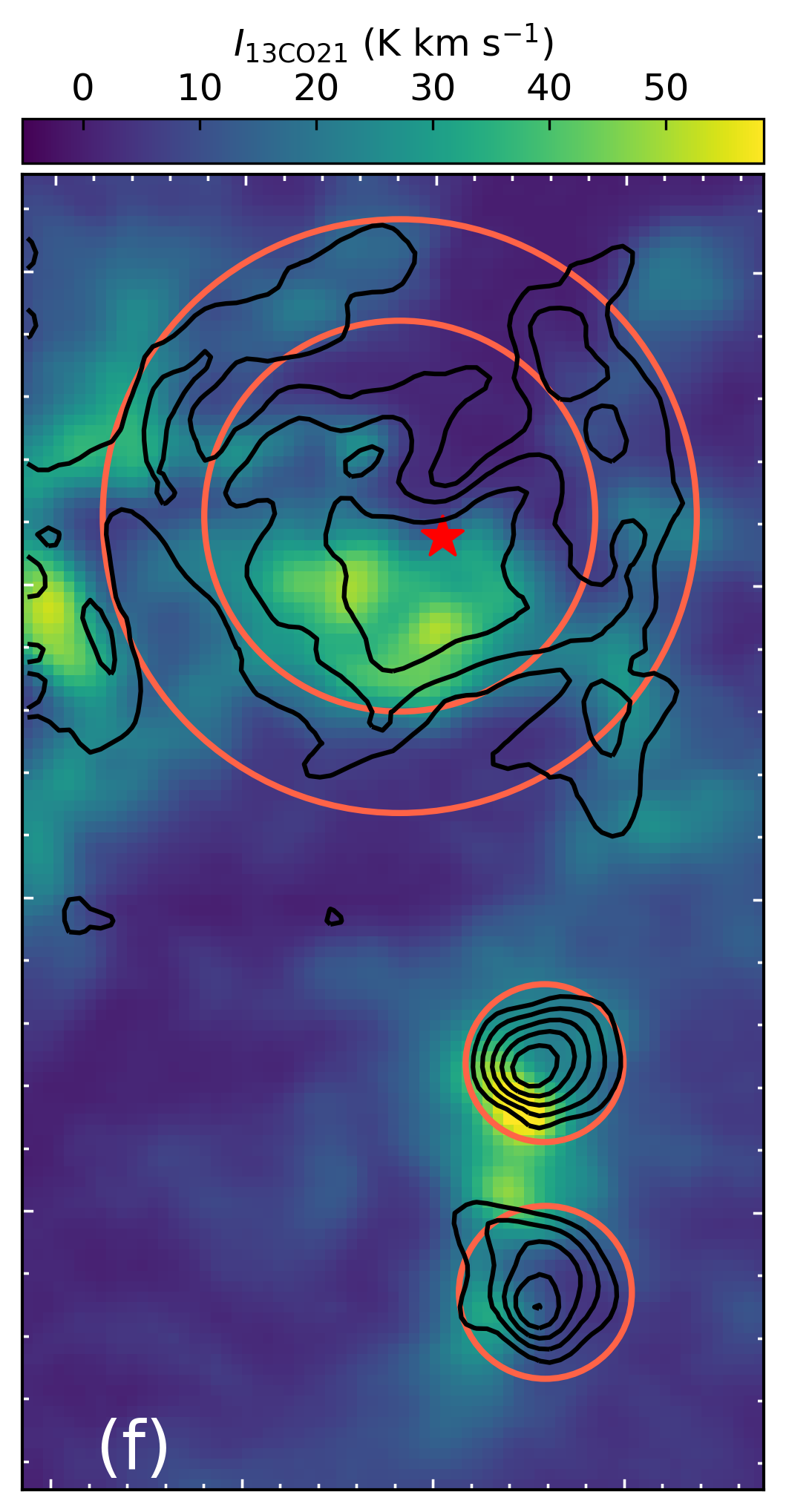}    \caption{Flux density maps of {\it WISE} $12$\microns and $22$\microns ((a) and (b), respectively),  flux density map of $1.4\,$GHz radio continuum 
   (c), and integrated ($-29 - -0.7$\,\kms) intensity maps of $^{12}$CO $2-1$ (d), $^{12}$CO $3-2$ (e), $^{13}$CO $2-1$ (f). The covered area is the same as on Figure~\ref{fig:cii_m0map}. Black contours are of \cii\ data at values of 15\%, 25\%, 40\%, 60\%, and 85\% of the peak \cii integrated intensity.  As in Figure~\ref{fig:cii_m0map}, red circles denote regions of interest (see Section~\ref{subsec:regions}) and the red filled star denotes the central ionizing source of S235MAIN. Much of the CO emission arises from the 
   close vicinity of the central ionizing source. The $12$\microns
   emission has a good spatial correlation with \cii, whereas the CO emission is more widespread than that of \cii, and the 22\,\microns and 1.4\,\ghz emission are more compact.\label{fig:int_maps}}
\end{figure*}

The CO moment maps have similar 
spatial distributions (see panels (d), (e) and (f) in Figure~\ref{fig:int_maps}). The molecular emission is more 
extended than that of \cii, but components associated with the ionized hydrogen zone and the PDR of S235MAIN are clearly visible.
The spatial distributions of the observed molecular gas tracers shows a  bridge between S235AB and S235C, especially in case of the denser gas tracer $^{13}$CO $2-1$. 
As these small \hii regions are essentially at the same velocity, this indicates that S235AB and S235C might be physically connected by a dense gas component.

In the S235PDR region the spatial correlation of the observed \cii and molecular gas is weaker than in the inner ionized hydrogen region.  The CO emission appears clumpy compared to the emission of the other tracers.
CO emission is absent toward the north. This ``emission-free'' part 
can also be seen in previous studies \citep[e.g.,][]{dewangan2016,dewangan2017}. 
This molecular deficit may be due to escaping ionized hydrogen gas in this direction \citep{dewangan2016}.  Although there is no indication that ionized hydrogen gas fills the evacuated space (see radio continuum map, panel (c) of Figure~\ref{fig:int_maps}), the gas may be too rarefied to produce strong radio continuum emission. 
%Investigation of the velocity channel maps in 
%\citet{dewangan2017} indicates that the $^{12}$CO gas does emit weakly in this cavity at $\sim -12\,$km s$^{-1}$, 
%Figure 1 in their paper). This velocity roughly coincides with the velocity component of $^{12}$CO($J=2-1$), we find throughout the entire region (see 
%Section~\ref{subsec:vel_comps}). Because the peak intensity of this $^{12}$CO($J=2-1$) velocity component is only a small fraction  of the main component 
%($\sim 14\%$ and $\sim 10\%$ in S235PDR and S235ION, respectively), it cannot be easily seen in the moment zero maps.
%\citet{dewangan2016} concluded that this cavity is a result of the negative feedback of the main ionizing source, namely that the ionizing photons can escape in 
%that direction. 
%Their theory is supported by their mid-infrared extinction map, which shows a broken part at the same spatial position. 
%Moreover, this region seems 
%to be associated with Br$\alpha$ emission surrounded by molecular hydrogen and/or PAH features, suggesting interaction between molecular and ionized gas in S235. 
%The spatial distribution of the observed molecular transitions follows our expectations, namely the lower density, the ISM material is more widely distributed.

\subsection{Velocity structure of \cii\ emission\label{sec:velocity}}
\begin{figure}
   \centering
\includegraphics[width=0.38\textwidth]{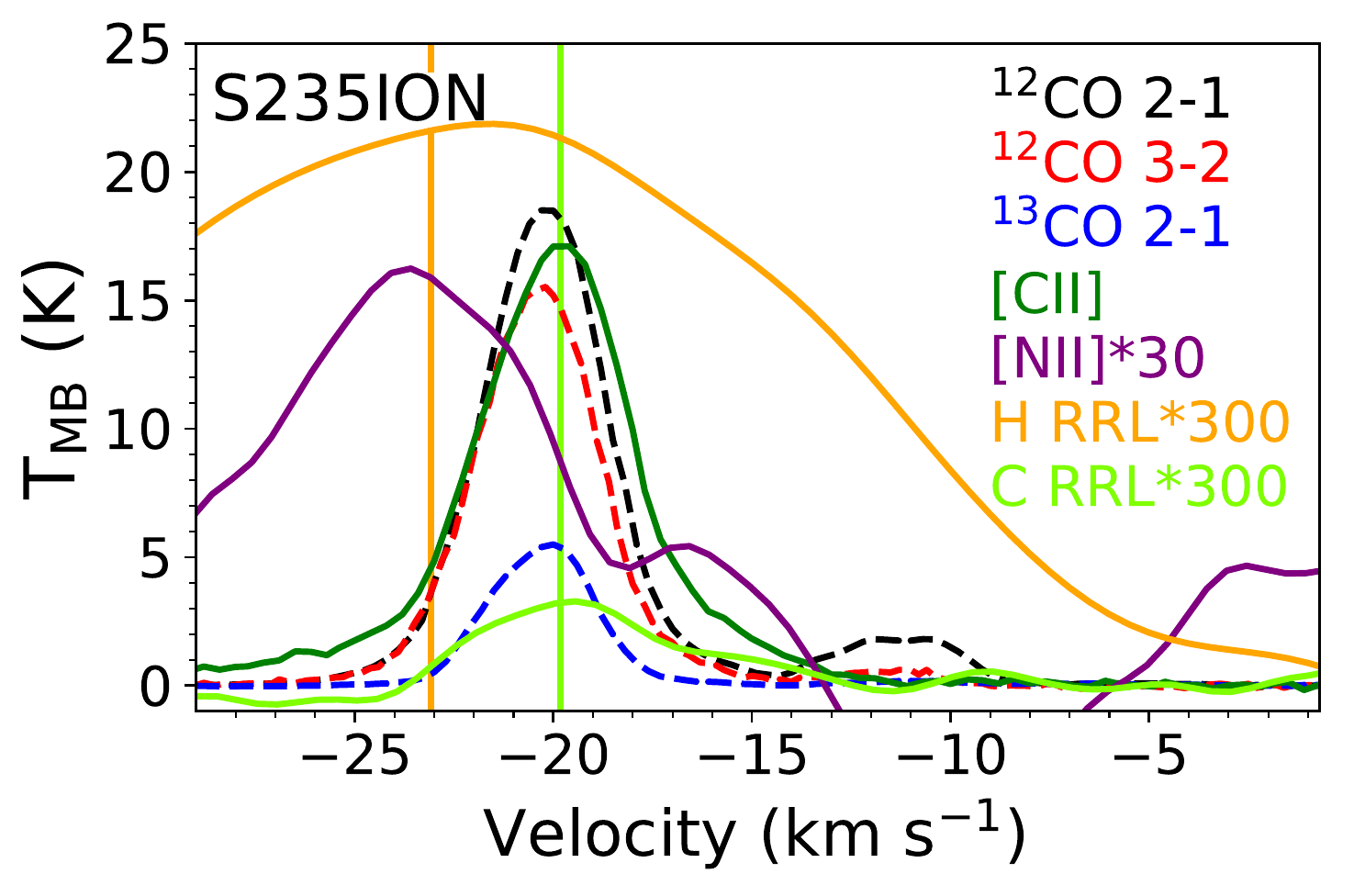}
   \includegraphics[width=0.38\textwidth]{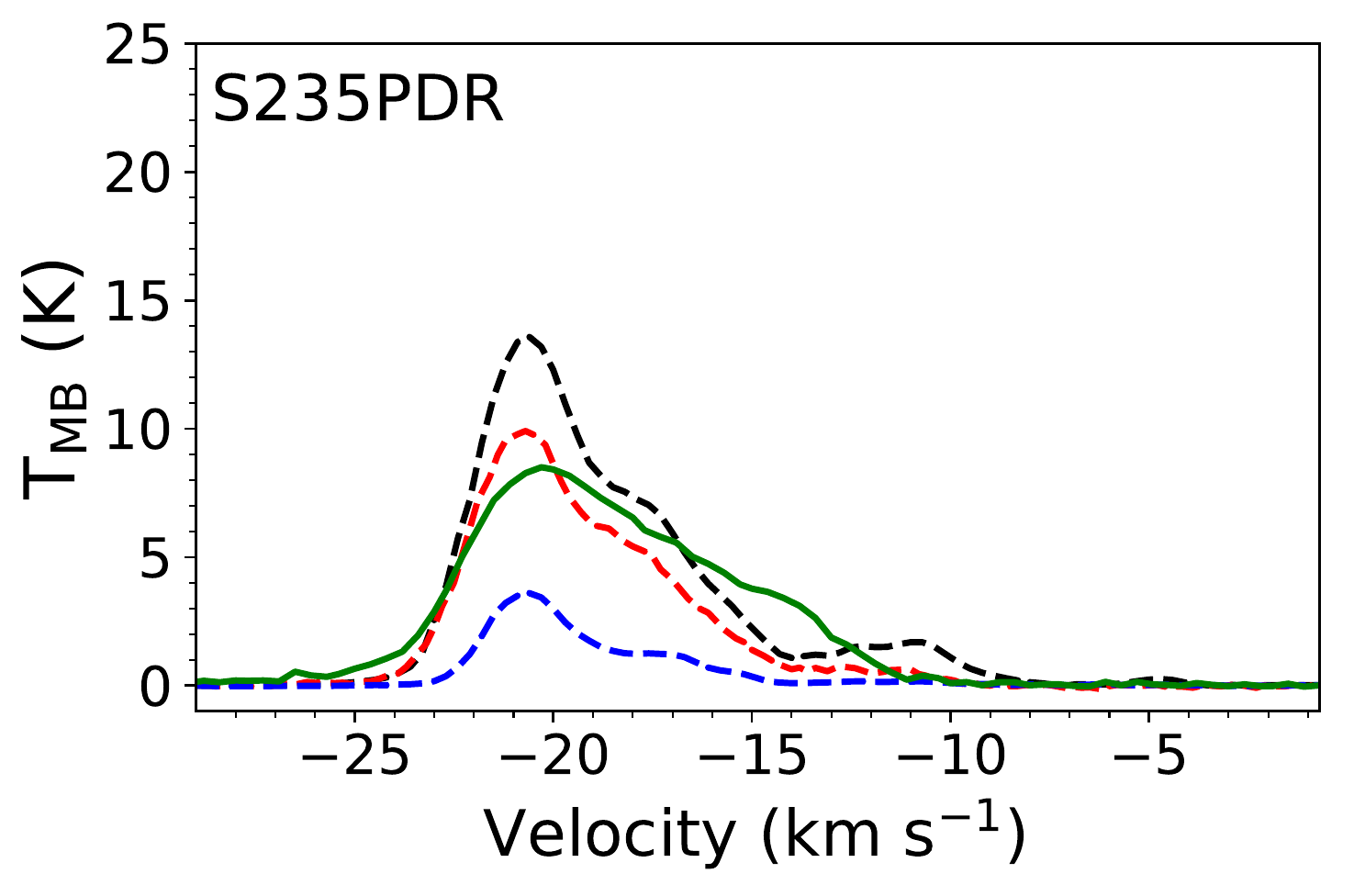}
      \includegraphics[width=0.38\textwidth]{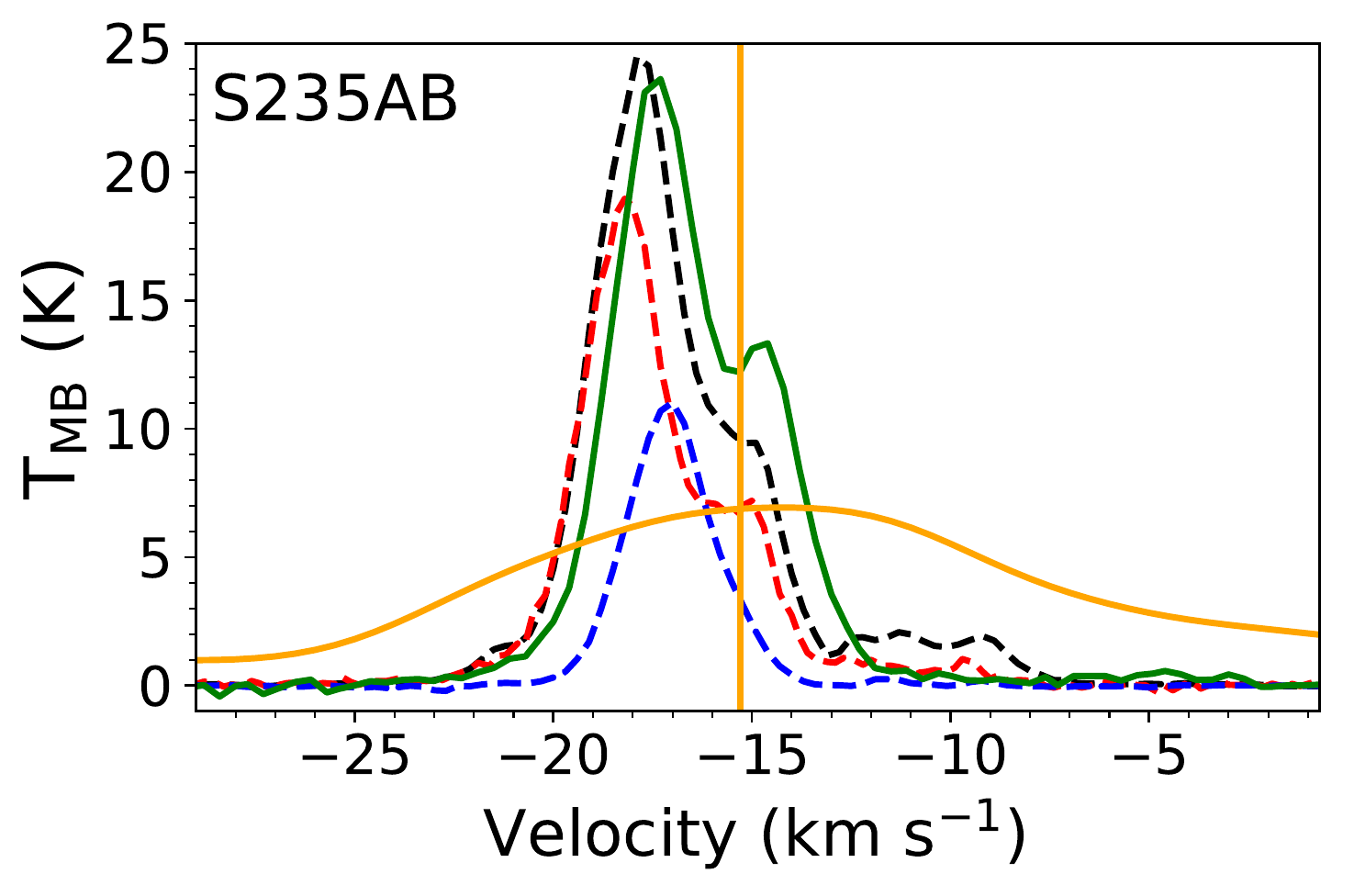}
   \includegraphics[width=0.38\textwidth]{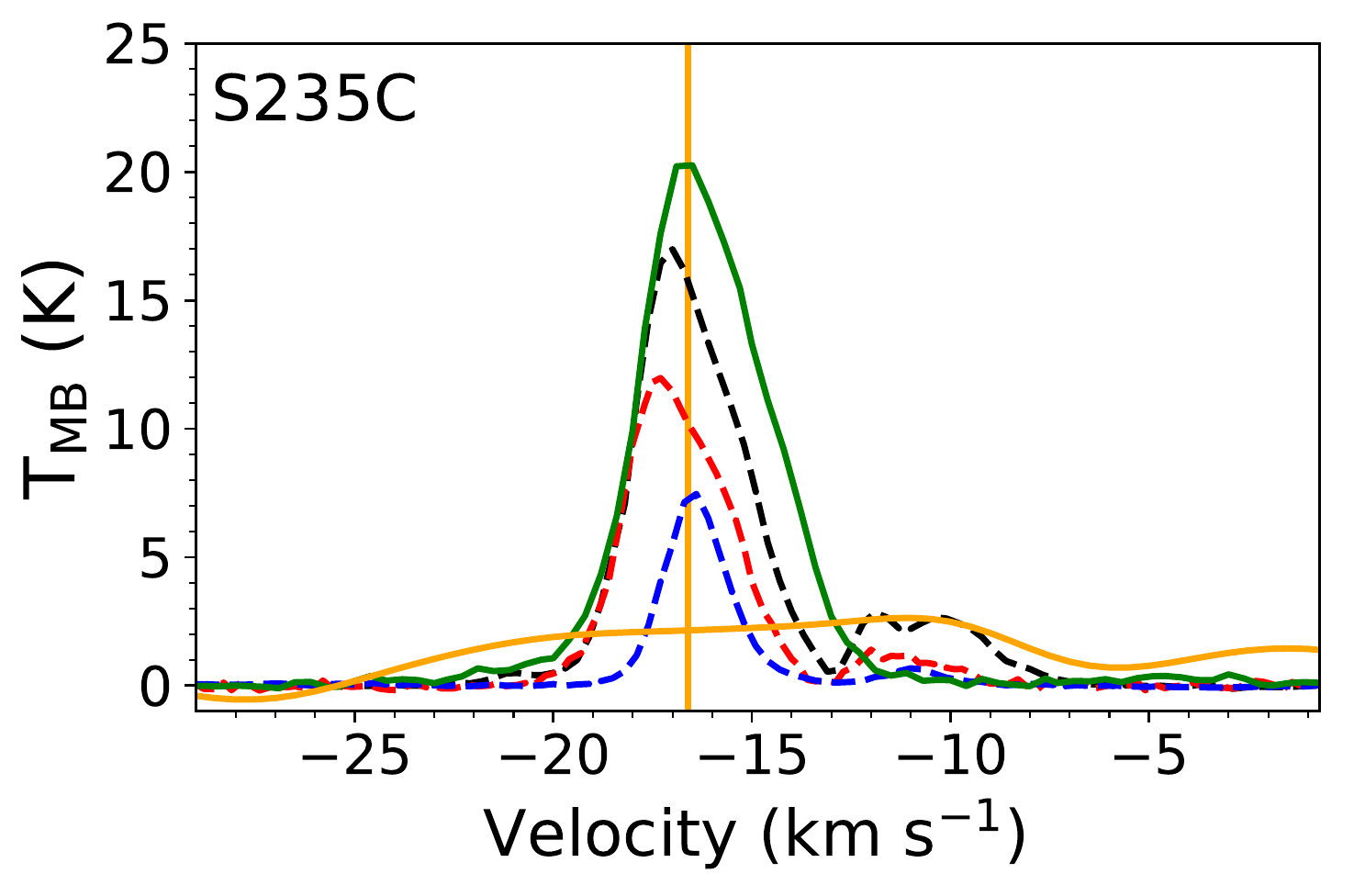}
   \caption{Spatially averaged spectra from the regions of interest for molecular gas tracers from \citet{bieging2016} ($^{12}$CO ($J=2-1$), $^{12}$CO ($J=3-2$), and $^{13}$CO ($J=2-1$) in black, 
   red and blue dashed lines, respectively), {\it SOFIA} \cii\ and \nii\ emission (green and purple curves, respectively), and GBT hydrogen and carbon RRLs (orange and light green curves, respectively).  We smoothed the \nii\ data with a 2.5\,\kms\ FWHM Gaussian.  The H and C RRL data have been smoothed with Gaussians of FWHM 5\,\kms\ and 2.5\,\kms, respectively.  Vertical orange and green lines show the peak hydrogen and carbon RRL velocities as found in the literature. \nii\ is only detected toward S235ION and the emission from S235PDR is not resolved in the RRL data so no RRL emission is shown in that panel. 
   %The main components of S235ION and S235PDR can be found in the velocity range 
   %$\sim [-22,-20]\,$km s$^{-1}$, while S235AB and S235C are at $\sim -17\,$km s$^{-1}$, in agreement with previous studies.
   \label{fig:av_specs}}
\end{figure}

We compare the spectra of \cii (Section~\ref{subsec:cii_obs}),  RRL (Section~\ref{sec:rrl}), %\form\ Section~\ref{sec:h2co}, 
and CO (Section~\ref{sec:co}) emission from the regions of interest in Figure~\ref{fig:av_specs}.
The \cii\ emission peaks near $-20$\,\kms\ for S235MAIN, and $-18\,\kms$ for S235AB and S235C.
All significant \cii\ emission is in the range $-29$ to $-12\,\kms$.
Hydrogen RRL emission is much broader than that of the other tracers.  It peaks near $-23\,\kms$ for S235ION an near $-16\,\kms$ for S235AB and S235C.  \citet{quireza06} found that H RRLs toward S235ION peak at $-23.07\,\kms$ and C RRLs peak at $-19.82\,\kms$  \citep[see also]{silverglate78, vallee87}.  \citet{anderson15} found that the RRL emission from S235A peaks at $-15.3\,\kms$, and that of S235C peaks at $-16.6\,\kms$.  The CO emission has a similar velocity profile to that of \cii, and carbon RRLs also peak near the velocities of strongest CO emission.  
We conclude that there is a real offset between the H RRL emission and that of other tracers in S235ION.  This offset is not seen in the other regions of interest.  
%Further analysis of the velocity structure in \cii\ data will be contained in a forthcoming paper (Kirsanova et al., 2018, in prep.).

%For all regions of interest, the \form\ profile can be decomposed into two components, one from $-23$ to $-12\,\kms$, and one from $-12$ to $-4\,\kms$.  Figure~\ref{fig:h2co_m0map} shows that the former component is associated with the $\sim-20\,\kms$ CO emission and the latter component is associated with the $\sim-12\,\kms$ CO emission (seen only in $^{12}$CO 2-1.  The velocity offset between H$_2$CO and CO is presumably caused by only portions of the molecular clouds contributing to the absorption signal.

\begin{figure*}
   \centering
\includegraphics[width=0.30\textwidth]{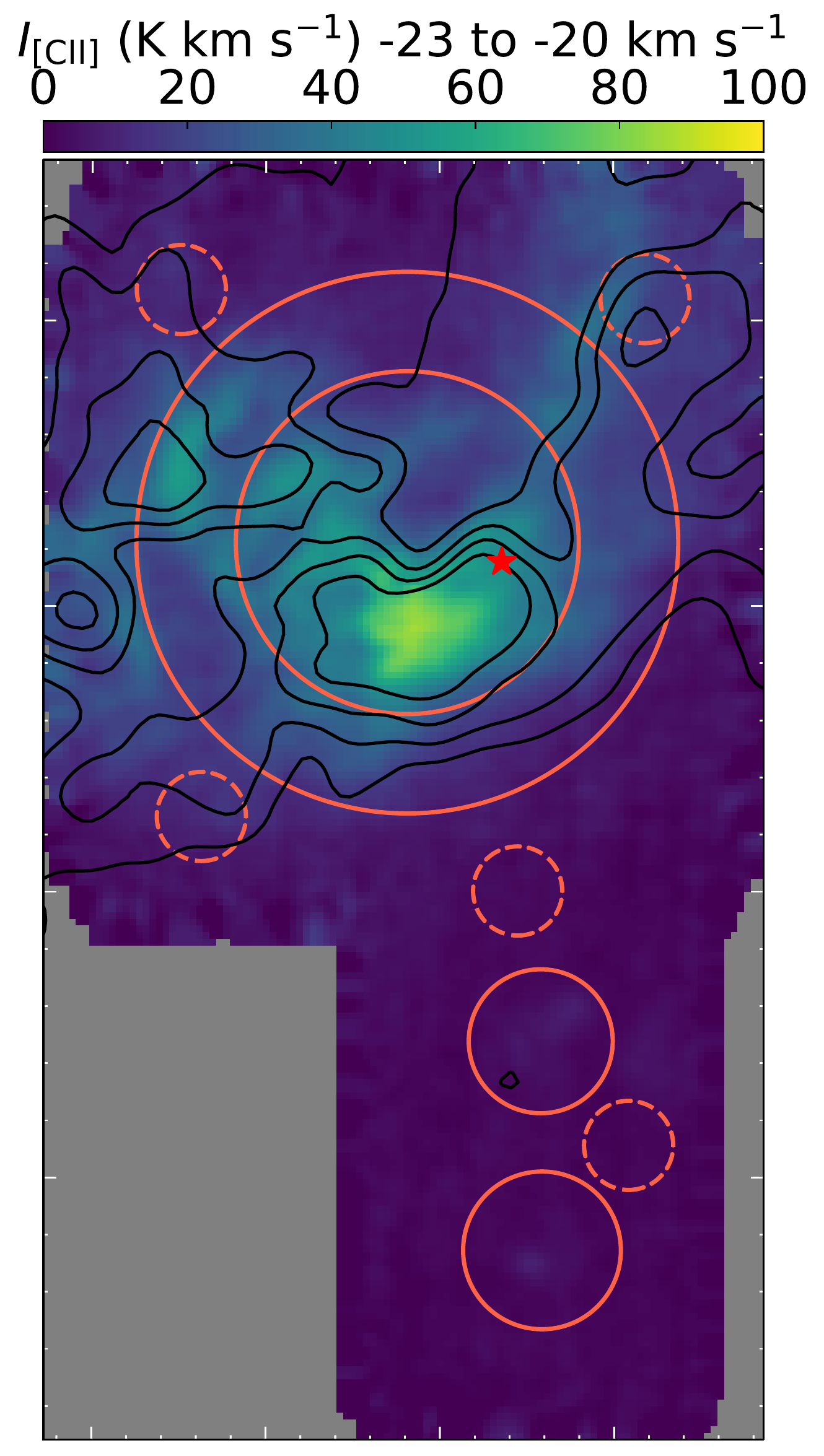}
   \includegraphics[width=0.30\textwidth]{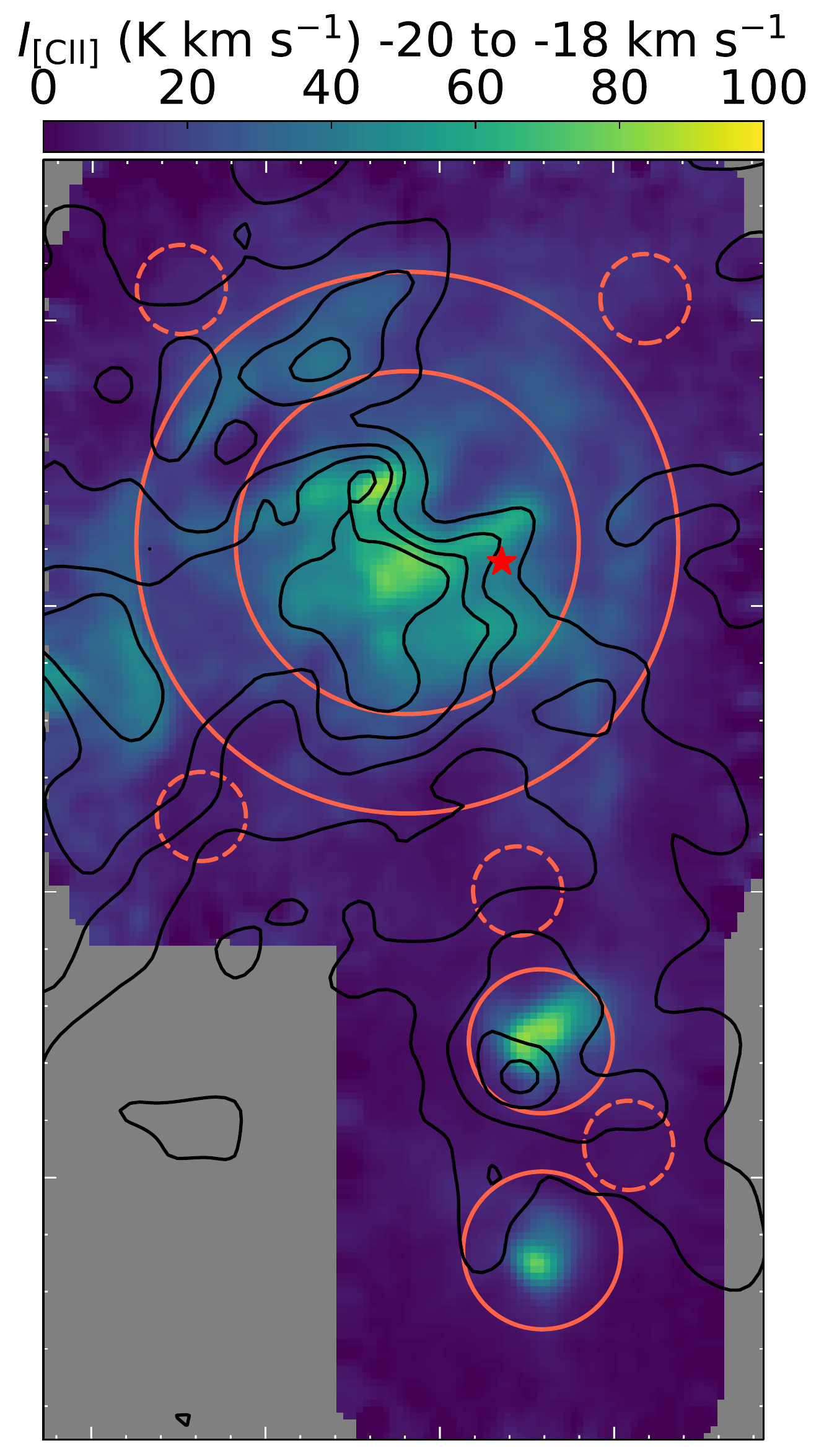}
      \includegraphics[width=0.30\textwidth]{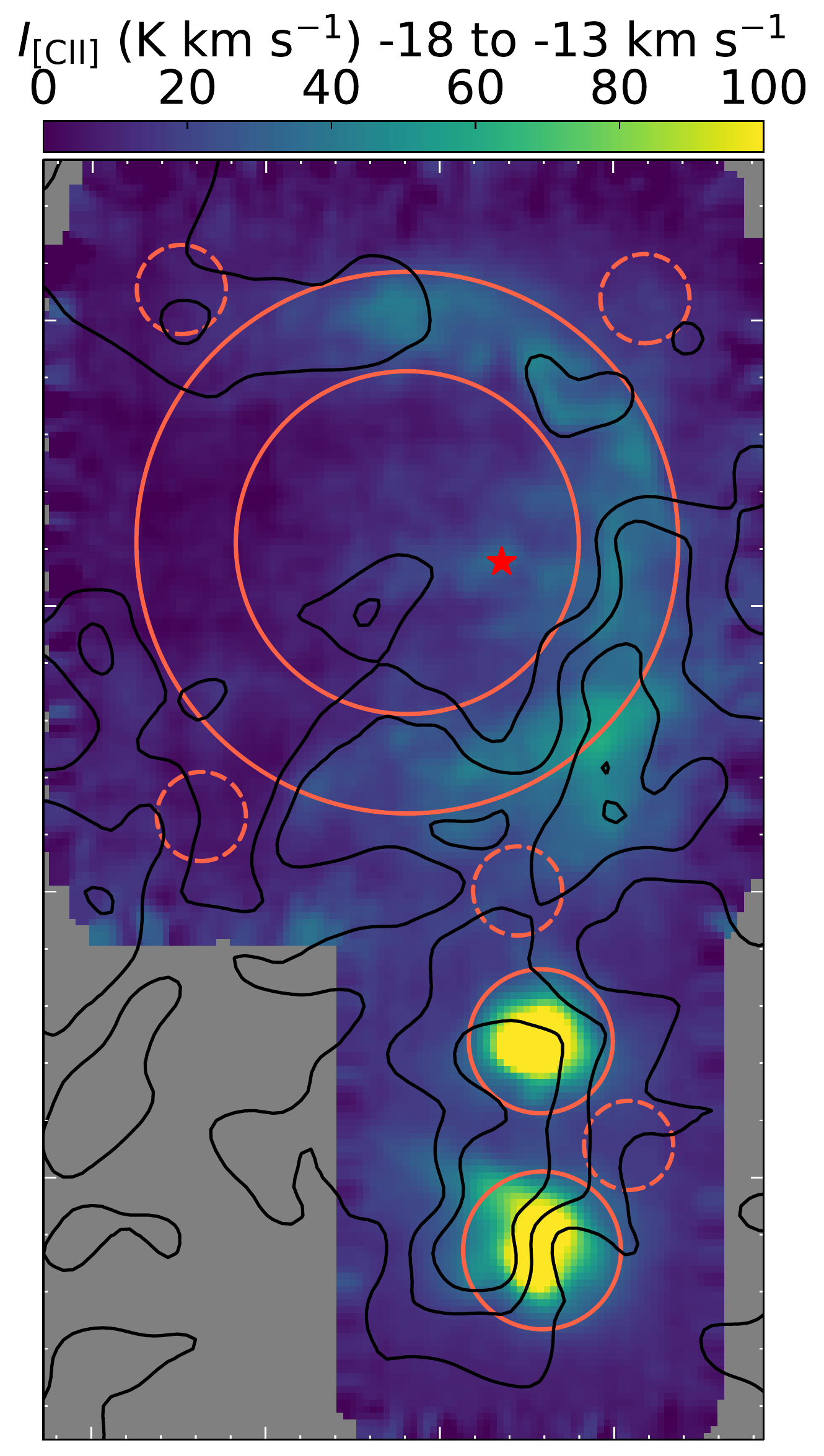}
\caption{Moment maps of \cii\ emission over three velocity ranges: $-23$ to $-20$\,\kms\ (left), 
$-20$ to $-18$\,\kms\ (middle), and $-18$ to $-13$\,\kms\ (right).  These three velocity ranges have distinct emission components. The field of view in each panel is the same as that of Figure~\ref{fig:cii_m0map}.    Contours show $^{12}$CO\,$2-1$ emission integrated over the same velocity ranges.  There is excellent spatial agreement between the \cii\ and CO emission. \label{fig:m0s}}
\end{figure*}

The \cii\ emission can be decomposed into three velocity ranges: $-23$ to $-20\,\kms$ (``low''), $-20$ to $-18$\,\kms\ (``middle''), and $-18$ to $-13$\,\kms\ (``high'').  These three ranges have distinct emission components.  We show moment maps of these three velocity ranges in Figure~\ref{fig:m0s}.  The low velocity range contains bright compact \cii\ emission in the ionized hydrogen region of S235MAIN, emission from the eastern PDR, and a ridge of emission extending to the northwest.  The middle velocity range contains multiple knots of compact \cii\ emission in the ionized hydrogen region of S235MAIN, and emission from the PDR to the northeast and to the west.  The high velocity range mainly contains \cii\ emission from the western PDR of S235MAIN, and also compact emission from S235AB and S235C.  Although S235AB and S235C are detected in the middle velocity range, this is just the blueshifted wing of emission (see Figure~\ref{fig:av_specs}).  All velocity components show excellent spatial agreement with $^{12}$CO\,$2-1$ emission.

%-------------------------------------------------------------------------------------------
\section{Correlations between \cii\ and ancillary data}
\label{sec:res}

Below, we determine correlations between \cii intensity and the intensity of the ancillary data sets.  We are dealing with emission integrated along the line of sight, and so in the direction of the ionized hydrogen gas we also detect emission from the front- and back-side PDRs.  This is especially important for the 12\,\microns emission, as it is a strong PDR tracer.  For example, the region S235ION contains dense molecular clumps S235 Central~E and Central~W \citep{kirsanova2014, dewangan2017}. These clumps are presumably do not reside within the ionized hydrogen volume, but nevertheless the emission associated with the clumps is contained in the S235ION region of interest.

The NVSS and RRL data both trace the emission from ionized gas in the region.  The spatial resolution of the NVSS is more than twice as fine as that of our RRL data.  We therefore use the NVSS for all spatial analyses, and use the RRL data for all spectral analyses.
  
%Considering that the S235PDR region is larger than S235ION by $\sim 23.5\%$, 
\begin{deluxetable*}{lcccccccc}
\tablecaption{Fluxes from regions of interest \label{tab:intensity}}
\tablehead{
\colhead{Region} &
\colhead{$F_{\cii}$} &
\colhead{$F_{12\,\micron}$} & 
\colhead{$F_{22\,\micron}$} &
\colhead{$F_{1.4\,\ghz}$} &
\colhead{$F_{\rm 12{\rm CO}21}$} &
\colhead{$F_{\rm 12{\rm CO}32}$} &
\colhead{$F_{\rm 13{\rm CO}21}$} &
\colhead{Area}\\
\colhead{} &
\colhead{$10^{-12}$ W\,m$^{-2}$} &
\colhead{Jy} & 
\colhead{Jy} &
\colhead{Jy} &
\colhead{$10^{-15}$ W\,m$^{-2}$} &
\colhead{$10^{-15}$ W\,m$^{-2}$} &
\colhead{$10^{-15}$ W\,m$^{-2}$} &
\colhead{sq. arcmin.}}
\startdata
S235MAIN   & 2.9 & 310 & 1000 &  1.9  & 5.7  &   14   & 1.0    & 71 \\
~~~S235ION & 1.5 & 160 & 520  &  1.9  & 2.7  &   6.9  & 0.54   & 31\\
~~~S235PDR & 1.4 & 150 & 520  &  0.022 & 3.0  &   6.8  & 0.46   & 40 \\
S235AB     &0.30 &  40 & 150  &  0.33  & 0.62 &   1.5  & 0.17   & 5.0\\
S235C      &0.27 &  33 & 110  &  0.051 & 0.41 &   0.86 & 0.088  & 6.0
\enddata
\end{deluxetable*}

\subsection{Total fluxes in regions of interest}
We give the total fluxes for the regions of interest in Table~\ref{tab:intensity}.  We find these values by integrating the emission over each aperture as defined in Table~\ref{tab:regions}.  For \cii and the molecular tracers, the data are in units of main beam temperature, $T_{\rm MB}$, so the integrated intensity $I_\cii = \int T_{\rm MB} dv$ has units of \K\,\kms.  We convert the \cii\ and CO data to the more useful quantity of flux in units of W\,m$^{-2}$ per pixel. 
%or total integrated intensity $I_{\rm \cii, total} = \int\int T_{\rm MB} dv d\Omega$ in units of K\,\kms\,deg.$^{2}$.  
For our {\it SOFIA} \cii\ maps with $7\arcsec$ pixels, the conversion between integrated intensity and flux for each pixel is 
\begin{equation}
    \frac{F_\cii}{\rm W m^{-2}} = 8.1\times 10^{-18}\frac{I_\cii}{\K\,\kms} \,.
\end{equation}
%We also give the luminosity in $\lsun$, using the distance of 1.56\,\kpc.
%Similarly, we convert the CO data to flux in units of ${\rm W m^{-2}}$, and give the values integrated over the regions of interest in 
%We give the flux above quantity summed over the regions of interest.
%For CO, we provide the total integrated intensity in units of K\,\kms\,deg.$^{2}$, integrated over the region of interest:
%\begin{equation}
%    I_{\rm total} = \int\int T_{\rm MB}\,dv\,d\Omega = \overline{T}_{\rm MB} \Omega\,,
%\end{equation}
%where $\overline{T}_{\rm MB}$ is the average main beam temperature within the region of interest.  
We subtract a local background value from the two {\it WISE} bands, computed as the median value surrounding the region, but perform no background correction for the other data.

The $I_\cii$ and $F_{12\,\micron}$ values appear in roughly the same ratio for all regions of interest.  This correlation is less strong for the other tracers.  We will explore this in more detail in Section~\ref{subsec:corr}.

We see from Table~\ref{tab:intensity} that about half the emission from S235MAIN comes from the S235PDR region of interest for \cii, the MIR bands, and all CO transitions.  The \cii\ emission is strong south of the ionizing source (see Figure~\ref{fig:cii_m0map}).  The \cii\ emission in a central zone $1.5\arcmin$ in radius around the ionizing source has an integrated intensity of $\sim0.6\times 10^{-12}$\,W\,m$^{-2}$, which is $\sim40\%$ that of S235ION and $\sim20\%$ that of the total emission from S235MAIN.

The values for all tracers are similar for S235AB and S235C, except for the radio continuum data.  S235AB has $\sim6$ times higher flux density than S235C, 
indicating a higher Lyman continuum photon production rate in S235AB, a higher optical depth in S235C, and/or stronger dust absorption in S235C. %That the \cii\ and radio continuum data give such different results hints that they are not tracing the same population of massive stars.  

%\citet{pineda2013} found that a small amount of \cii\ emission 
%comes from ``cold'' ionized \ldacom{temp?} and atomic gas. \ldacom{This is not the conclusion reached by Pabst et al. 2017 
%Cite new references and summarized the results known so far, even if they do not agree within each others} Some \cii emission originates from the warm ionized medium and diffuse clouds  \citep[e.g.,][]{pabst2017}, although given the relatively small field of view, our \cii data are largely not sensitive to these components. 
%The small clump visible in the north-west part in the 
%S235PDR is identified as ``S235 North-West'' by \citet{kirsanova2008}

%The integrated intensity of $^{12}$CO and $^{13}$CO contains two concentrations in the S235ION region. This emission likely arises from dense clumps, first reported 
%by \citet{kirsanova2008}. They observed strong emission from CS($J=2-1$ and $^{13}$CO($J=1-0$) near to young stellar populations (located to the south of the main 
%ionizing star), indicating that those young stars are still embedded in the parental molecular cloud. These two dense clumps are coincident with the two bright infrared sources IRS1 and IRS2.

%The average integrated CO intensity is higher in S235ION than in S235PDR by 
%$\sim 14\%$, $\sim 26\%$ and $\sim 37\%$ for $^{12}$CO($J=2-1$), %$^{12}$CO($J=3-2$) and $^{13}$CO($J=2-1$), respectively. 
For S235MAIN, roughly half of the integrated intensity in all CO transitions is coming from the S235ION region, and half from the S235PDR region.
In all the observed molecular transitions, the S235AB region is marginally brighter than S235C.
%, by $\sim 55\%$, $\sim 74\%$ 
%and $\sim 90\%$ (mean), and by $\sim 45\%$, $\sim 42\%$ and $\sim 40\%$ (total) for $^{12}$CO($J=2-1$), $^{12}$CO($J=3-2$) and $^{13}$CO($J=2-1$), respectively. 

\subsection{\cii\ emission from front- and back-side PDRs}

Because of emission from front- and back-side PDRs along the line of sight, it is difficult to accurately determine the percentage of \cii\ emission from the entire PDR of S235.  Based on results from previous studies of \hii regions in \cii emission, we expect that nearly all the \cii\ emission arises from dense PDRs \citep[e.g.,][]{pabst2017}.
%, but these studies were done on objects lacking PDRs along the line of sight toward the ionized gas.
%We attempt to estimate the percentage of emission spatially coincident with the volume of ionized hydrogen using the fact that the \cii\ is centrally peaked around the ionizing source of S235.  Because line of sight PDRs would not reproduce the intensity of this emission, we assume that the emitting ionized carbon is cospatial with the ionized hydrogen volume.  
%We attempt to determine the fraction of the emission seen toward the hydrogen ionized gas that is from front- and back-side PDRs by analyzing position-velocity (p-v) diagrams.  

\begin{figure*}
\includegraphics[width=6.5in]{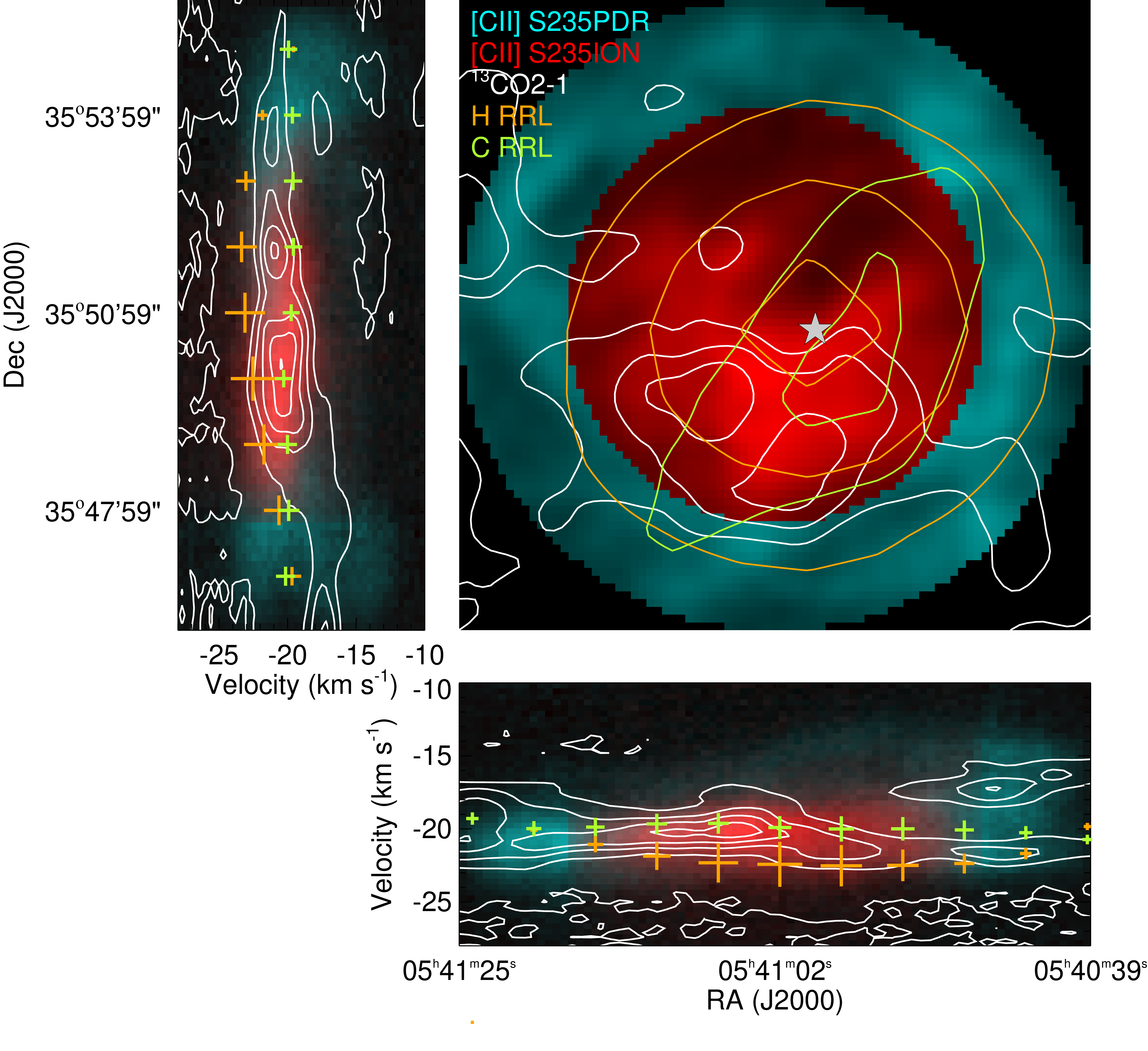}
\caption{Position-velocity analysis of \cii\ emission toward S235MAIN.  The top right panel shows the moment~0 integrated intensity map from Figure~\ref{fig:cii_m0map}, color coded with S235ION in red and S235PDR in cyan.  White contours are of $^{13}$CO\,$2-1$ moment~0 emission (Figure~\ref{fig:int_maps}) and green contours are of RRL moment~0 emission (Figure~\ref{fig:rrl}).  The gray star symbol shows the location of the ionizing source.  The other two panels show p-v diagrams with the same color scheme and $^{13}$CO\,$2-1$ contours.  The green crosses show the velocity derived from Gaussian fits to RRL data integrated along a spatial direction; the size of the crosses indicate the amplitude. We plot H RRL points from the PDR even though this emission is not resolved; these points rather represent emission from the edge of the ionized hydrogen volume. \cii\ emission seen in the direction of S235ION is blueshifted relative to the ionized hydrogen gas, whereas that of S235PDR is found at a broader range of velocities.   We find no evidence of significant \cii\ emission from the ionized hydrogen volume.  \label{fig:pv}}
\end{figure*}

In Figure~\ref{fig:pv} we show position-velocity (p-v) diagrams for S235MAIN.  These diagrams show \cii\ emission from the S235ION region in red and from S235PDR in cyan.  
%In H-$\alpha$ observations, \citet{lafon83} found a strong gradient in the ionized gas such that it is $\sim -15\,\kms$ in the south-east of the region and $\sim -25\,\kms$ in the north-west.  
We create the orange (from H RRL data) and green (from C RRL data) crosses in this figure by integrating the RRL data along its position axes, and then fitting Gaussians spaxel by spaxel.

\begin{figure}
   \centering
\includegraphics[width=0.45\textwidth]{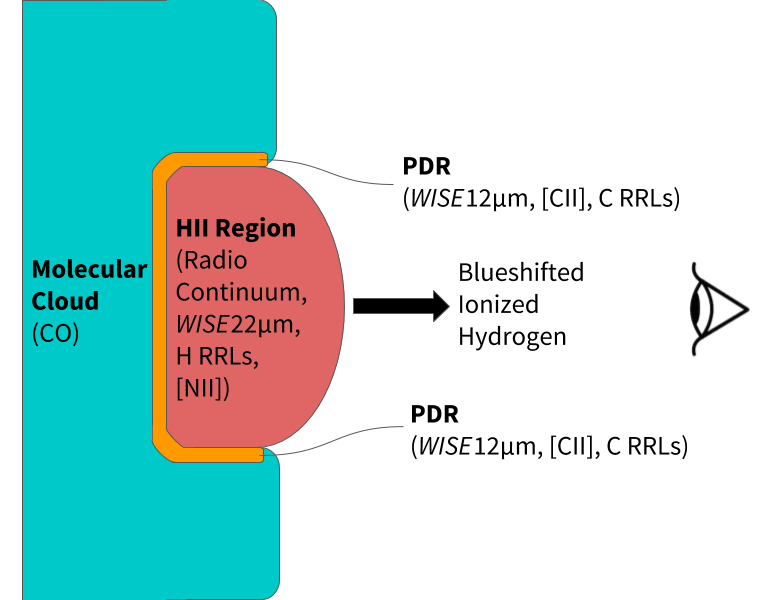}
\caption{Cartoon model we suggest as an explanation of the velocity offset between \cii\ and H RRL emission.  We propose that the molecular cloud associated with S235 is behind the main region, and that the \cii\ emission comes from the PDR boundary between the molecular cloud and the ionized gas. \label{fig:cartoon}}
\end{figure}

Based on its coincidence with CO and the velocity offset of the ionized hydrogen gas with respect to all other tracers,
%which \citet{dewangan2017} argued is foreground to the %ionized hydrogen, 
we believe that the \cii\ emission seen toward the ionized hydrogen is from line-of-sight PDRs and not from the ionized hydrogen volume.  
The peak velocity of \cii, CO, and C RRL emission toward the ionized hydrogen gas is redshifted relative to that of the ionized hydrogen gas itself, whereas \cii\ and CO emission from the PDR is at a similar velocity to that of the ionized hydrogen gas.  In support of this, we note the excellent spatial agreement between \cii\ and CO emission in the three velocity ranges shown in Figure~\ref{fig:m0s}.
This is consistent with an expansion of the ionized hydrogen gas preferentially toward us.  In this scenario, the \cii, CO, and C RRL emission are therefore due to back-side PDRs, as we draw in the cartoon model of Figure~\ref{fig:cartoon}.  This explanation is consistent with the fact that minimal absorption is seen across the face of the region in H$-\alpha$ maps \citep[see Figure~\ref{fig:s235complex}; also][their Figure~10b]{dewangan2017}.  If there were dense foreground PDR material, we would see H-$\alpha$ absorption across the face of the region, which we do for \hii\ regions RCW120 \citep[see ][their Figure~1]{anderson15} and M20 \citep[the Trifid Nebula; see ][their Figure~1]{rho06}. Our picture of S235 is similar to the model of Orion in \citet[][cf. their Figure~3]{pabst19}.  The main difference is that they observe a shell of \cii\ expanding toward us, whereas in S235 we see no evidence for such a feature. 
%, in disagreement with the findings of \citet{dewangan2017}.
%We see no evidence for \cii\ emission from back-side PDRs and find no evidence of \cii\ emission from within the ionized gas volume.

%Our observations of S235MAIN show that about half of the integrated intensity for \cii\ and all CO species comes from the inner, ionized hydrogen region, and about half comes from the PDR.
%Front- and back-side PDRs along the line of sight contribute to these values, so the percentage coming from the ionized hydrogen region is less than 50\%.
%, so we suggest that the actual values may be more discrepant.  
%There is, however, strong \cii emission surrounding the ionizing source that cannot be explained by PDRs along the line of sight.
%Considering the closeness of the ionizing star and the embedded infrared sources, IRS1 and IRS2, 
%The bright \cii\ emission from the inner region may be explained by the presence of H$_{2}$ illuminated by FUV photons \citep{pineda2013}.

We can also use the measured \cii\ and \nii\ intensities to estimate the fraction of \cii\ emission that arises from neutral hydrogen regions.  \citet{croxall17} found that the \cii\ to \nii\ intensity ratio has a value of $I_\cii/I_\nii\simeq 4$ for ionized gas, independent of electron density.  Values of $I_\cii/I_\nii>4$ indicate \cii\ emission from regions of neutral hydrogen.  From the S235ION region, we measure $I_\nii \simeq 2\times10^{-13}$\,W\,m$^{-2}$, which gives us $I_\cii/I_\nii\simeq10$ (see Table~\ref{tab:intensity}).  Using Equation~1 from \citet{croxall17}, this results in a fraction of \cii\ emission from neutral hydrogen regions $f_{\rm \cii, Neutral} \simeq 0.5$.  Because of the low quality of our \nii\ data, this value is only approximate.  Our intensity ratio, however, is roughly consistent with that observed in the external galaxy sample of \citet{croxall17} and the study of \citet{rigopoulou13}.  If half of the \cii\ emission towards S235ION comes from locations of neutral hydrogen, just 25\% of the emission from all of S235MAIN is from the ionized hydrogen volume.

\subsection{Correlations with \cii intensities}
\label{subsec:corr}

We investigate the correlation between \cii intensity and that of the other tracers to help determine the origin of the \cii emission.  We plot pixel-by-pixel correlations of the regridded data
 (Figures~\ref{fig:corr_w3}, \ref{fig:corr_nvss}, \ref{fig:corr3}, and \ref{fig:corr1}) to examine the relationships 
between the integrated intensity of \cii\ and that of the other tracers in the S235MAIN (S235ION and S235PDR), S235AB, and S235C regions. The bottom panel of each figure contains all data points, one per pixel of the regridded maps.
%, including data from all five background regions. 

To quantify the relationships between \cii\ and other observed tracers, we fit linear regressions of the form
$I_{\cii} = A + B \times I_\lambda$, where $I_{\cii}$ is the integrated intensity of \cii, $I_\lambda$ is the integrated intensity of CO or the flux density for the IR and radio continuum data.  We use a robust least-squares method to determine the fit parameters, with a Cauchy loss function.  The robust least squares fit minimizes the influence of outliers, which otherwise can skew the results.  For data with few outliers, the choice of loss function has minimal impact on the fit parameters.  
%For data with significant outliers, the inverse is true.  
We have left the data in their natural units, so $B$ has units of K\,\kms/mJy for the {\it WISE} and NVSS data, and is unitless for the CO data.  We compute the coefficient of determination $R^2$ and list these values and the fit parameters in Table~\ref{tab:correlations}.

Based on previous results, we expect that this linear relationship can approximate the correlation between \cii\ and the other tracers, as \citet{pabst2017} found between \cii\ and 8.0\,\micron\ emission in L1630.  For example, \citet{malhotra97} show that the \cii/IR ratio for galaxies is roughly constant at $2–7 \times 10^{-3}$ for $L_{\rm IR} <10^{11} L_{\odot}$.  \citet{delooze2011} claim that the $L_{\cii}/_{\rm FIR}$ ratio is typically linear within normal galaxies, but that it shows non-linearities for ultraluminous galaxies.  Although more complicated forms of the relationship are also found in the literature \citep[e.g., $\log(L_{\cii}) = A \times \log(L_{\rm IR}) + B$;][]{ibar15}, we prefer the simplicity of the linear relationship.

\begin{deluxetable}{llrrc}
\tablecaption{Correlations\tablenotemark{a} between \cii\ and other tracers  \label{tab:correlations}}
\tablehead{
\colhead{Tracer} & \colhead{Regions} & \colhead{$A$\tablenotemark{b}} & \colhead{$B$} &\colhead{$R^{2}$} \\
\colhead{} & \colhead{} & \colhead{\K\,\kms} & \colhead{\K\,\kms/\mjy or None} &\colhead{}}
\decimals
\startdata
$F_{12\,\mu m}$ & S235ION & $-5.01$ & $0.89 \pm 0.02$ & $0.70$ \\ 
& S235PDR & $-12.44$ & $1.02 \pm 0.02$ & $0.59$ \\ 
& S235AB & $0.64$ & $0.71 \pm 0.03$ & $0.51$ \\ 
& S235C & $-3.86$ & $0.61 \pm 0.02$ & $0.85$ \\ 
& S235BG & $-9.71$ & $0.59 \pm 0.02$ & $0.57$ \\ 
& Global & $-6.99$ & $0.89 \pm 0.01$ & $0.67$ \\ 
$F_{22\,\mu m}$ & S235ION & $20.44$ & $0.38 \pm 0.01$ & $0.44$ \\ 
& S235PDR & $-3.61$ & $0.89 \pm 0.02$ & $0.64$ \\ 
& S235AB & $14.95$ & $0.22 \pm 0.01$ & $0.54$ \\ 
& S235C & $-13.94$ & $0.86 \pm 0.01$ & $0.68$ \\ 
& S235BG & $-5.73$ & $0.79 \pm 0.13$ & $0.04$ \\ 
& Global & $16.19$ & $0.45 \pm 0.00$ & $0.55$ \\ 
$F_{\rm 1.4\,GHz}$ & S235ION & $32.34$ & $10.91 \pm 0.30$ & $0.36$ \\ 
& S235PDR & $36.35$ & $19.36 \pm 0.89$ & $0.15$ \\ 
& S235AB & $28.55$ & $14.86 \pm 0.68$ & $0.74$ \\ 
& S235C & $26.69$ & $62.88 \pm 2.47$ & $0.76$ \\ 
& S235BG & $7.42$ & $15.93 \pm 6.66$ & $0.04$ \\ 
& Global & $36.74$ & $11.14 \pm 0.20$ & $0.46$ \\ 
$I_{\rm 12CO21}$ & S235ION & $19.80$ & $0.62 \pm 0.01$ & $0.56$ \\ 
& S235PDR & $9.39$ & $0.33 \pm 0.02$ & $0.07$ \\ 
& S235AB & $2.86$ & $0.46 \pm 0.21$ & $0.18$ \\ 
& S235C & $1.75$ & $0.58 \pm 0.16$ & $0.00$ \\ 
& S235BG & $3.00$ & $0.02 \pm 0.03$ & $0.04$ \\ 
& Global & $14.15$ & $0.37 \pm 0.02$ & $0.23$ \\ 
$I_{\rm 13CO21}$ & S235ION & $26.25$ & $0.78 \pm 0.02$ & $0.57$ \\ 
& S235PDR & $14.75$ & $0.38 \pm 0.02$ & $0.12$ \\ 
& S235AB & $-4.99$ & $0.78 \pm 0.23$ & $0.20$ \\ 
& S235C & $8.87$ & $0.76 \pm 0.20$ & $0.00$ \\ 
& S235BG & $5.18$ & $-0.06 \pm 0.11$ & $0.01$ \\ 
& Global & $16.08$ & $0.47 \pm 0.02$ & $0.26$ \\ 
$I_{\rm 12CO32}$ & S235ION & $30.36$ & $1.44 \pm 0.05$ & $0.45$ \\ 
& S235PDR & $18.66$ & $1.06 \pm 0.06$ & $0.09$ \\ 
& S235AB & $11.58$ & $0.97 \pm 0.39$ & $0.04$ \\ 
& S235C & $19.53$ & $0.82 \pm 0.38$ & $0.00$ \\ 
& S235BG & $3.39$ & $0.02 \pm 0.04$ & $0.01$ \\ 
& Global & $21.84$ & $1.37 \pm 0.05$ & $0.23$ \\ 

\enddata
\tablenotetext{a}{Fits made using  $I_\cii=A + B\times I_\lambda$.}
\tablenotetext{b}{The A-values for the {\it WISE} correlations are uncertain due to the absolute flux calibration of the {\it WISE} data.}
\end{deluxetable}

%\citet{Stacey1991} report that there is a good correlation between the FIR continuum intensity and the [C II] line intensity, but that this correlation is significantly nonlinear such that the [C II] line-to-FIR continuum ratio decreases as the FIR intensity increases.  They posit that the increase in the [C II] line intensity with the FIR continuum intensity in gas-rich galaxies is not simply due to an increase in the number of [C II] and FIR-emitting clouds within the telescope beam.  This behaviour is usually referred to as the '[CII] deficit'. 
%\citet{Ibar2015} study the [C II]/IR ratio in galaxies out to $z=0.2$ and approximate the correlation with a simple linear regression in log-space: $log(L_{[C II]}) = A \times log(L_{IR}) + B$.

\subsubsection{\cii and MIR data}
\begin{figure}
   \centering
   \includegraphics[height=8in]{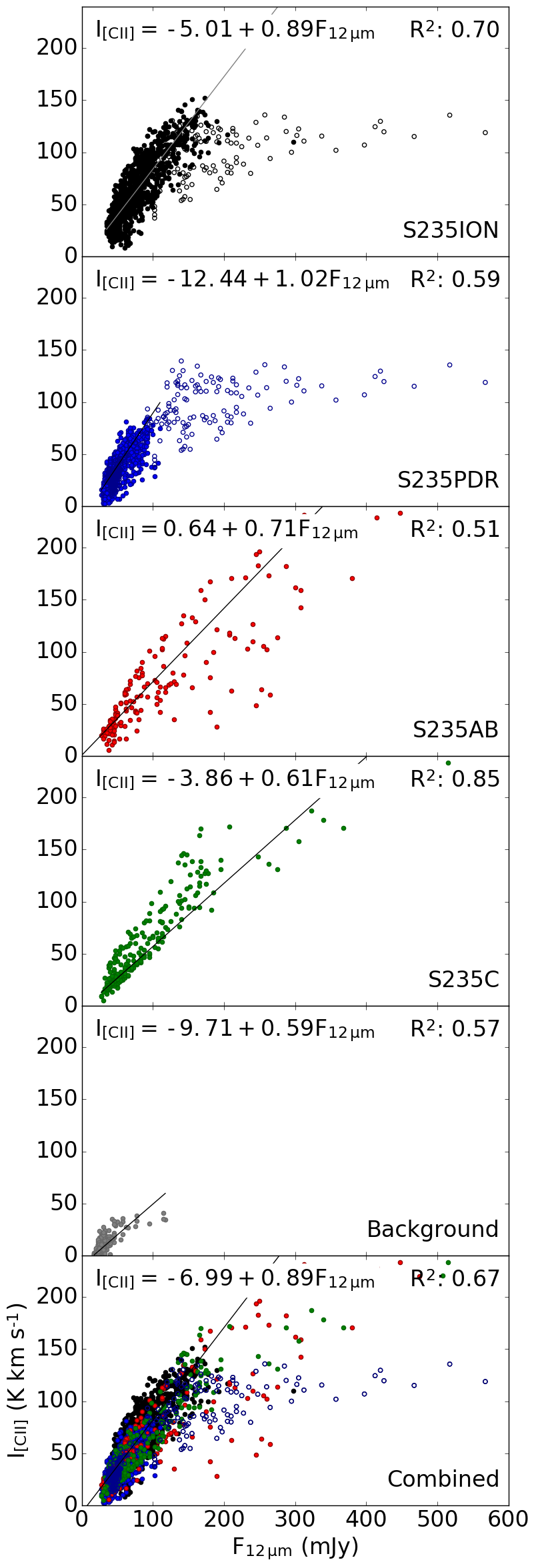}
   \caption{Correlations between \cii\ and $12$\microns emission in the regions of interest.  The S235ION region is spatially coincident with the ionized gas of S235, S235PDR with the PDR of S235, S235AB with blended sources S235A and S235B, and S235C with \hii\ region S235C (see Figure~\ref{fig:cii_m0map}).  Open circles in the top two panels are from locations of point sources; these data points are not used in the fits.
   %We perform linear regression fit to the data points and we represent the relationship between the dependent and independent variables in the 
   %appropriate panel. We also mark the R$^{2}$ values that gives impression about the goodness of the fitting. The top panels includes both the 
   %S235ION (blue) and S235PDR (yellow) regions. 
   The bottom panel shows data  from all investigated regions. %, with
   %and a fit, i.e. global fit, to 
   %those data points. 
   %filled black squares %representing data from the five %background regions. 
   We find good correlations between \cii\ and $12$\microns 
   flux density for all regions of interest.
   %and we observe that \cii\ intensity increase faster than $12$\microns intensity, especially in the PDR region.
   \label{fig:corr_w3}}
\end{figure}

The strongest correlations are between \cii\ and {\it WISE} 12\,\microns emission  (Figure~\ref{fig:corr_w3}), for which $I_{\cii}$\,[\K\,\kms] $\simeq 0.89 F_{12\,\microns}$ [\,mJy] (because of the background correction, the $A$ parameter is uncertain). The relationship is similar, and similarly strong for all investigated regions of interest, including for the five background regions.  We do not fit data points spatially coincident with IRS1, IRS2, the ionizing source of S235, or the star found in the north-east of the S235 PDR; data points from these locations are shown as open circles.
%correlations between the intensities of \cii\ and  $12$\microns emission (the other correlation plots can be seen in 
%Appendix~\ref{appdx:correlations}). 
Similarly, S235AB and S235C have contributions from point sources, and we suggest that these contributions lower the slope of the fit line.  
%If points with 12\,\microns flux densities $>7\,\mjy$ are excluded, the slope is the same as that of the S235ION and S235PDR regions.  
In support of this, the same trend is seen for the excluded locations coincident with IRS1 and IRS2 in S235MAIN (open circles in Figure~\ref{fig:corr_w3}).  
%We note that the correlation breaks down at small values of the MIR flux.
%It is also noticeable that \cii\ intensity increases faster than the $12$\microns intensity, especially in the PDR region. %This trend can be 
%observed in the smaller \hii regions (S235AB and S235C) as well, therefore this trend holds for the entire investigated S235 region (see bottom panel in 
%Figure~\ref{fig:corr_w3}). However, the S235AB shows some higher degree of scatter, which might be caused by the fact that this region contains more than one nebula. 
%The 
%global fit (bottom panel in Figure~\ref{fig:corr_w3}) shows that, in the lower intensity regime (up to $I_\cii \sim 70\,\K\,\kms$ and 
%$I_{12\microns} \sim 5\,\mjy$, the \cii\ emission is well correlated with the 12\,\microns emission \ldacom{R$^2 = xxx$}.
%with $\mathrm{I(12\microns) \simeq 0.048 \times I([CII]) + 0.441}$.

The slope of the \cii/12\,\microns relationship can be used as a proxy for the hardness of the radiation field.
The 12\,\microns emission is caused by PAHs excited by photons with energies $\sim\!5\,\ev$ \citep{voit92}, and carbon requires photons of energy $>11.3\,\ev$ to be ionized.  Therefore, 
softer radiation fields will have lower \cii/12\,\microns ratios.  We find that the slope of the \cii/12\,\microns fit in the background regions is lower than that found toward the \hii regions, which is indicative of a softer UV radiation field.  A softer UV field can be produced by moderate dust absorption in the diffuse medium.

The $22$\microns emission is reasonably well correlated with the \cii\ emission for S235AB and S235C, but is less strongly correlated for S235ION (Figure~\ref{fig:corr3}).  As stated previously, the 22\,\micron\ emission from \hii\ regions is due to small dust grains heated by stellar radiation, but young stellar objects and evolved stars are also bright at 22\,\micron. As is the case for the 12\,\microns correlation, the \cii/$F_{22\,\microns}$ correlation weakens at high values of $F_{22\,\microns}$; we again suggest that this is due to contributions to the 22\,\microns flux density from point sources.

\subsubsection{\cii and radio continuum data}
\begin{figure}
   \includegraphics[height=8.0in]{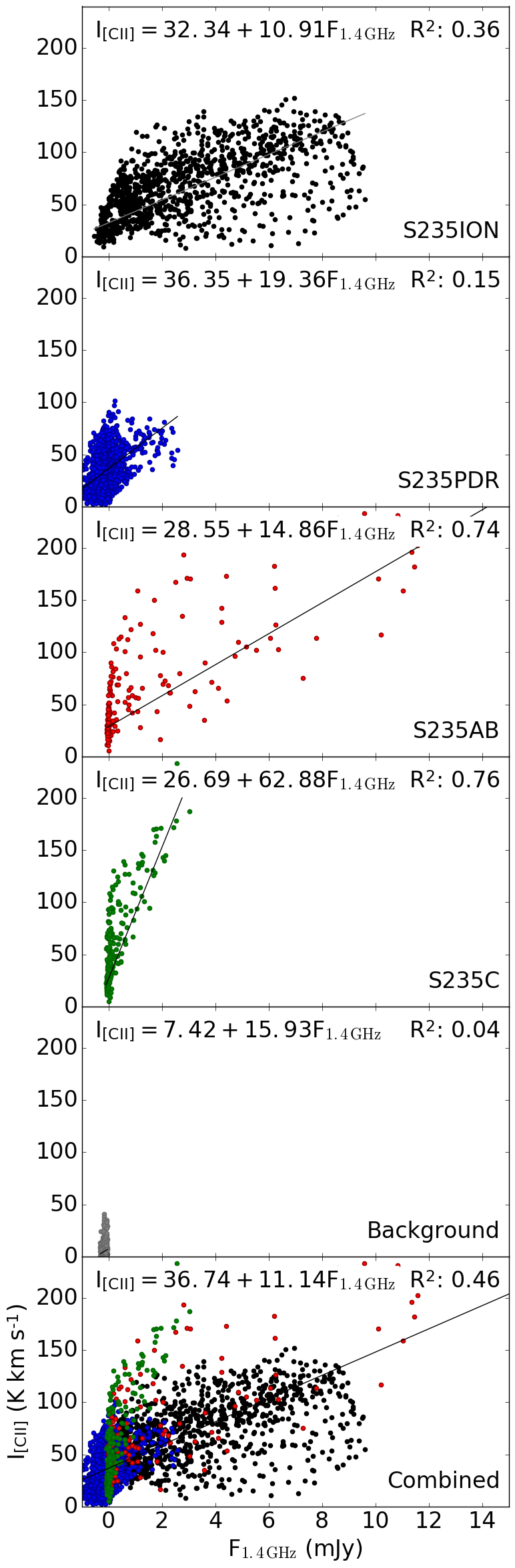}
\caption{Correlations between 1.4\,\ghz\ intensity and that of \cii.  Panels are the same as that of Figure~\ref{fig:corr_w3}.
The radio continuum emission shows a weak correlation with \cii\ intensity, which indicates that they trace different ionized phases of 
   the ISM in the S235 star-forming region.\label{fig:corr_nvss}}
\end{figure}

Similar to the \cii/22\,\micron\ correlation, the \cii/1.4\,\ghz correlation is weak in the hydrogen-ionized zone and largely absent outside it (Figure~\ref{fig:corr_nvss}).
Importantly, the weakness of the \cii and radio continuum correlation for S235ION indicates that the strong correlation between \cii and 12\,\microns emission here is not due to the ionized hydrogen gas itself.  Rather, this suggests that it is PDRs along the line of sight that are responsible for the 12\,\microns emission.

Despite having similar \cii\ intensities and reported ionizing sources, S235AB has much higher 1.4\,\ghz\ peak flux density than S235C.  The peak RRL and radio continuum emission in S235AB is much higher than that of S235C.  %We therefore suggest that the 1.4\,\ghz flux in S235AB has a high optical depth and 
%is caused by hydrogen ionized by photons with energies $>13.6\,\ev$, and C$^+$ is ionized by photons with energies $>11.3\,\ev$ this may suggest that S235AB has a harder radiation field than S235C, We cannot be sure of this conclusion, however.  At 1.4\,\ghz, \hii\ regions, especially compact ones, can be optically thick, so the 1.4\,\ghz\ fluxes 
%may not reliably trace the full ionized hydrogen content.

\subsubsection{\cii and molecular gas}
There is a weak correlation between the strength of \cii\ emission and that of molecular gas for the S235ION region ($R^2$ values near 0.5), but this correlation is weak for all other investigated regions (see Figures~\ref{fig:corr1}). The two smaller \hii regions (S235A and S235C) show weak correlations between \cii integrated intensity and the intensity of the observed molecular gas. Evidently, the strong correlation between \cii and molecular gas observed for entire galaxies does not hold for at the pc-scale for individual \hii\ regions.  For such small regions, even \cor\ may be optically thick, which would decrease the significance of the correlation. The strength of the galaxy-wide correlation is likely due to the fact that both \cii\ and molecular gas trace star formation.
The correlation at large scales may reflect the general relation between molecular gas and star-formation rather than local feedback physics.

%does not show as good correlation as with $12$\microns. The higher scatter in the inner, ionized 
%region (blue data points in the top panel) support the assumption that the scattered data represent the emission from the main ionizing source and the dense clumps as, 
%basically, all the emitted radiation seems to arise from these sources (see panel (d) in Figure~\ref{fig:int_maps}). Similar trend can be noticed, namely the \cii\ 
%emission increases faster than the $22$\microns dust emission. We found that the ionized carbon is correlated with the mixture of PAHs and VSGs with 
%$\mathrm{I(22\microns) \simeq 0.133 \times I([CII]) - 2.434}$.

%The \cii\ and the observed molecular gas tracers show lack of good correlations  In the ionized part, 
%the observed transitions of the $^{12}$CO and its isotopologue $^{13}$CO, and \cii\ indicating an almost unity ratio with large uncertainty. The emission from the 
%surrounding PDR region (yellow data points) is well isolated in the correlation plots.  The overall relationships between $^{12}$CO and $^{13}$CO and \cii\ are significantly weaker than between dusts 
%($12$\microns and $22$\microns) and \cii. This indicates that a significant portion of the \cii\ emission have no counterpart with the carbon-monoxid.

\subsection{Difference maps}
\label{subsec:map_ratios}

\begin{figure*}
   \centering
\includegraphics[width=0.25\textwidth]{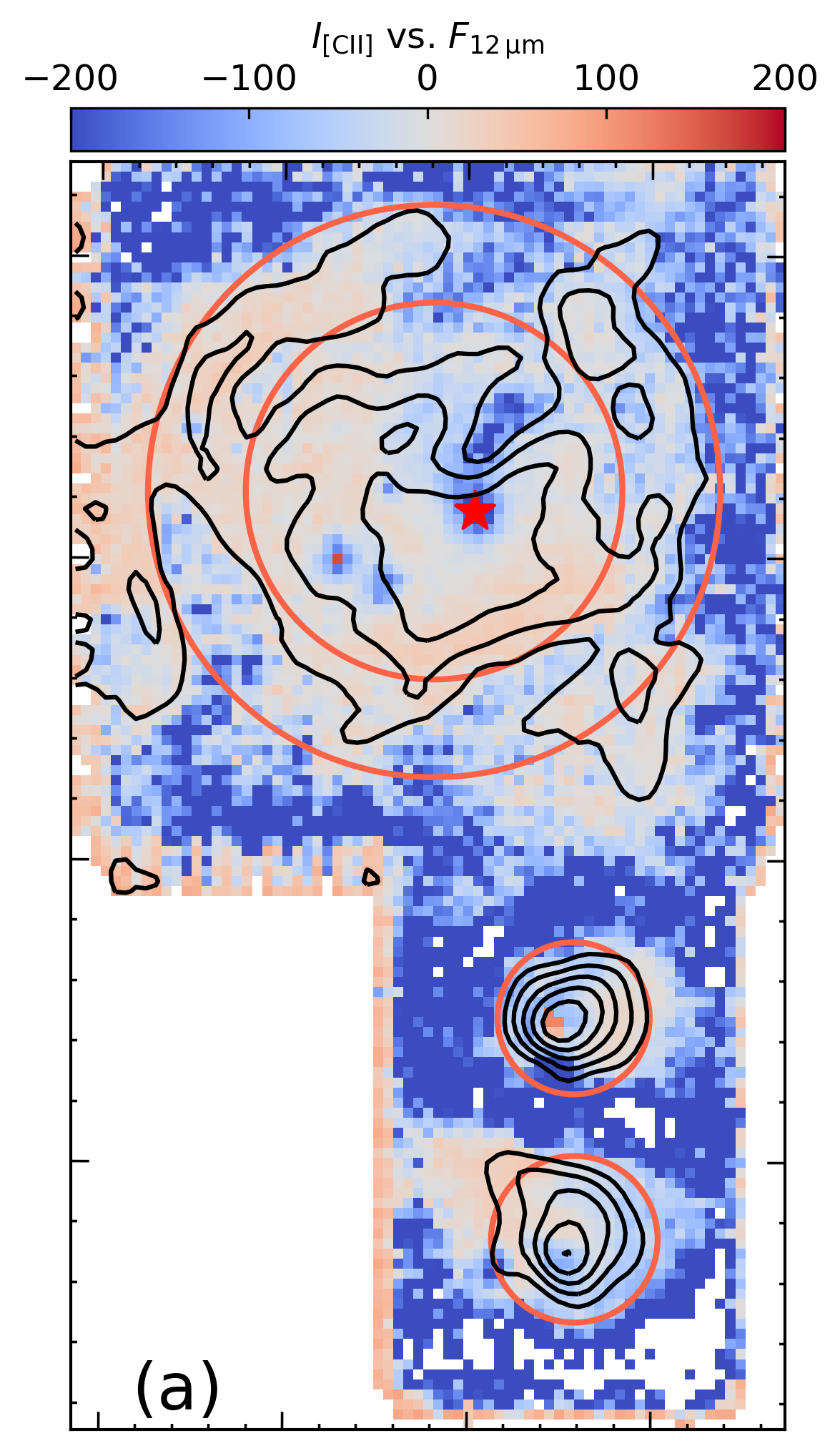}
   \includegraphics[width=0.25\textwidth]{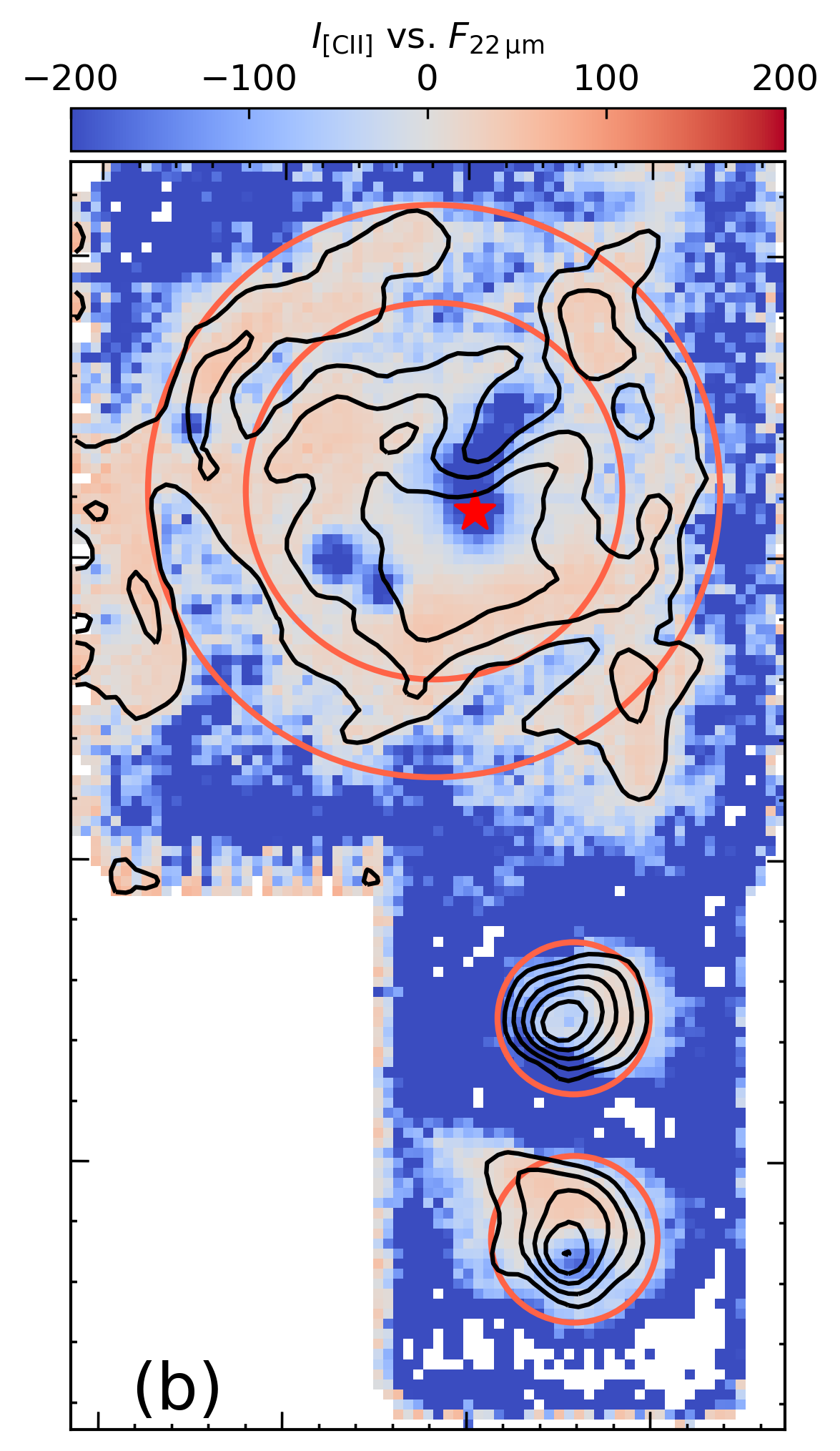}
   \includegraphics[width=0.25\textwidth]{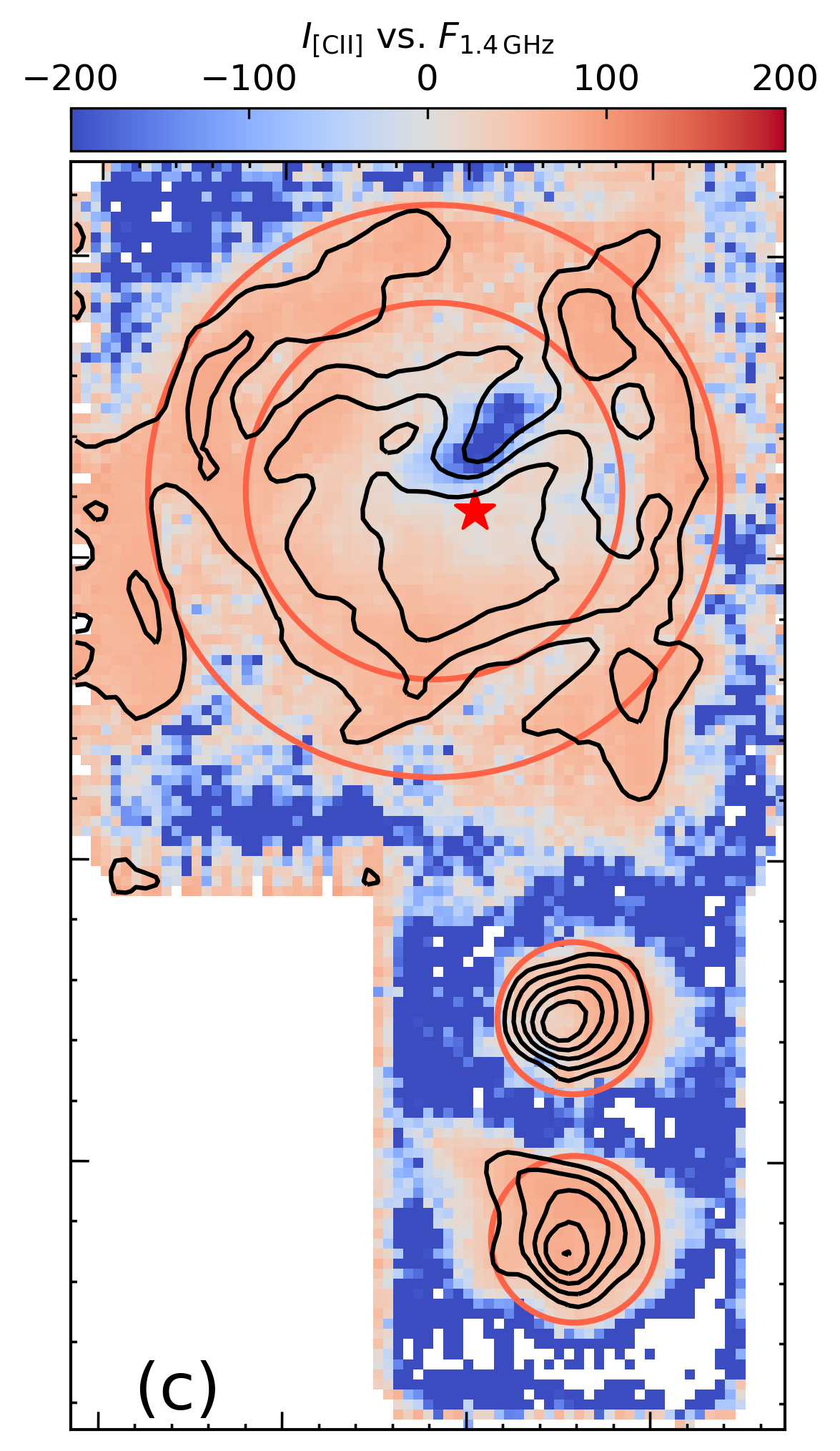}
      \includegraphics[width=0.25\textwidth]{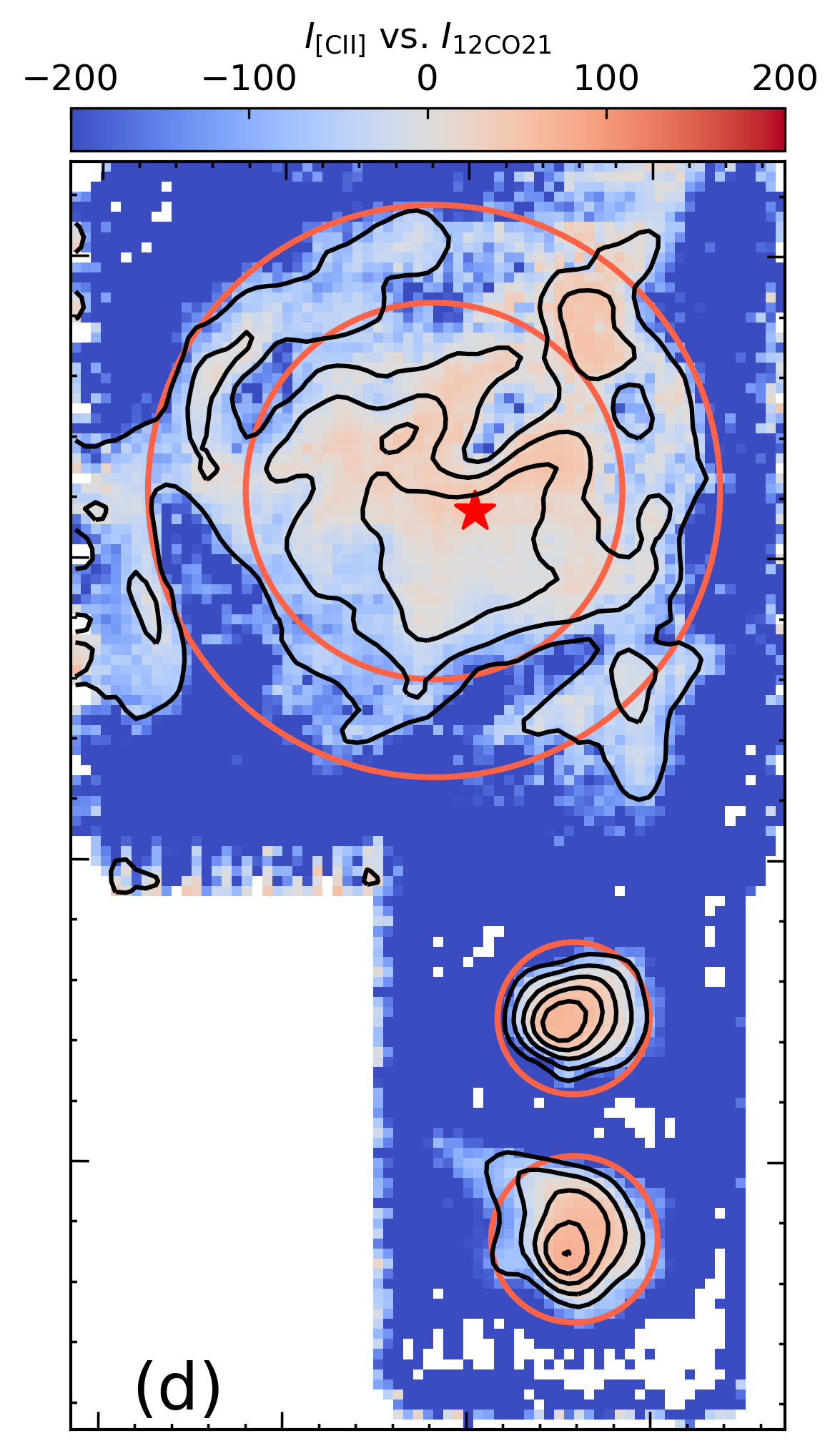}
   \includegraphics[width=0.25\textwidth]{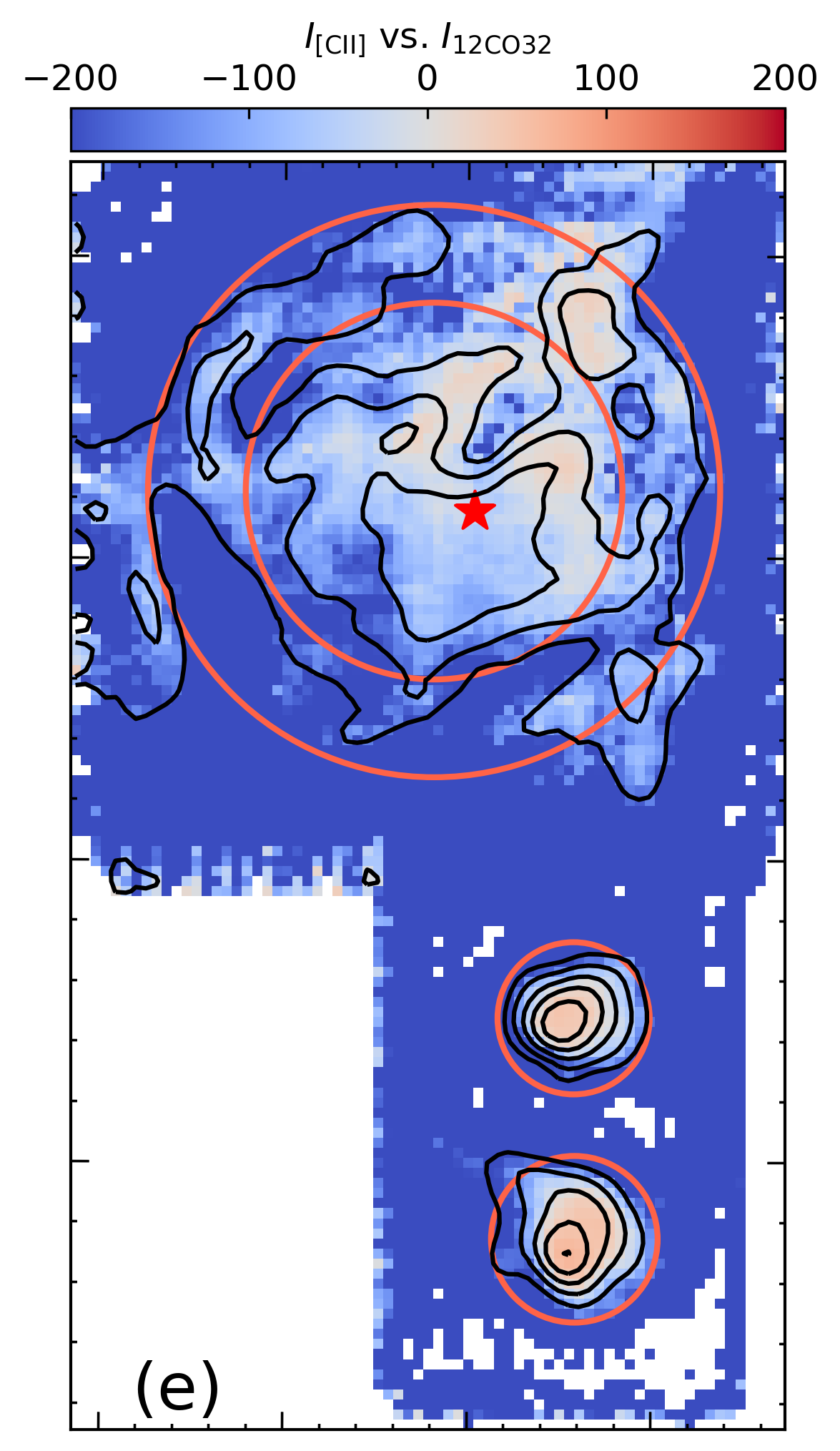}
   \includegraphics[width=0.25\textwidth]{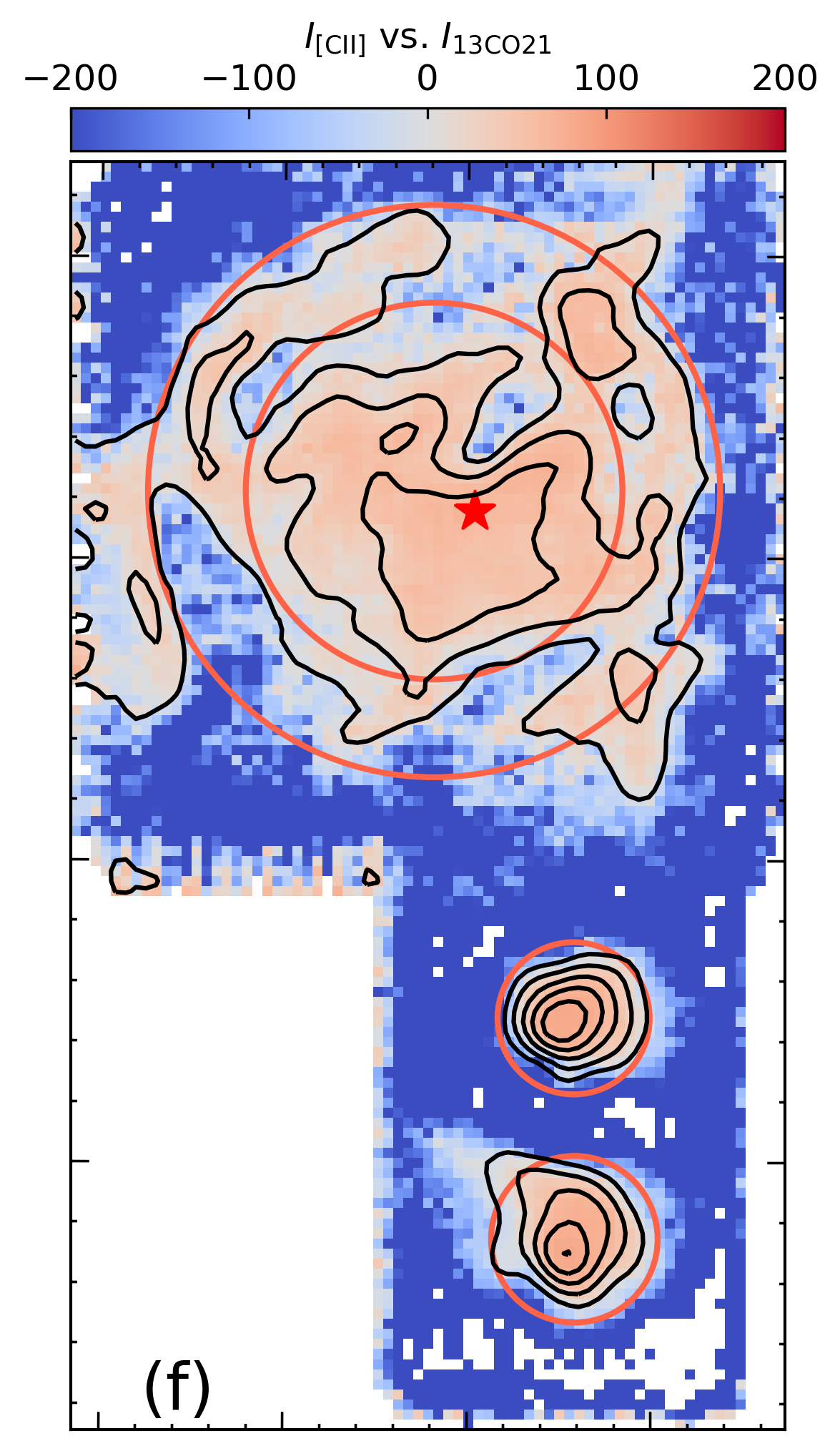}
   \caption{Intensity difference maps of \cii\ and 
   $12$\microns (a), $22$\microns (b), $1.4\,$\ghz radio continuum (c), $^{12}$CO($J=2-1$) (d), $^{12}$CO($J=3-2$) (e), and $^{13}$CO($J=2-1$) (f). 
     The covered area is the same as on 
   Figure~\ref{fig:cii_m0map}.  The maps show locations where the \cii\ emission is bright relative to the other tracers (red), and those where the \cii\ emission is faint relative to the other tracers (blue). Black contours represent \cii\ integrated intensity. The stars again mark the location of the ionizing source for S235. \label{fig:ratio_maps}}
\end{figure*}

We investigate spatial differences in the correlation between \cii\ emission and that of the other tracers in the percentage difference maps of Figure~\ref{fig:ratio_maps}.  The maps give the difference between the actual \cii\ intensity and the \cii\ intensity that would be expected from the other tracers if they strictly followed the correlation measured for S235ION (Table~\ref{tab:correlations}), normalized by the measured \cii\ integrated intensity. We do, however, adjust the offset (the ``A'' parameter in the fits) so that emission from S235ION has an average difference of zero.  These maps show locations where the \cii\ emission is bright relative to that expected from the other tracers (red) and where the \cii\ emission is faint compared to that of the other tracers (blue).  The values of the maps are therefore less important than the distribution of values.

These maps support our results from the preceding subsections.  The \cii/12\,\microns difference is nearly uniform across the map, indicating that there is little change in the relative intensities in the field.  The \cii\ versus 22\,\microns difference map shows that the relative intensities in the ionized hydrogen and PDR zones differ.  Locations within S235ION of bright IR emission from point sources (IRS1, IRS2, and the ionizing source) appear as \cii deficits in these panels.  For the \cii/radio continuum difference map, the ionized hydrogen zone again has different values compared with the rest of the map.  Compared with \cii, there is a negligible amount of CO gas in the cavity located to the north. In the 
ionized hydrogen region (S235ION) and most of the PDR region (S235PDR), the CO difference maps do not show a large difference between the ionized and molecular gas components.

\subsection{Implications for extragalactic studies of \cii}
We can use the correlations established above to investigate the implications for studies of galaxy-wide \cii\ emission.  The 158\,\microns\ \cii\ emission line is known to be an excellent tracer of ongoing star formation, as are CO, radio continuum, and MIR emission.  Therefore, we would expect a correlation between all these measures of active star formation.

We create a catalog of galaxies observed in \cii, CO, radio continuum, and {\it WISE} emission.  We take \cii\ emission line data from \citet{Brauher2008}, who observed 227 galaxies in far-IR continuum and \cii\ line emission using the Infrared Space Telescope.  These \cii\ data have an angular resolution of $\sim85\arcsec$.  We supplement the \cii\ data set with \cii\ observations by \citet{stacey1991} using the Kuiper Airborne Observatory, which have an angular resolution of $\sim55\arcsec$.  We compile $^{12}{\rm CO} (1-0)$ data from the FCRAO Extragalactic CO Survey \citep{Young1995}, which have an angular resolution of $46\arcsec$. For galaxies with more than one observed CO position, we choose the position closest to that of \cii. We complement our CO and \cii\ data with NVSS 1.4\,GHz continuum emission data (at a resolution of $45\arcsec$) and \emph{WISE} 12 and 22\,$\mu$m data (at resolutions of $6\arcsec$ and $12\arcsec$), both compiled through spatial searches of the public point source catalogs.  We attempt to remove the stellar contribution to the 12 and 22\,\micron\ fluxes by subtracting the flux at 3.5\,\micron, since for stars the 3.5\,\micron\ band flux is similar to that in the 12 and 22\,\micron\ bands \citep{nikutta14}.  
%This survey includes CO data for 1412 positions in 300 galaxies taken with the Five College Radio Astronomy Observatory. 
We determine galaxy types and angular diameters using the SIMBAD database.  

The galaxies have a range of diameters from $\sim30\arcsec$ to $\sim 1\degree$.  Observations of each galaxy will therefore be sensitive to emission from different galactic components.  For the largest galaxies, the data will contain emission only from the central region.  This emission may be dominated by contributions from an older stellar population or the presence of a black hole.  Observations of smaller galaxies will contain a better sampling of the total emission.  To ensure that our comparisons deal with similar portions of all galaxies, we restrict our sample to those with diameters $<150\arcsec$.

Our combined data set includes 32 galaxies: 9 active galaxies (which we call ``AGN''; Simbad classifications ``LIN,'' ``Sy1,'', ``Sy2,'', and ``SyG''), 3 \hii\ and starbursting galaxies (which we call ``\hii''; Simbad classifications ``H2G'' and ``SBG''), 19 normal spiral galaxies, (which we call ``Gal''; Simbad classifications ``G,'' ``GiC,'' ``GiG,'' ``GiP,'' and ``IG'') and 1 of other type (Simbad classification ``EmG'').  Nearly all galaxies in our sample have measured intensities for all tracers.

\begin{figure*}
   \centering
   \includegraphics[width=0.45\textwidth]{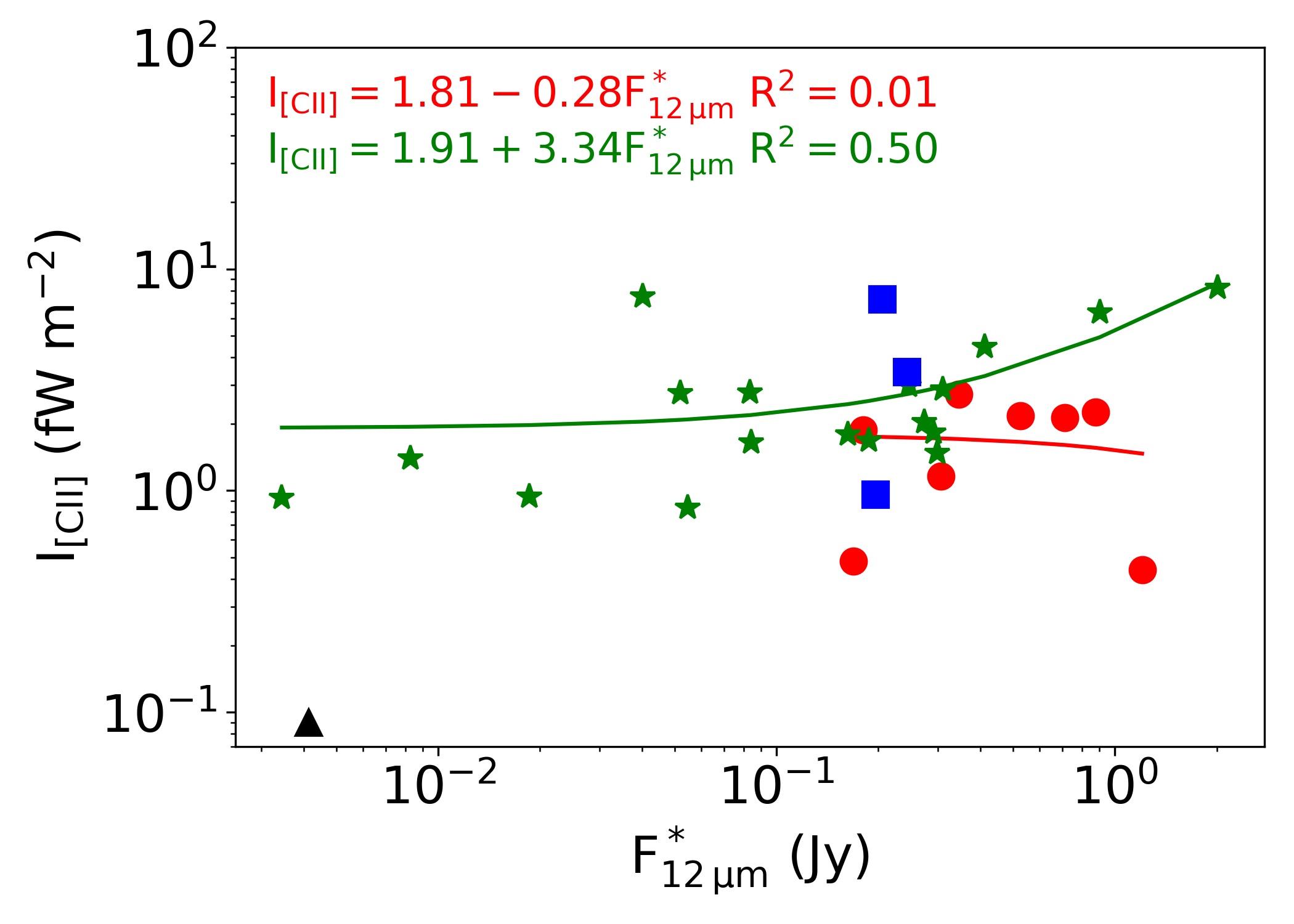}
   \includegraphics[width=0.45\textwidth]{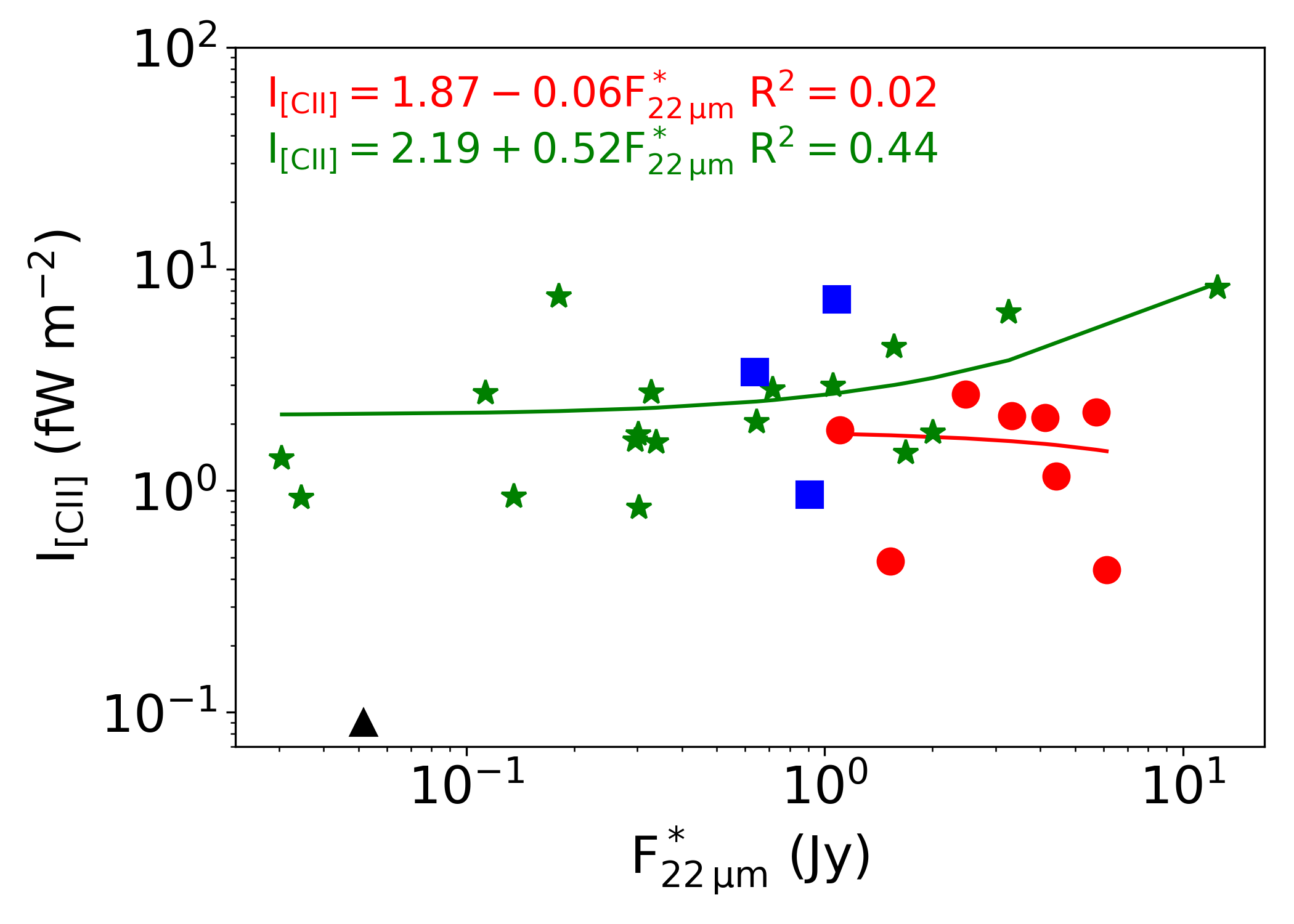}
      \includegraphics[width=0.45\textwidth]{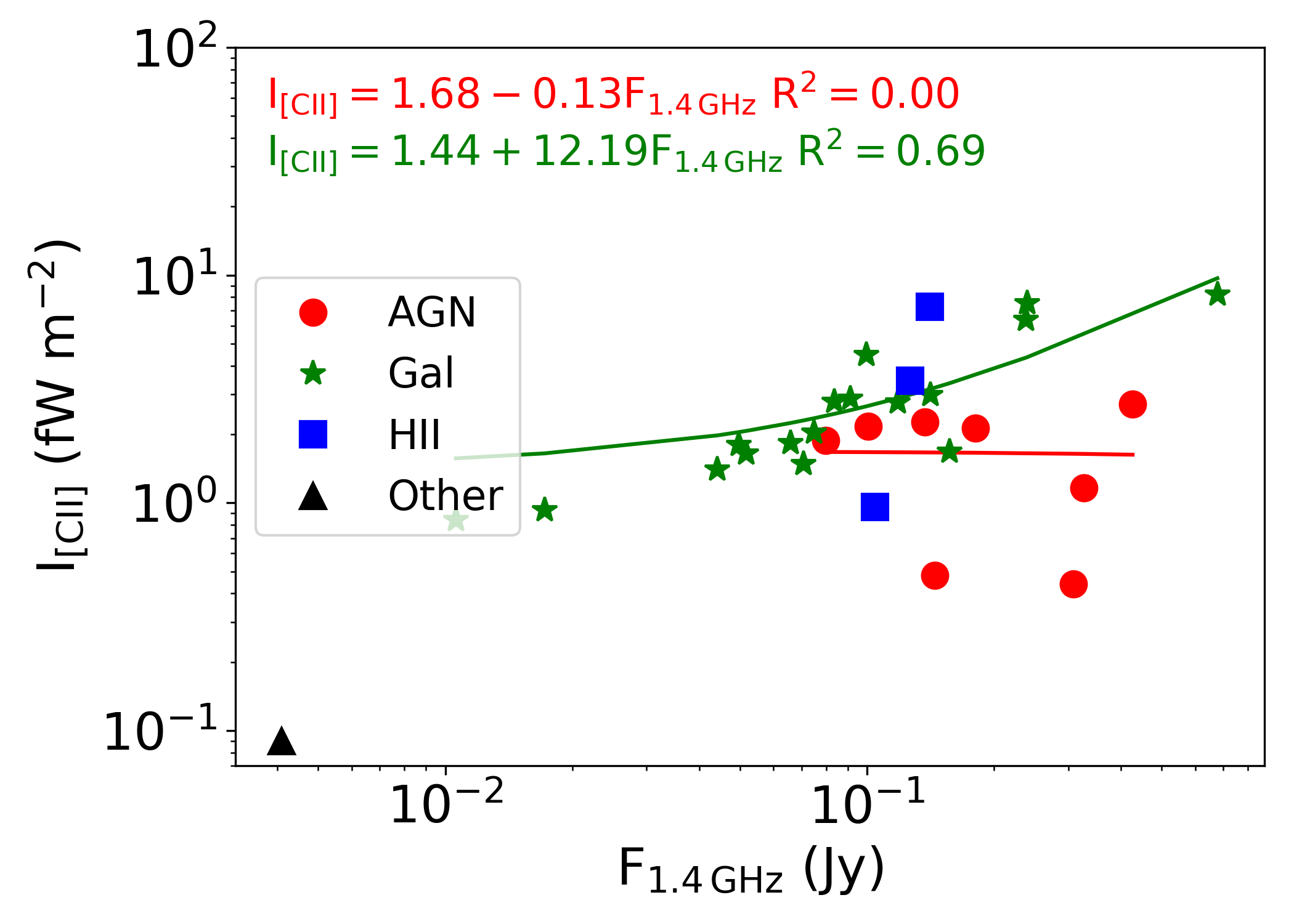}
   \includegraphics[width=0.45\textwidth]{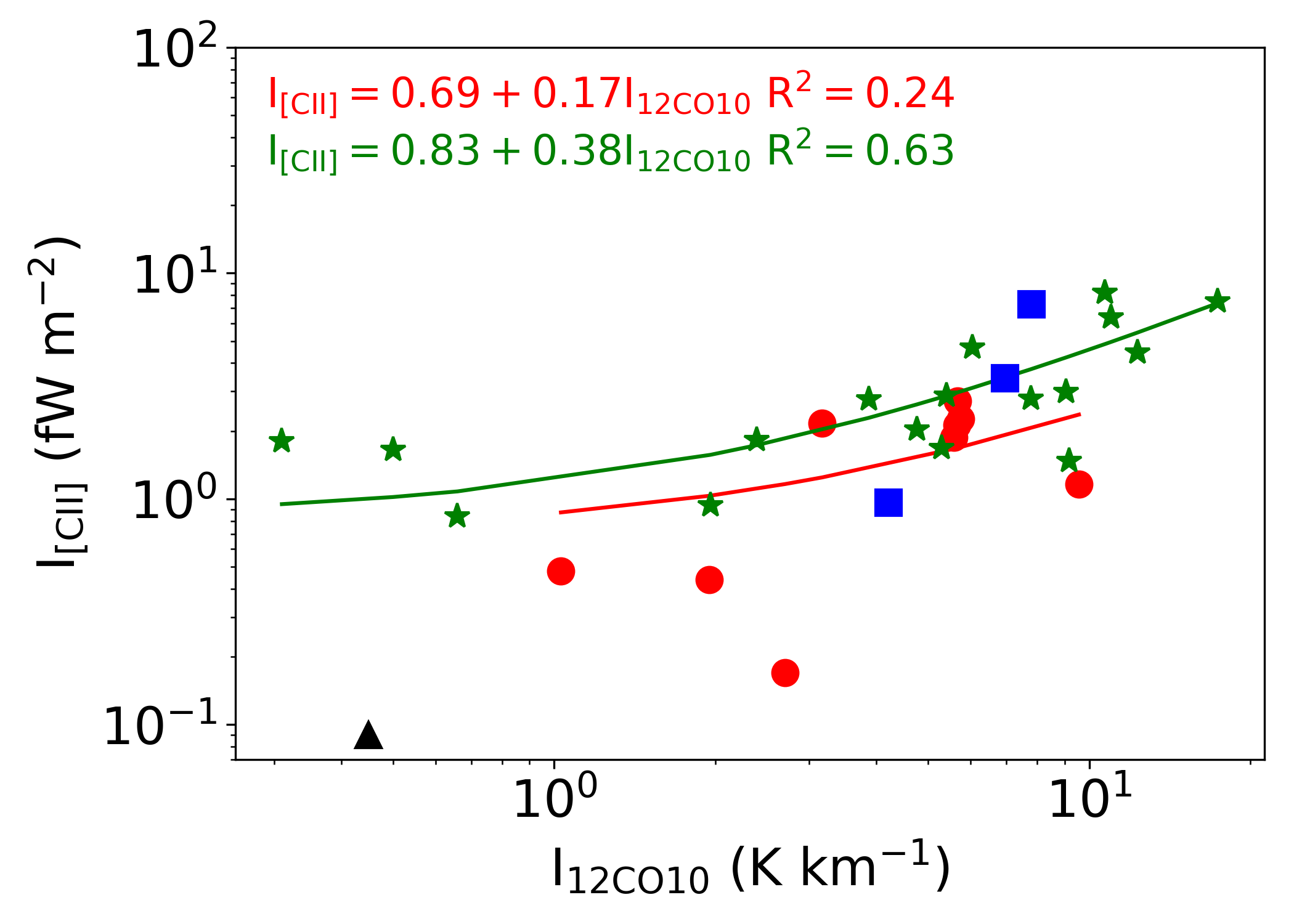}
\caption{Correlations between \cii\ and {\it WISE} 12\,\micron\ emission (top left) {\it WISE} 22\,\micron\ emission (top right) NVSS 1.4\,\ghz\ emission (bottom left), and $^{12}{\rm CO}\,(1-0)$ emission (bottom right) for various galaxy types.  For normal and starbursting galaxies, the correlations are similar to what was found for S235. \label{fig:galcorr}}
\end{figure*}

We fit a linear regression to the compiled data of the same form as before ($I_\cii = A + B I_\lambda)$ using an orthogonal distance regression (ODR), and show the results in Figure~\ref{fig:galcorr}.  An ODR minimizes the summed squared distances between the fit and the data points, and therefore simultaneously takes into account errors on the dependent and independent variables.  We perform this fit individually for AGN and normal spiral galaxies.  In these figures, the asterisk on the 12 and 22\,\micron\ data represents that it is point-source corrected.  We expect that the \cii\ emission from \hii\ galaxies would show good correlations with all tracers if there were sufficient data for a fit, since the \cii\ emission traces ongoing star formation.  The normal spirals should also have good correlations, and due to the potentially intense emission from the active nuclei themselves, the AGN correlations should be poorer.

These hypotheses are only partially born out.  There are positive correlations between \cii\ and the other tracers for the normal galaxies.  The strongest such correlations are between \cii\ and NVSS or \cii\ and CO.  There is no correlation between \cii\ strength and that of the other tracers for the AGN sample.  There are only three \hii\ galaxies, so although we do not perform fits, the trends appear consistent with our expectations.

%For the normal spiral galaxies, the slope of the fits is similar to that found for S235, providing some support to the notion that results for individual \hii\ regions may be extrapolated to entire galaxies.  
All the tracers provide a measurement of star formation activity.  The {\it WISE} bands, however, may contain contributions from stars that are not removed in our method, the 1.4\,\ghz\ flux densities are potentially contaminated by the emission from central black holes (even in galaxies that are not active, like the Milky Way), and the CO integrated intensities may include contributions from CO that is not currently in the process of forming stars.  That all these tracers are correlated with the integrated \cii\ emission is not surprising.  It is encouraging, however, that we find similarly strong correlations for the single high-mass star forming region S235 as we do for the integrated intensities from galaxies.

We are comparing results from our \cii observations of a single \hii\ region complex at sub-pc resolution with those integrated over entire galaxies.  Results obtained at such different spatial scales should not be over-interpreted.  It does, however, appear that correlations between \cii\ and other SFR tracers may only exist because they all trace star formation when averaged over large areas, not because the small-scale emission properties are related. 

%-------------------------------------------------------------------------------------------
\section{Summary}
\label{sec:sum}

We compared \textit{SOFIA} observations of \cii\ $158$\microns emission with {\it WISE} MIR 
$12$\microns and $22$\microns data, NVSS $1.4\,$GHz radio continuum data, and molecular data ($^{12}$CO\,($2-1$), $^{12}$CO\,($3-2$), $^{13}$CO\,($2-1$)) toward the star formation complex Sh2-235. The main purpose of the present study is to determine where 
 \cii\ emission arises, relative to the emission from tracers of different phases of the ISM. We analyzed emission from four regions of interest representing 
the inner ionized hydrogen in the main \hii\ region S235, the PDR of S235, and the two smaller \hii regions in the field, S235A and S235C. We also sampled the diffuse background emission in five locations.

Roughly half of the \cii\ emission in S235 originates in the direction of the ionized hydrogen region, mainly close to the location of the central ionizing star.  The remaining \cii\ emission in the S235 region arises from the filamentary PDR surrounding the ionized region.  This direction toward the ionized hydrogen also includes potential contributions from front- and back-side PDRs along the line of sight.  We find little evidence that the \cii\ emission seen in the direction of the ionized hydrogen is actually from the ionized hydrogen volume; the data instead are consistent with \cii\ emission seen in the direction of the ionized hydrogen originating in the back-side PDR.  Our results therefore are similar to those of \citet{pabst2017} and \citet{goicoechea2015a} that no more than 10\% of the \cii\ emission is cospatial with the ionized hydrogen gas.
The two additional small \hii\ regions in the field have similar \cii\ emission strengths, despite very different radio continuum strengths.

The \cii\ emission is strongly correlated with {\it WISE} $12$\microns intensity across the entire field.  The correlation holds for all three \hii\ regions, and is also reasonably strong for the background regions.  This indicates that 12\,\microns intensity can predict \cii intensity from \hii regions at various evolutionary stages ionized by O and early B-type stars having a range of ionizing fluxes, and also from diffuse C$^+$ gas.
The UV radiation field found in \hii\ region PDRs excites emission in the {\it WISE} 12\,\micron\ band and also ionizes carbon, leading to \cii\ emission.  The 12\,\microns\ band traces emission from PAHs and warm dust.  PAHs fluoresce when hit by soft-UV photons \citep[$E\gtrsim 5\,\ev$;][]{voit92}, and ionized carbon is created by photons with $E>11.3\,\ev$.
Both \cii and 12\,\microns emission trace emission from surfaces irradiated by ultraviolet photons.  They are therefore surface rather than volume tracers.
 We expect the same relationship between \cii\ and  {\it Spitzer} 8.0\,\microns data \citep{pabst2017}, since the {\it WISE} 12\,\microns and {\it Spitzer} 8.0\,\microns band emission are themselves strongly correlated for \hii\ regions \citep{anderson2014, makai17}. 

Radio continuum and {\it WISE} 22\,\micron\ emission are both centrally peaked surrounding the ionizing source of S235.  The correlation 
between \cii\ emission and {\it WISE} $22$\microns emission is weaker that that found for {\it WISE} 12\,\micron\ data.  The correlation between \cii\ emission and 1.4\,GHz radio continuum is only apparent toward the ionized hydrogen gas, and is weak toward the PDR.  This further supports the idea that the \cii/12\,\microns correlation found toward the ionized gas of S235 is in large part caused by line-of-sight PDRs.
%At this wavelength, 90\% of the emission comes from the vicinity of the central star and the 
%infrared sources.
The $1.4\,$GHz radio continuum traces emission from ionized hydrogen. The lack of good spatial correlation with the \cii\ emission indicates that we observe 
emission from different ionized gas components in S235. %Essentially all the radio continuum ($\sim 98\%$) is emitted from a small and compact region that contains 
%the ionizing star and IRS1, and IRS2.

%The {\it SOFIA} \cii\ emission shows good correlation with molecular gas intensity toward the ionized hydrogen region of S235, but poor correlation outside of it.  As there is little molecular gas within the ionized region itself, the origin of the correlation is unclear.
%Much of the  ionized carbon emission does not arise from regions that are predominantly molecular.  

Averaged over entire spiral galaxies, we see similar trends; the intensity of \cii and {\it WISE} data are correlated, although weakly so.  We find modest correlations between \cii\ and NVSS 1.4\,\ghz\ intensities, as well as between \cii\ and $^{12}{\rm CO} (1-0)$ intensities, for normal spiral galaxies.  Galaxies hosting AGN show poor correlations.
 
Therefore, although many tracers are correlated with the strength of \cii\ emission, only {\it WISE} 12\,\micron\ emission is correlated on small scales within S235, as well as averaged over entire galaxies.  Further \cii\ observations of a larger sample of Galactic \hii\ regions would allow us to determine if our results for S235 are representative.

%Considering all data points (i.e., global fit to the combined data set), we find that the \cii\ intensity is increasing faster by roughly $2\times$, $2\times$,  
%$6\times$, $21\times$, $7\times$ and $2\times$ than the intensity of the $^{12}$CO($J=2-1$), $^{12}$CO($J=3-2$), $^{13}$CO($J=2-1$), $F_{12\microns}$, 
%$F_{22\microns}$ and $F_{1.4\,GHz}$ emission, respectively. We find the best correlation between \cii\ and the PAH tracer $12$\microns emission.

\acknowledgements

%We thank the anonymous referee for his/her comments, which 
%improved the paper.
This work is based on observations made with the NASA/DLR Stratospheric Observatory for Infrared Astronomy 
({\it SOFIA}). {\it SOFIA} is jointly operated by the Universities Space Research Association, Inc. (USRA), under NASA 
contract NAS2-97001, and the Deutsches SOFIA Institut (DSI) under DLR contract 50 OK 0901 to the University 
of Stuttgart.  We thank West Virginia
University for its financial support of GBT operations, which
enabled some of the observations for this project. 
This research has
made use of NASAs Astrophysics Data System Bibliographic Services and
the SIMBAD database operated at CDS, Strasbourg, France.  This
publication makes use of data products from {\it WISE}, which is a joint
project of the University of California, Los Angeles, and the Jet
Propulsion Laboratory/California Institute of Technology, funded by
the National Aeronautics and Space Administration. 

Financial support for this work was provided by NASA through awards $\#04-0132$ and $\#05-0061$ issued by USRA to LDA.
VO was supported by Deutsche Forschungsgemeinschaft (DFG) within the Collaborative Research Center 956, project C1 - project ID 184018867.
NS acknowledges support by the french ANR and the German DFG through the project ``GENESIS'' (ANR-16-CE92-0035-01/DFG1591/2-1) 
and funding by the Bundesministerium f\"ur Wirtschaft
und Energie through the grant MOBS (50OR1714).
MSK was supported by Russian 
Foundation for Basic Research, grant number 18-32-20049. 
AMS was supported by Russian Foundation for Basic Research, grant 18-02-00917 and by the Act 211 Government of the Russian Federation, agreement No. 02.A03.21.0006.  

\facility{{\it SOFIA}, GBT}

\software{AstroPy \citep{astropy2013, astropy2018}, AplPy \citep{aplpy2012,aplpy2019}, IDL}
%-------------------------------------------------------------------------------------------
\bibliographystyle{aasjournal}
\bibliography{CIIpaper}

%-------------------------------------------------------------------------------------------
\clearpage

\appendix
\renewcommand\thefigure{\thesection.\arabic{figure}}

\section{Correlations between \cii\ 22\micron\ emission and between \cii\ and molecular gas emission}
\label{appdx:correlations}
\setcounter{figure}{0} 

Here, we show correlations between the intensity of integrated \cii\ emission and the intensities of {\it WISE} 22\,\micron\ and CO emission.  The parameters of the fits shown in the figures are given in Table~\ref{tab:correlations}.
For all figures in this appendix, the symbols, panels and fits are similar to those in Figures~\ref{fig:corr_w3} and \ref{fig:corr_nvss}. 

\begin{figure*}[!h]
   \centering
   \includegraphics[height=6.5in]{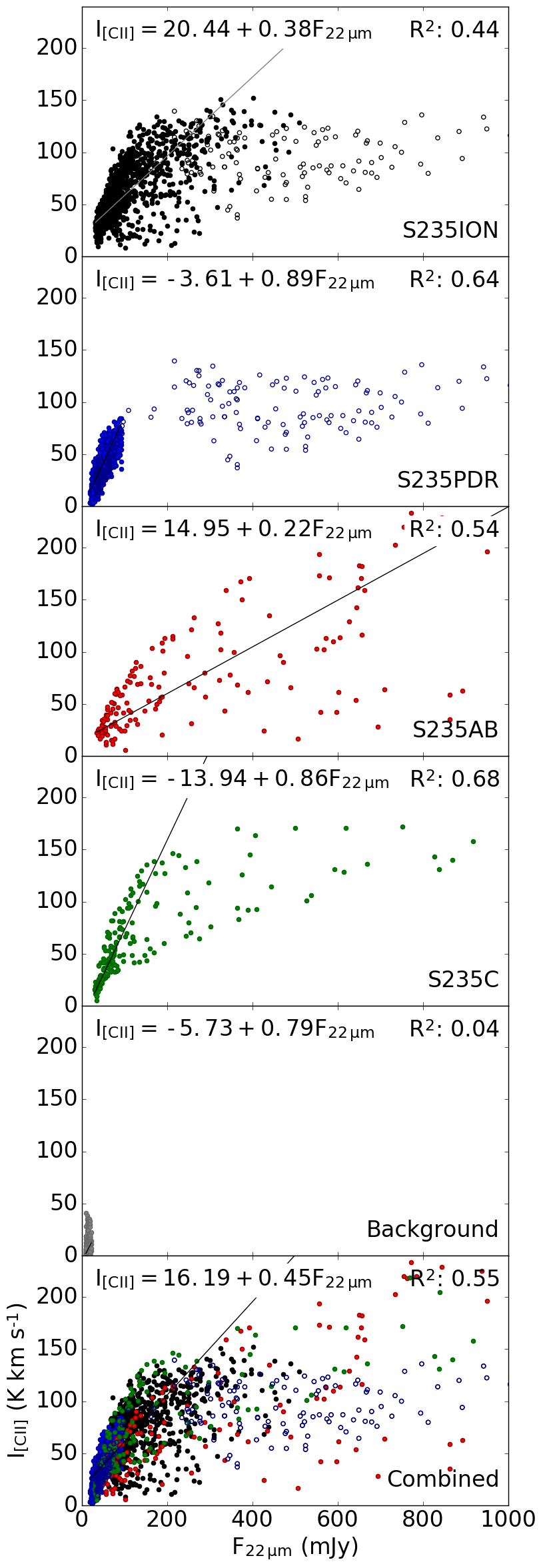}
   \caption{Correlations between 22\,\micron\ intensity and that of \cii.  
   The S235ION region is spatially coincident with the ionized gas of S235, S235PDR with the PDR of S235, S235AB with blended sources S235A and S235B, and S235C with \hii\ region S235C (see Figure~\ref{fig:cii_m0map}). The $22$\microns emission shows a good correlation with the \cii\ emission, albeit with larger spread than the $12$\microns 
   emission (Figure~\ref{fig:corr_w3}).  As seen in Figure~\ref{fig:int_maps}, much of the $22$\microns emission arises near to the central ionizing star.\label{fig:corr3}}
\end{figure*}

\begin{figure*}[!h]
   \centering
   \includegraphics[height=6.5in]{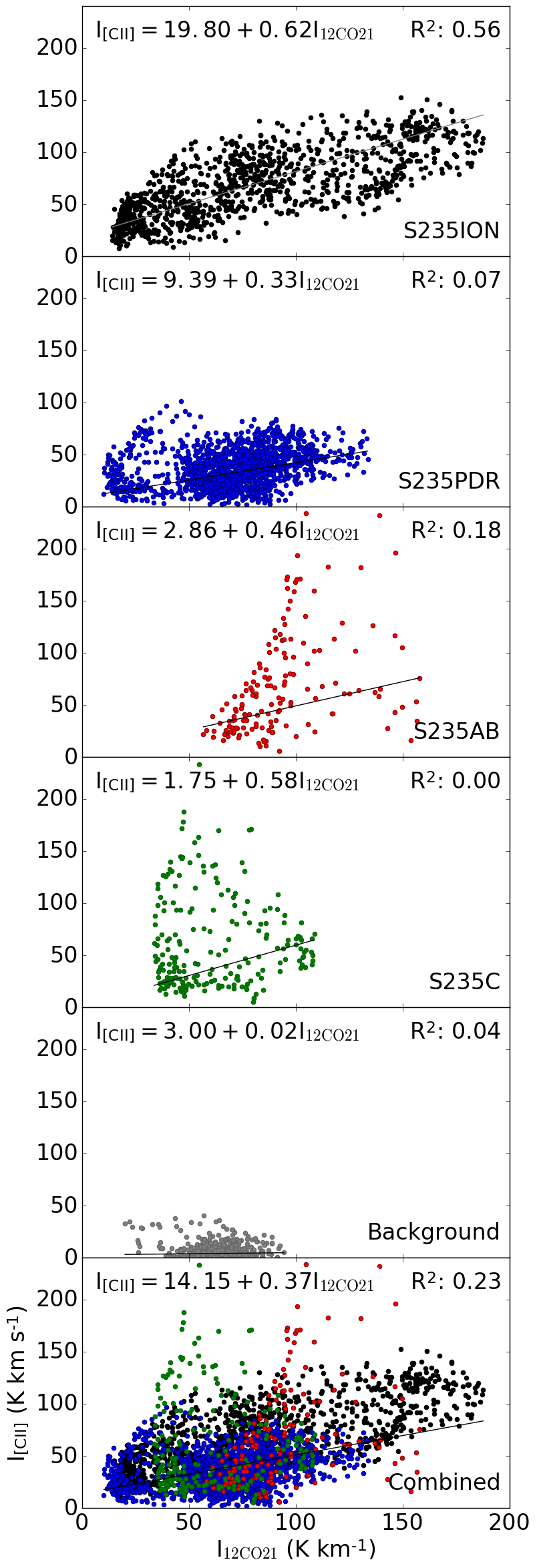}
   \includegraphics[height=6.5in]{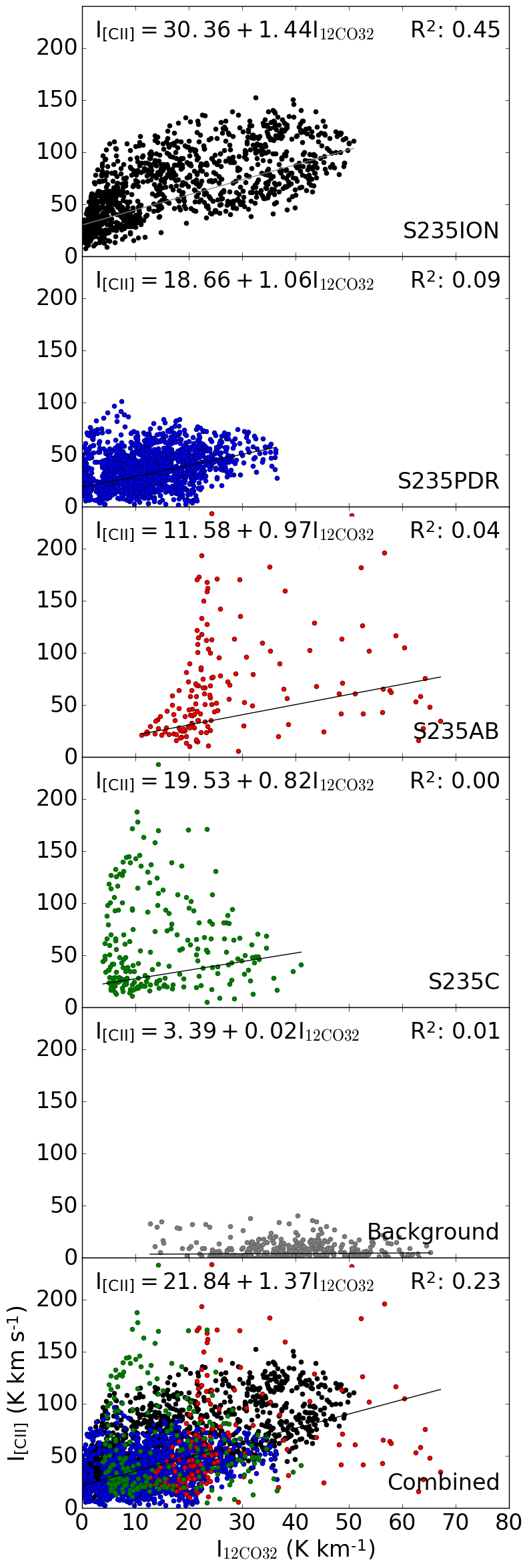}
   \includegraphics[height=6.5in]{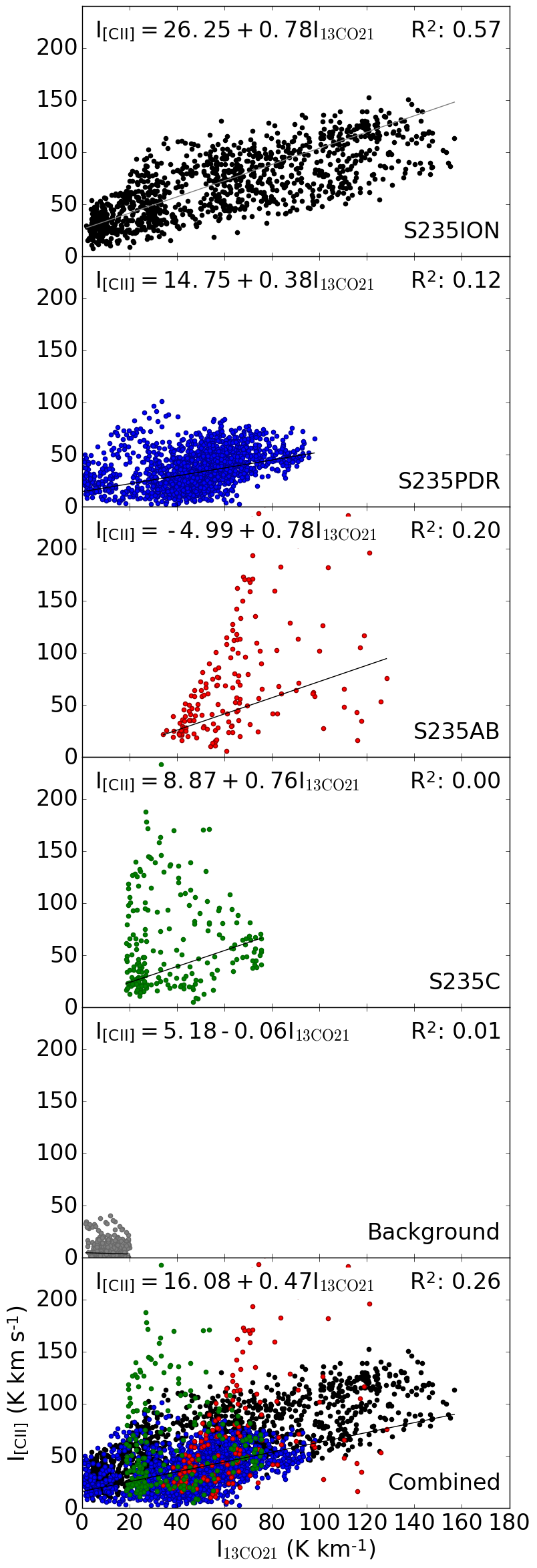}
   \caption{Correlation between CO intensity and that of \cii, for $^{12}$CO $2-1$ (left), $^{12}$CO $3-2$ (middle), and $^{13}$CO $2-1$ (right).  The S235ION region is spatially coincident with the ionized gas of S235, S235PDR with the PDR of S235, S235AB with blended sources S235A and S235B, and S235C with \hii\ region S235C (see Figure~\ref{fig:cii_m0map}).  The large spread of the data points may indicate that a significant amount of the ionized carbon is not associated with CO 
   gas in the S235 star formation complex. \label{fig:corr1}}
\end{figure*}

\end{document}